\pdfoutput=1

\documentclass[useAMS,usenatbib,usegraphicx]{mn2e}

\include{journal}
\usepackage{graphicx}
\usepackage{color}
\usepackage{times}
\usepackage{natbib}
\usepackage{setspace}
\usepackage{rotating}

\newif\ifAMStwofonts
\AMStwofontstrue
\definecolor{red}{rgb}{1,0.,0.}

\title[Combining Physical models with radio observations to constrain the SFRs of high-$z$ dusty SF galaxies]{Combining Physical galaxy models with radio observations to constrain the SFRs of high-$z$ dusty star forming galaxies}
\author[B.~Lo Faro et al.]{B. Lo Faro$^{1,2}$\thanks{E-mail:
barbara.lofaro@lam.fr}, L. Silva$^{3}$, A. Franceschini$^{1}$, N. Miller$^{4}$, A. Efstathiou$^{5}$\\
$^{1}$Dipartimento di Fisica e Astronomia, Universit${\grave{a}}$ di Padova, vicolo Osservatorio, 3, 35122 Padova, Italy\\
$^{2}$Aix-Marseille Universit${\grave{e}}$, CNRS, LAM (Laboratoire d’Astrophysique de Marseille) UMR7326, 13388, France\\
$^{3}$INAF-OATs, Via Tiepolo 11, I-34131 Trieste, Italy\\
$^{4}$Stevenson University, 1525 Greenspring Valley Rd, Stevenson, MD 21153, USA\\
$^{5}$School of Sciences, European University Cyprus, Diogenes Street, Engomi, 1516 Nicosia, Cyprus}

\begin{document}

\date{Received...;Accepted}

\pagerange{\pageref{firstpage}--\pageref{lastpage}} \pubyear{}

\maketitle
\label{firstpage}

\begin{abstract}
We complement our previous analysis of a sample of z $\sim$ 1-2 luminous and ultra-luminous infrared galaxies ((U)LIRGs), by adding deep VLA radio observations at 1.4 GHz to a large dataset from the far-UV to the sub-mm, including \textit{Spitzer} and \textit{Herschel} data. 
Given the relatively small number of (U)LIRGs in our sample with high S/N radio data, and to extend our study to a different family of galaxies, we also include 6 well sampled near IR-selected BzK galaxies at z $\sim$ 1.5.
%
From our analysis based on the rad-tran spectral synthesis code GRASIL, we find that, while the IR luminosity may be a biased tracer of the star formation rate (SFR) depending on the age of stars dominating the dust heating, the inclusion of the radio flux offers significantly tighter constraints on SFR. 
Our predicted SFRs are in good agreement with the estimates based on rest-frame radio luminosity and the Bell~(2003) calibration.
The extensive spectro-photometric coverage of our sample allows us to set important constraints on the SF history of individual objects. For essentially all galaxies we find evidence for a rather continuous SFR and a peak epoch of SF preceding that of the observation by a few Gyrs. This seems to correspond to a formation redshift of z$\sim$ 5-6.
We finally show that our physical analysis may affect the interpretation of the SFR-M$_{\star}$ diagram, by possibly shifting, with respect to previous works, the position of the most dust obscured objects to higher M$_{\star}$ and lower SFRs. 
\end{abstract}


\section{Introduction}
\label{intro}
In star-forming galaxies, stars, gas and dust are mixed in a very complicated way, and dust obscuration strongly depends on their relative geometrical distribution. This is particularly related to the ages of stellar populations, as young stars embedded in dense molecular clouds (MCs) are more extinguished than older stars (e.g. Calzetti, Kinney \& Storchi-Bergmann~1994; Silva et al.~1998; Charlot \& Fall~2000; Poggianti et al.~2001).

New data combining deep optical and far-IR photometric imaging, as discussed in \citeauthor{LoFaro2013}~(2013, BLF13 hereafter)
among many others, then force us to consider relatively realistic models of galaxy synthetic spectra, with radical complications with respect to previous modelling, occurring at two levels. On one side dust
extinction can not be neglected and should be considered as a function of the age of the stellar populations in the galaxy.
In addition, the differential extinction in age and dusty environments entails that geometrical effects in the distribution of stars and dust play a fundamental role in determining the UV to sub-mm spectral energy distribution (SED), and have to be carefully modelled.

By analysing the far-UV to sub-mm properties of a sample of 31 dusty star forming (U)LIRGs at $z\sim$1-2, BLF13 have in fact demonstrated that very idealized approaches, e.g. based on the assumption of a homogeneous foreground screen of dust, single extinction law, and optical-only SED-fitting procedures, may produce highly degenerate model solutions and galaxy SEDs unable to energetically balance the dust reprocessed IR emission from the galaxy. For the most dust obscured objects, as observed by BLF13, this can result in severely underestimated stellar masses. Another crucial aspect of the BLF13 analysis concerns the estimates of the star-formation rates (SFRs) for high-$z$ dusty star forming galaxies.
The classical Kennicutt~(1998) calibration, widely used in literature to estimate SFRs from the rest-frame 8-1000 $\,\mu$m total IR luminosity, is based on the assumption that the bolometric luminosity produced by a 100 Myr constant SF is all emitted in the IR.
The BLF13 typical galaxies, instead, appeared to include significant contributions to the dust heating by intermediate-age stellar populations, i.e. older than the typical escape time of young stars from their parent MCs (in the range between $\sim $\,3 Myr and $\sim$\,90 Myr), and powering the \textit{cirrus} emission. 
This brings to a factor of $\sim$\,2 lower calibration on average of SFR from the total $L_{IR}$, for moderately star-forming galaxies, with respect to the Kennicutt's calibration.
Therefore, although the FIR emission is surely related to the recent star formation rate in a galaxy,
it also includes contributions by all populations of stars heating the dust
(i.e., the cirrus component; \citealt{Helou1986,LonsdalePerssonHelou1987,Hirashita2003,Bendo2010,Li2010}),
as actually confirmed by our analysis. The idea that some or most of the far-infrared and submillimetre emission of submilimetre galaxies may be due to cirrus was first discussed by \citeauthor{ERR03}~(2003; ERR03). ERR03 actually also used the radio data and the far-ir/radio correlation in the fitting. \cite{ESib09} explored this idea further by modelling a sample of sub-mmillimetre galaxies with Spitzer spectroscopy and redshifts and confirmed the idea that cirrus is contributing significantly in the far-infrared.
In addition, a fraction of the UV-optical emission escapes the galaxy and is not registered by the IR luminosity.

The estimates of masses and SFRs for high-$z$ galaxies provide fundamental constraints for models of galaxy formation and evolution.
Notably, these quantities bear on the position of galaxies on the SFR vs stellar mass plane, with galaxies
on the Main Sequence (MS) (e.g. Rodighiero et al.~2011) being interpreted as steadily evolving, in contrast to outliers (off-MS objects) probably undergoing starburst episodes.
Therefore, further independent tests of the star formation histories (SFH) and recent SFR are then required in order to check our interpretation of high-$z$ galaxy SEDs particularly for what concerns the contribution of intermediate age stars, affecting both the estimated stellar mass and recent SFR.
Especially suited to this aim is the radio spectral range.

The radio emission from normal star forming galaxies is usually dominated by the non-thermal component (up to $\sim$\,90\% of the radio flux) which is due to the synchrotron emission from relativistic electrons accelerated into the shocked interstellar medium, by core-collapse Supernova (SN) explosions \citep{Condon1992}. When considering high redshift highly star forming galaxies and starbursts usually the contribution from free-free emission becomes more important and dominates at $\sim$ 30 GHz rest-frame. By looking for spectral slope variations in a sample of sub-mm galaxies at z$\sim$2-3, \cite{Ibar2010} found the same $\alpha \sim 0.7$ between 610 MHz and 1.4 GHz emphasizing that apparently the optically thin synchrotron emission was still dominant at those redshifts.

Only stars more massive than $\sim\,8 M_{\odot}$ produce the galactic cosmic rays responsible for the non-thermal emission. These stars ionize the HII regions as well contributing to the thermal component. 
Radio emission is therefore a probe of the recent star formation activity in normal galaxies.

The inclusion of radio data into our SED-fitting procedure is therefore crucial as the radio continuum offers a further independent way to estimate and constrain the SFR in galaxies completely unaffected by extinction. Moreover as the radio emission probes the SFH on a different timescale 
with respect to the IR emission, this analysis is also useful for our understanding of the SFH of galaxies.
At the same time, including new data directly related to the rate of formation of new stars can help to better constrain the average dust extinction, by comparison with the far-IR and UV spectral information.

Despite the strong effort devoted to comparing the different SFR indicators from X-ray to radio (see e.g. \citealt{Daddi2007a, Daddi2010, Murphy2011b, Kurczynski2012} among the others) and probing the FIR-radio correlation up to high redshift (e.g.\citealt{Ivison2010,Sargent2010,Mao2011,Pannella2014}) there are only few works including radio emission modelling in galaxy evolution synthesis with the aim of unveiling the nature of different galaxy populations. Most of them are based on the more classical and semi-empirical multi-component SED-fitting procedure where the different parts of the SED (optical-NIR, MIR-FIR, radio) are modelled individually assuming for the radio the FIR-radio correlation and with the main aim of disentangling the AGN contribution from that of SF dominated sources (see e.g. \citealt{Vivian2012}).   

The work by \cite{Bressan2002} was mainly focused on understanding the nature of the FIR-Radio correlation, particularly for a local sample of obscured starbursts, by making use of state-of-the-art models of star forming galaxies. They used GRASIL to model the IR emission of galaxies and extended its predictions beyond the sub-millimetre regime, up to 316MHz. They showed that the delay of the non-thermal emission with respect to the IR and thermal radio can give rise to observable effects in the IR to radio ratio as well as in the radio spectral slope, potentially allowing the analysis of obscured starbursts with a time resolution of a few tens of Myr, unreachable with other SF indicators.

\cite{Vega2008} also included radio modelling (with at least 3 data points in the radio regime) in their analysis of a sample of 30 local (U)LIRGs, with the aim of studying their starburst nature and disentangling the possible AGN contribution. The well-sampled radio spectra allowed them to put strong
constraints on the age of the burst of star formation. 

This work represents the first attempt to apply this sophisticated physical analysis, including the modelling of radio emission, to well sampled (from far-UV to radio) SEDs of high-$z$ dust obscured SF galaxies. We investigate the effect of accounting for radio data in the SED-fitting procedure on the derived main physical properties of galaxies, focusing in particular on the constraints on the current SFR, SFH and consequently M$_{\star}$ of galaxies. Our analysis points towards the radio flux as an essential information for interpreting star forming galaxies at high redshifts and for recovering reliable SFH.

With the above purposes, in this work we have complemented the mid IR-selected sample studied in BLF13 by including into our SED analysis new Very Large Array (VLA\footnote{The National Radio Astronomy Observatory is a facility of the National Science Foundation operated under cooperative agreement by Associated Universities, Inc.}) observations of the integrated $1.4$ GHz emission of galaxies in the GOODS-South field. Both because of the small number of (U)LIRGs in our sample with high S/N radio data (4/31), and to extend our analysis also to a different family of galaxies, we have included in our study the 6 well sampled near IR-selected BzK galaxies by Daddi et al.~(2010) at $z\sim 1.5$ in the GOODS-North field.
The best fit model for each observed galaxy is searched within a GRASIL-generated library of $\sim 10^6$ models, covering a huge range of star formation histories and dust model parameters, with model SEDs covering self-consistently the UV to radio spectral range.
The FIR and radio luminosities as star formation indicators are then discussed by making use of the best fit SFRs from our thorough SED modelling, and by comparing to the most widely adopted relations by Kennicutt~(1998), derived from stellar population modelling, and Bell~(2003), based on the FIR-radio correlation.
Finally we discuss the consequences of our inferred values of stellar mass and SFR on the Main Sequence relation.

The paper is organized as follows.
In Section \S \ref{data} we describe the (U)LIRGs and BzK data sample we use for our analysis, including the new VLA data.
The physical model and SED-fitting procedure are presented in Section \S \ref{grasiltop}.
The results concerning the best-fit far-UV to radio SEDs are discussed in Section \S \ref{results}, while we compare with the observed FIR-radio relation
in Section \ref{qtirsection}. In Section \S \ref{radioSFR} the radio and FIR as SFR indicators are discussed, and the implications of our models for the SFR-stellar mass relation
are presented in Section \ref{sfr-m-section}. Our Summary and Conclusions are presented in Section \S \ref{conclusions}.

The models adopt a $0.1-100$ $M_\odot$ Salpeter~(1955) IMF. Where necessary, and indicated on the text, we scale to a Chabrier~(2003) IMF by dividing masses by a factor $1.7$.
The cosmological parameters we adopt assume H$_{0}$=70 km s$^{-1}$ Mpc$^{-1}$, $\Omega_{\Lambda}$=0.7, $\Omega_{M}$=0.3.

\section{Data Sample \& Observations}
\label{data}

This work is intended to be complementary to the physical analysis performed in our previous paper (BLF13) and aimed at testing BLF13 estimates, in particular those concerning $M_{\star}$ and $SFR$, against observations through the investigation of the radio emission of high-$z$ star forming galaxies. For this reason we first apply our analysis to the same sample of 31 high-$z$ (U)LIRGs whose far-UV to sub-mm properties have been deeply investigated in BLF13 and then, in order to strengthen our conclusions, we extend the analysis by including a small but well representative sample of 6 BzK-selected star forming main sequence (MS) galaxies at z$\sim$1.5 for which also high S/N radio observations are available. These two sample benefit from a full multiwavelength coverage from far-UV to Radio including Spitzer and Herschel (PACS+SPIRE) data. The two sample are briefly presented in Subsection~\ref{faddadata} and \ref{bzk}.

\subsection{The (U)LIRG data sample}
\label{faddadata}

The sample of high-$z$ (U)LIRGS whose radio properties are
investigated here was originally selected from the sample of
IR-luminous galaxies in GOODS-S presented by
\citeauthor{Fadda2010}~(2010 F10 hereafter). It includes the
faintest $24$ $\mu$m sources observed with the Spitzer Infrared
Spectrograph (IRS) ($S_{24} \sim 0.15-0.45$ mJy) in the two
redshift bins ($0.76-1.05$ and $1.75-2.4$). These galaxies have
been selected by F10 specifically targeting LIRGs at $z\sim1$
and ULIRGs at $z\sim2$. This sample is therefore crudely
luminosity selected and F10 did not apply any other selections.

As emphasized by F10, except for having higher dust
obscuration, these galaxies do not have extremely deviant
properties in the rest-frame UV/optical compared to galaxies
selected at observed optical/near-IR band. Their observed
optical/near-IR colors are very similar to those of extremely
red galaxy (ERG) populations selected by large area K-band
surveys. Moreover, all the $z\,\sim$2\, ULIRGs are IR-excess
BzK ((z-B) vs (K-z):\cite{Daddi2004}) galaxies and most of them
have $L_{\mathrm{FIR}}$/$L_{\mathrm{1600 \AA}}$ ratios higher than those of
starburst galaxies at a given UV slope. The ``IR excess'' is
mainly due to the strong PAH contribution to the mid-IR
luminosity which leads to overestimation of $L_{\mathrm{IR}}$
and to underestimation of the UV dust extinction.

The sample studied here includes only galaxies powered by star
formation. All the objects which were classified by F10 as
AGN-dominated, on the basis of several indicators such as broad
and high ionization lines in optical spectra, lack of a $1.6\
\mu$m stellar bump in the SED, X-ray bright sources, low mid-IR
$6.2\ \mu$m EW and optical morphology (see also
\citealt{Pozzi2012}), were excluded by BLF13 from the original
sample. With the further requirement of a full complementary
optical/near-IR photometry provided by the multiwavelength
MUSIC catalogue of \citep{Santini2009} the final sample
consists of 31 (U)LIRGs among which 10 are at $z\sim1$ (all in
the nominal LIRG regime) and 21, mostly ULIRGs, are at
$z\sim2$.

All the 31 sources in the GOODS-S field have the advantage of a
full FIR coverage from $70$ to $500$ $\mu$m from both
\textit{Herschel} SPIRE and PACS instruments with the SPIRE and
PACS data taken, respectively, from the Herschel Multi-tiered
Extragalactic Survey (HerMES: \citealt{Oliver2012}) and the
PACS Evolutionary Probe (PEP: \citealt{Lutz2011}) programs.
Typical noise levels are $\sim 1$ mJy for PACS $70-160$ $\mu$m
and $\sim 6$ mJy for SPIRE $250-500$~$\mu$m, including
confusion.

\subsubsection{Radio observations}
\label{radioulirgs}

In addition to the availability of a spectroscopic redshift and complete FIR coverage we also looked for VLA detections at 1.4 GHz. As already pointed out above, the inclusion of radio emission into our procedure has been revealed to be fundamental to constrain the `current' SFR in galaxies, mostly contributed by very young stars, and their SFH.

\begin{figure}
\centerline{
\includegraphics[width=8.cm]{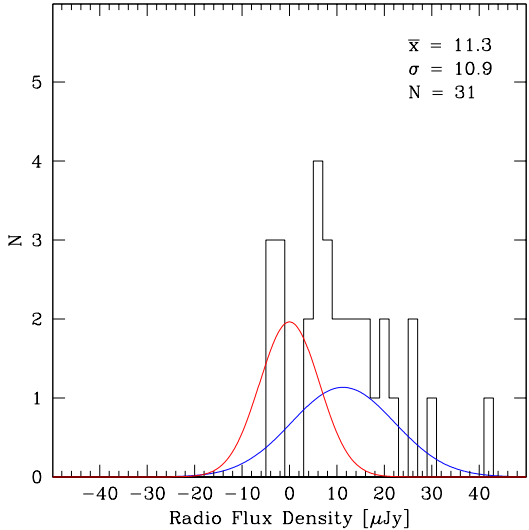}}
\caption[]{Histogram showing the measured radio flux densities at the indicated positions of (U)LIRGs. It clearly shows that the expectation of a Gaussian centred on zero and with a dispersion of the image noise (red histogram) is very different from the actual measured distribution.}
\label{histoMiller}
\end{figure}

The publicly available VLA catalogue in GOODS-S of \cite{Miller2013} includes only three of the 31 objects of our (U)LIRG sample. Inclusion in this catalogue required a point-source detection of $> 5 \sigma$, a fairly conservative but well-established limit for radio catalogues. We have extracted radio flux densities for the remaining sources directly from the released radio mosaic image, adopting the flux density at the position of the (U)LIRG coordinates. This measurement, in units of $\mu$Jy per beam, is a reasonable representation of the flux density of any point source at the location corresponding to the (U)LIRG coordinates. The majority of the (U)LIRGs are associated with positive flux densities at lower S/N than required for the formal published catalog, as seen in Figure~\ref{histoMiller}. In fact, we find one additional source for which S/N $> 4 \sigma$ and we include this source in our detections presented in Table~\ref{miller} which contains the coordinates of the (U)LIRGs, the measured flux densities in $\mu$Jy at those coordinates, and the RMS of the radio mosaic at those coordinates. 

As a variation on the histogram shown in Figure~\ref{histoMiller}, we have also performed a radio stacking analysis on the 31 (U)LIRGs. This involved taking cutouts of the radio mosaic image centered on each of the 31 (U)LIRG coordinates and combining them to evaluate a single statistical representation of the full population. When performing the stack using an average with rejection of the single highest and single lowest value (``minmax'' rejection), we recovered a significant detected source having a peak flux density of 11.6 $\mu$Jy and an RMS of 1.2 $\mu$Jy. A stack produced using a straight median instead of the average yielded consistent results. To check the validity of this stack detection, we created a list of 1000 random positions in the vicinity of the (U)LIRG coordinates and evaluated a stack based on these coordinates. At the central pixel of this stack, the measured flux density was 8 pJy per beam (i.e., effectively zero) and the RMS in the stack image was 0.2 $\mu$Jy. Thus, we are confident that the (U)LIRG population represents real radio emission with flux densities only occasionally above the threshold used in the formal \cite{Miller2013} catalogue.

\begin{table}
\begin{tabular}{|r|l|r|r|r|r|r|}
\hline
  \multicolumn{1}{|c|}{ID} &
  \multicolumn{1}{c|}{name} &
  \multicolumn{1}{c|}{$z$} &
  \multicolumn{1}{c|}{RA} &
  \multicolumn{1}{c|}{DEC} &
  \multicolumn{1}{c|}{$S_{1.4 GHz}$} &
  \multicolumn{1}{c|}{RMS} \\
\hline
    &      &        &                  &                  & [$\mu$Jy] & [$\mu$Jy]\\ 
\hline    	
  8 & U4812 & 1.930 & 53.19827 & -27.74786 & 42.6 & 6.3\\
  14 & U5152 & 1.794 & 53.05226 & -27.71833 & 26.7 & 6.5\\
  17 & U5652 & 1.618 & 53.07268 & -27.83420 & 25.2 & 6.3\\
  33 & L5420 & 1.068 & 53.02496 & -27.75204 & 29.6 & 6.2\\
\hline\end{tabular}
\caption{Measured VLA 1.4 GHz flux densities at the coordinates of
the 4 (U)LIRGs detected at S/N $> 4 \sigma$.}
\label{miller}
\end{table}

When modelling the radio data we have used the following
prescriptions to deal with the errors: (i) for the objects
having $S_{1.4 GHz}/RMS \ga 3\,\sigma$ we have used the same
error as given in the catalogue, usually of the order of
$6.1-6.3\, \mu Jy$; (ii) for those objects having instead
$2\,\sigma\,\la S_{1.4 GHz}/RMS \la 3\,\sigma$ we have assumed
a $2\,\sigma$ error-bar associated to the measured flux
density; (iii) finally we consider as $3\,\sigma$ upper limits
all the sources having negative average flux densities measured
at the position of the source and/or $S_{1.4 GHz}/RMS \la
2\,\sigma$.

We note that the presence of a single data point in the
radio part of the spectrum compared to the larger number of
data available in the far-UV to FIR regime does not affect
significantly the determination of the best-fit model in terms
of its $\chi^{2}$ value but it plays a crucial role in
constraining the SFH of the galaxy. What we are interested in, in
fact, is determining if our best-fit solutions are able to
reproduce, within the uncertainties, the observed radio flux
densities. Thus, we intend to verify if the amount of massive young
stellar populations dominating the most recent SFR is well
constrained by our model as well as the fraction of stars
outside the MCs contributing to cirrus heating.

\subsection{BzK data sample}
\label{bzk}

The six BzK galaxies analysed here have been selected from the original sample studied in \citeauthor{Daddi2010}~(2010, D10 hereafter). 
These are NIR-selected galaxies, in GOODS-N, with K $<$ 20.5 (Vega
scale; or K $<$ 22.37 AB), to which the BzK color selection
criterion of \cite{Daddi2004} has been applied, together with
the requirement of a detection in deep Spitzer $24\,\mu\,m$
imaging, in order to select a sample of star forming galaxies
at $1.4 < z < 2.5$. This small sample of BzK galaxies has the
advantage of having, in addition to the availability of a
spectroscopic redshift, also a very rich photometric dataset
including full far-UV to NIR coverage, Spitzer (IRAC, MIPS + 16
$\mu\,m$ InfraRed Spectrograph peak-up image
\cite{Teplitz2011}, \cite{Pannella2014}) and Herschel (from both PACS and SPIRE)
observations and high S/N radio detections at 1.4 GHz \citep{Morrison2010}.
All the BzKs have redshift in the range 1.4\,$\la$\,$z$\,$\la$\,1.6, that is in the lowest redshift regime probed by BzK selection. This is due to the requirement, by \cite{Daddi2010}, of a radio detection coupled with the choice of observing the CO[2-1] transition. Its rest-frame frequency of 230.538 GHz can be observed with the PdBI only up to $z=1.87$.

Among the six targeted galaxies, five redshifts were obtained through the GOODS-N campaigns at Keck using DEIMOS (D. Stern et al. 2010). The redshift for \textit{BzK-12591}
was instead derived by \cite{Cowie2004}. This galaxy, showing a strong bulge in the Hubble Space Telescope (HST) imaging, has also a possible detection of $[NeV]\lambda\,3426\,\AA$ emission line, suggesting the presence of an AGN. 

For these objects, thanks to the availability of data from MIR to mm wavelengths, \cite{Magdis2012} have recently provided robust estimates of their dust masses, ($LogM_{dust}$ in the range 8.52-9.11),
based on the realistic models by \cite{DraineLi2007}, independent estimates of the CO-to-H$_{2}$ conversion factor ($\alpha_{CO}$), molecular gas masses and SFEs
by exploiting the correlation of gas-to-dust mass with metallicity ($M_{gas}$/$M_{dust}-Z$).
The SFRs have been computed using the information coming from the full coverage from MIR to sub-mm offered by Herschel data and therefore
do not suffer from the same bias of SFR estimates from 24 $\mu\,m$ only. It has been shown, in fact, that the $L_{IR}$ from 24 $\mu\,m$ can
be up to a factor $\sim$\,2 higher than the `real' $L_{IR}$ from Herschel (see e.g. \citealt{Oliver2012}, Canalog et al.~2013 in prep.).

The rich suite of empirical estimates of the main physical parameters of these galaxies, provided by both D10 and M12, allows us to put strong constraints on our best-fit solutions.
In particular in this paper we will focus only on the results concerning the radio emission, the SFR and the SFH of these objects. A full physical characterization of their molecular gas and dust properties including a detailed comparison of our predictions to empirical estimates based on observations is matter of a forthcoming paper (Lo Faro et al., in prep.).


\section{SED modelling with GRASIL}
\label{grasiltop}

The approach used here to physically characterize these galaxy populations and in particular their radio emission, is based on galaxy evolution synthesis technique.

When modelling the SEDs of star forming galaxies, dust effects
become crucial, particularly at high redshift, so we need to
include an appropriate dust model accounting for both the
absorption and thermal re-emission from dust. Several works dealt with the radiative transfer (RT) in spherical geometries, mainly aimed at modelling starburst galaxies (e.g. Rowan-Robinson~1980; Rowan-Robinson \& Crawford~1989; Efstathiou, Rowan-Robinson \& Siebenmorgen~2000; Popescu et al.~2000; Efstathiou \& Rowan-Robinson~2003; Takagi, Arimoto \& Hanami~2003a; Takagi, Vansevicius \& Arimoto~2003b; Siebenmorgen \& Krugel~2007; Rowan-Robinson 2012). Early models of this kind did not include the evolution of stellar populations. Silva et al.~(1998) were the first to couple radiative transfer through a dusty ISM and the spectral (and chemical) evolution of stellar populations. 

To model the emission from stars and dust consistently in order to get
reliable estimates of the main physical parameters of galaxies
(stellar mass, average extinction, SFR etc..), we need to solve
the radiative transfer equation for idealized but realistic
geometrical distributions for stars and dust as well as taking
advantage of a full multiwavelength coverage from far-UV to
radio. The GRASIL spectrophometric code (\citealt{Silva1998,Silva1999,Silva2011})
satisfies all these requirements. For a detailed description of the code we defer
the reader to the original works. Here we provide a brief
summary of its main features relevant to our work.

\subsection{GRASIL main features} 
\label{grasil}

GRASIL is a self-consistent physical model able to predict the SEDs of galaxies from far-UV to radio including a state-of-the-art treatment of dust
extinction and reprocessing based on a full radiative transfer solution. It computes the radiative transfer effects for three different dusty environments:
%
(i) dust in interstellar HI clouds heated by the general interstellar radiation field (ISRF) of the galaxy (the ``cirrus'' component),
(ii) dust associated with star-forming molecular clouds and HII regions (dense component) and
(iii) circumstellar dust shells produced by the windy final
stages of stellar evolution.
%

It accounts for a realistic geometry where stars and dust are
distributed in a bulge and/or a disk profiles. In the case of
spheroidal systems, a spherical symmetric distribution with a
King profile is adopted for both stars and dust, with core
radius $r_{c}$.


Disk-like systems are modelled by using a double exponential of
the distance from the polar axis, with scale radius $R_{d}$,
and from the equatorial plane with scale height $z_{d}$.


In the following, we assign the same scale lengths to the
stellar and dust distributions.


The clumping of young stars and dust within the diffuse medium
together with the accounting for a realistic geometrical
distribution for stars and dust, gives rise to an age-dependent
dust attenuation which is one of the most important feature of
this approach as widely discussed in BLF13 (see also
\citealt{Granato2000, Panuzzo2007}). The dust model consists of
grains in thermal equilibrium with the radiation field, and
small grains and polycyclic aromatic hydrocarbon (PAH)
molecules fluctuating in temperature.

\subsection{Input Star Formation Histories}
\label{chehisto}

The input star formation histories are computed with $CHE\_EVO$
(Silva 1999), a standard chemical evolution code which provides
the evolution of the SFR, M$_\mathrm{gas}$ and metallicity,
assuming an IMF, a SF law $SFR(t)=\nu_{\mathrm{Sch}} \cdot
M_{\mathrm{gas}}(t)^{\mathrm{k}} + f(t)$, (i.e. a Schmidt-type
SF with efficiency $\nu_{\mathrm{Sch}}$ and a superimposed
analytical term to represent transient bursts), and an
exponential infall of gas ($dM_{\mathrm{inf}}/dt \propto
\exp(-t/\tau_{\mathrm{inf}})$).

\begin{figure}
\centerline{
\includegraphics[width=8.3cm]{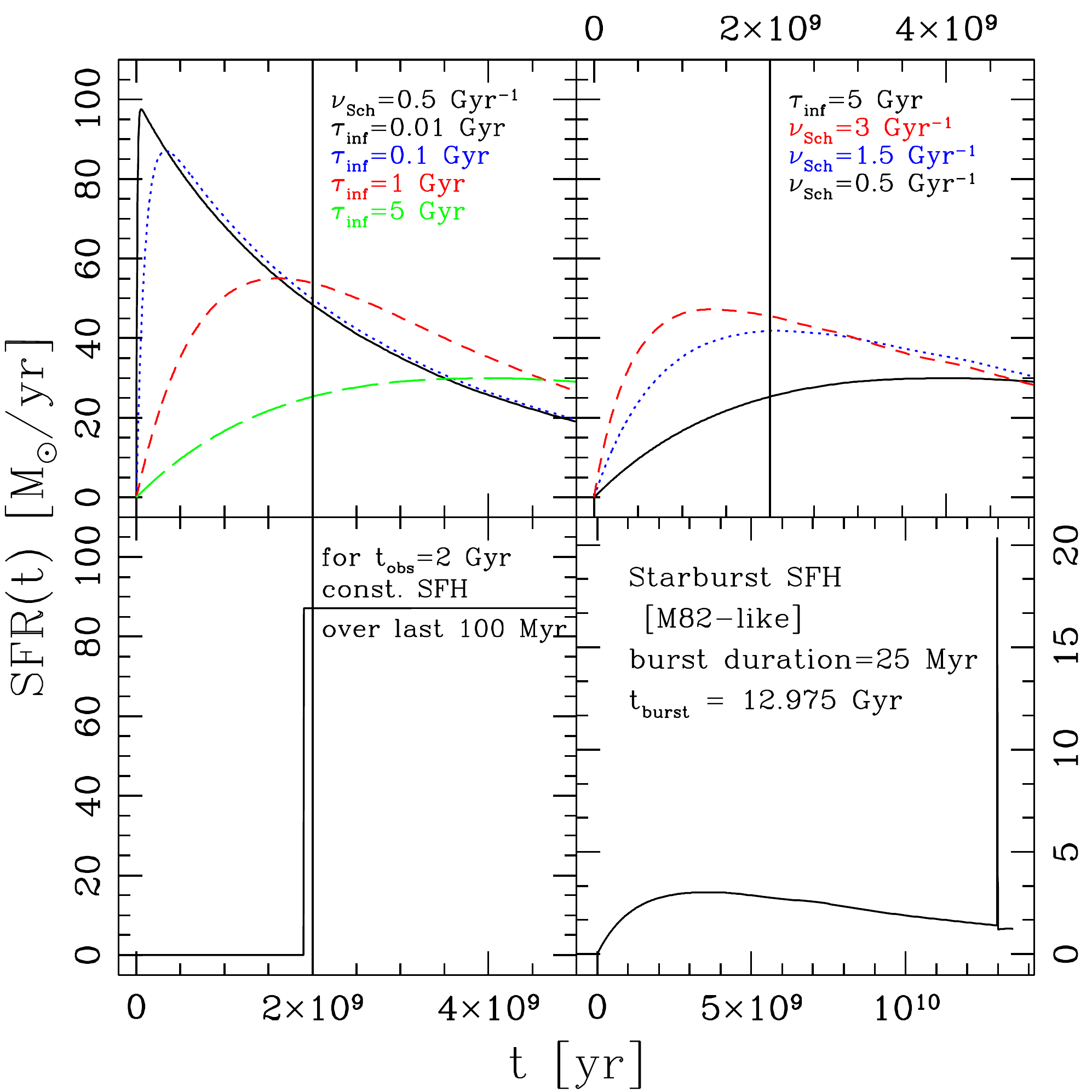}}
\caption{SFHs from chemical evolution code: from top-left to bottom-right we show, as examples several SFHs obtained with our chemical evolution code. Top: exponentially declining SFHs at fixed $\nu_{Sch}$ for increasing values of $\tau_{inf}$ (left), then at fixed $\tau_{inf}$ and decreasing values of $\nu_{Sch}$ (right). When we consider longer infall timescales and lower efficiencies our SFHs assume the form of a classical delay-$\tau$ model (see text for details). When, instead a shorter infall timescale is considered, a more rapidly declining SFH resembling the so-called $\tau$-models, is obtained. Bottom-left panel shows a constant SFR while the bottom-right panel shows a typical `starburst' SFH. Solid black vertical lines highlight an assumed observing time of 2 Gyr. In the bottom-right panel the time at which the galaxy is observed is highlighted in the figure.}
\label{sfhcheevo}
\end{figure}

By varying the two parameters, $\nu_{\mathrm{Sch}}$ and
$\tau_{\mathrm{inf}}$ we are able to recover a wide range of
different SFHs, from smooth ones for large values of
$\tau_{inf}$, to `monolithic-like' ones characterized by very
short infall timescales. A very short $\tau_{inf}$ can be used
to have the so-called close box chemical evolution model, which
ensures that the gas going to form the galaxy is all available
at the beginning. Figure \ref{sfhcheevo} shows some examples of
the possible SFHs which can be implemented in the chemical
evolution code.

In the Figure it can be seen that by coupling the gas accretion
phase with the depletion due to star formation we get SFHs
which closely resemble, in their functional form, the so-called
delay $\tau$-models of \cite{Lee2010} (see also the seminal
work by \cite{Sandage1986}). These are in fact usually
characterized by an early phase of rising SFRs with late-time
decay ($SFR \propto t^{\beta} e^{-t/\tau}$ according to
Lee+2010 formalism) and are probably the most suitable to
explain the SEDs of high-redshift galaxies and also some local
galaxies \citep{Gavazzi2002}.

The average SFHs of the 31 (U)LIRGs whose radio properties are analysed here were first presented in our previous work by BLF13 and are now discussed in more detail in Sec.~\ref{sedandsfhs} and Fig.~\ref{sfhulirgs} where the best-fit SFH of each object is shown.
The SFHs of the six SF BzK galaxies at $z\sim1.5$ are instead discussed in Sec.~\ref{sedsfhBzK}.

In the following we adopt $k=1$, $f(t)$ exponential, and a Salpeter IMF which is the default choice for the chemical evolution code. When required, we convert to the Chabrier IMF by dividing by $1.7$.

Our reference library of SSPs is from \citet{Bressan1998, Bressan2002}, which directly includes the effects of dusty envelopes around AGB stars and the radio emission, as described
below.

\subsection{Radio emission in GRASIL}
\label{radioemingrasil}

In the SSP models by \citeauthor{Bressan2002}~(2002, B02 hereafter) and \cite{Vega2008}, which we adopt here, the radio emission is assumed to be the sum of two different contributions: thermal emission from free electrons in HII regions, and synchrotron emission from relativistic electrons accelerated by Core-Collapse Supernova (CCSN) explosions \citep{Condon1990, Condon1992}.

The thermal component is assumed to be proportional to the number of H ionizing photons directly derived from the stellar populations and thus dependent on their age, metallicity and IMF, and scales as $\approx \nu^{-0.1}$ (see Eq.~1 in B02).

The non-thermal (NT) radio emission is assumed to be proportional to the CCSN rate (see Section~3 and Equations 3, 5 and 17 in B02 for the demonstration of this assumption). It is computed by calibrating the total NT emission/CCSN (type II SNe) rate relation (proportional to the average luminosity per supernova event) on the observed properties of our Galaxy, thus using the MW type II SN rate and total synchrotron luminosity, and accounting for the $\approx 6\%$ contribution by SN remnants (SNR) (Equations 5 and 17 in B02). The CCSNe rate is directly provided by the SSPs, since it represents the death rate of stars more massive than 8 $M_{\odot}$, i.e. $\nu_{CCSNe}=\Phi(m_{d}) \cdot dm_{d}/dt$, with $m_{d}(t)$ being the initial mass of the dying star in the stellar population of given age t.

As the calibration of B02 depends on the time sampling of the SSPs via $dm_{d}/dt$, this was later increased by a 30\% due to a finer re-sampling of the time steps, in order to maintain the same radio emission under the same conditions (private communication).

The model has been shown to reproduce very well the FIR-to-radio correlation of normal star forming galaxies, namely $q_{1.49 GHz}=2.3$, as well as the radio emission of local starburst and (U)LIRGs.

We emphasize that, in our models, the SFR depends linearly on the gas fraction (Schmidt law), while the FIR emission depends on the SF, gas fraction and metallicity history, and the SN rate on the recent SFR. The consistency between FIR emission, radio emission and supernova rate is thus remarkable and should be considered as a successful test of the model (Bressan et al. 2002).

\section{RESULTS}
\label{results}

The models used to interpret the observed SEDs are selected from different libraries self-consistently generated with GRASIL. Each library includes thousands of spectra corresponding to different combinations of model parameters. The total number of models available is $\sim 10^{6}$. Depending on the geometry (spheroid or disc) and on the assumed SFH (with or without burst), the number of free parameters, (described in detail in BLF13), typically ranges between 6 and 9. The number of photometric data-points is always larger than 18.

The best-fit solution is obtained through a standard $\chi^2$ minimization procedure by comparing the total observed SED of a given galaxy to the set of model spectra: 

\begin{equation}
\chi^2=\frac{1}{n-1}\sum_{i=1}^{n} \frac{(F_{mod}(i)-F_{obs}(i))^{2}}{\sigma^{2}(i)}
\end{equation}

where $F_{mod}(i)$, $F_{obs}(i)$, and $\sigma^{2}(i)$ are the model and observed flux densities and the observational errors, respectively. $n$ is the number of photometric data-points used for the fit. Since $n$ is always greater than the number of free parameters of the model, the best fit SED is always well constrained.
We want to stress here that based on the $\chi^2$ minimization procedure we get our best-fit to the total SED, not just a part of it. In this way those wavelengths whose fluxes are characterized by very low uncertainties (as in the optical-NIR up to 24 micron) weight more in the fit in the sense that if the fit is globally very good but around the observed 24 micron, where you can have the effect of PAH on the MIR SED, the fit is not perfect the resulting value of the $\chi^2$ is larger. Within our procedure we require the fit  to be good over the entire wavelength range, not just in one region of the observed SED, without applying any different weight to the different photometric bands.

When spectral information is also available, as for example the
IRS spectra in our case, the best-fit solution to the
photometry only is then compared to the spectrum and if
necessary the specific parameters influencing that region of
the spectrum are fine-tuned in order to reproduce both the
photometry and the spectrum. This is done for all the solutions
having a reliable $\chi^{2}$ value.

The best-fit model provides the physical parameters of the
galaxies, in particular their gas, dust and stellar mass,
instantaneous SFR and SFH, optical depth and attenuation.

\subsection{Interpretation of $z\,\sim\,1$ LIRGs and $z\,\sim\,2$ (U)LIRGs}
\label{interpretation}

As clearly stated in Sec.~\ref{radioulirgs} our major interest here is determining whether the best-fit solutions obtained for the overlapping galaxy sample, (z$\sim$1-2 (U)LIRGs), in our previous paper (BLF13) are able to reproduce, within the uncertainties, the observed radio flux densities. In other words we want to check if the amount of massive young stellar populations dominating the most recent SFR is well constrained by our model as well as the fraction of stars outside the MCs contributing to cirrus heating.
Given the dominant role of cirrus emission to the $L_{\mathrm{IR}}$ of BLF13 (U)LIRG SEDs, it is important to understand if the predictions for the FIR are consistent with those from the radio emission and if the conspicuous cirrus component contributing to the FIR is an `effect' of a poor parameter exploration or if it is real, and required in order to reproduce the NIR-to-FIR properties of our galaxies.

So given the best-fit SEDs computed in BLF13 using the procedure described in Sec.~\ref{results} and corresponding to the set of model parameters discussed in BLF13  we have simply added to the photometric data-points the radio flux densities and extended the fit to the radio.

Below we give a very brief summary of the main results obtained in BLF13.

According to BLF13 all these galaxies appear to require
massive populations of old ($>1$\ Gyr) stars and, at the same
time, to host a moderate ongoing activity of SF with typical
SFRs $\la$ 100\ $M_{\odot}$/yr. The bulk of the stars appear to
have been formed a few Gyr before the observation in
essentially all cases (see also Fig.~\ref{sfhulirgs}). 
Their average extinctions and stellar masses are found
to be higher with respect to estimates based on optical-only
SED-fitting procedures. In particular the stellar mass
difference is found to be larger for the most dust obscured
(U)LIRGs at $z\sim2$ for which it reaches a factor of $\sim$6
for the $z\sim1$ and $z\sim2$ (U)LIRGs, respectively). The
predicted SFRs are found to be lower than those computed from
$L_{\mathrm{IR}}$ using the Kennicutt relation due to the
significant contribution to the dust heating by
intermediate-age ($t_{obs}\,\ga\,10-90\,Myr$) stellar
populations through `cirrus' emission ($\sim$73\% and
$\sim$66\% of total $L_{\mathrm{IR}}$ for $z\sim1$ and $z\sim2$
(U)LIRGs, respectively).
 

\subsubsection{Best-fit far-UV-to-Radio SEDs \& SFHs}
\label{sedandsfhs}
Figure~\ref{best-fit-radiol1} shows the GRASIL best-fits to the far-UV-to-Radio SEDs of our $z\,\sim\,1$ LIRGs and $z\,\sim\,2$ (U)LIRGs listed according to their ID. The $3\sigma$ upper limits are indicated as red arrows while the insets report the fit to their IRS MIR spectra (discussed in detail in BLF13). 
The fit obtained in our previous work reproduces, well within a factor of two, the far-UV to radio emission for almost all the (U)LIRGs into our sample without re-fitting. 28/31 (U)LIRGs show modelled radio fluxes within data error-bars.

\begin{figure*}
\centerline{
\includegraphics[width=19.5cm,height=21.cm]{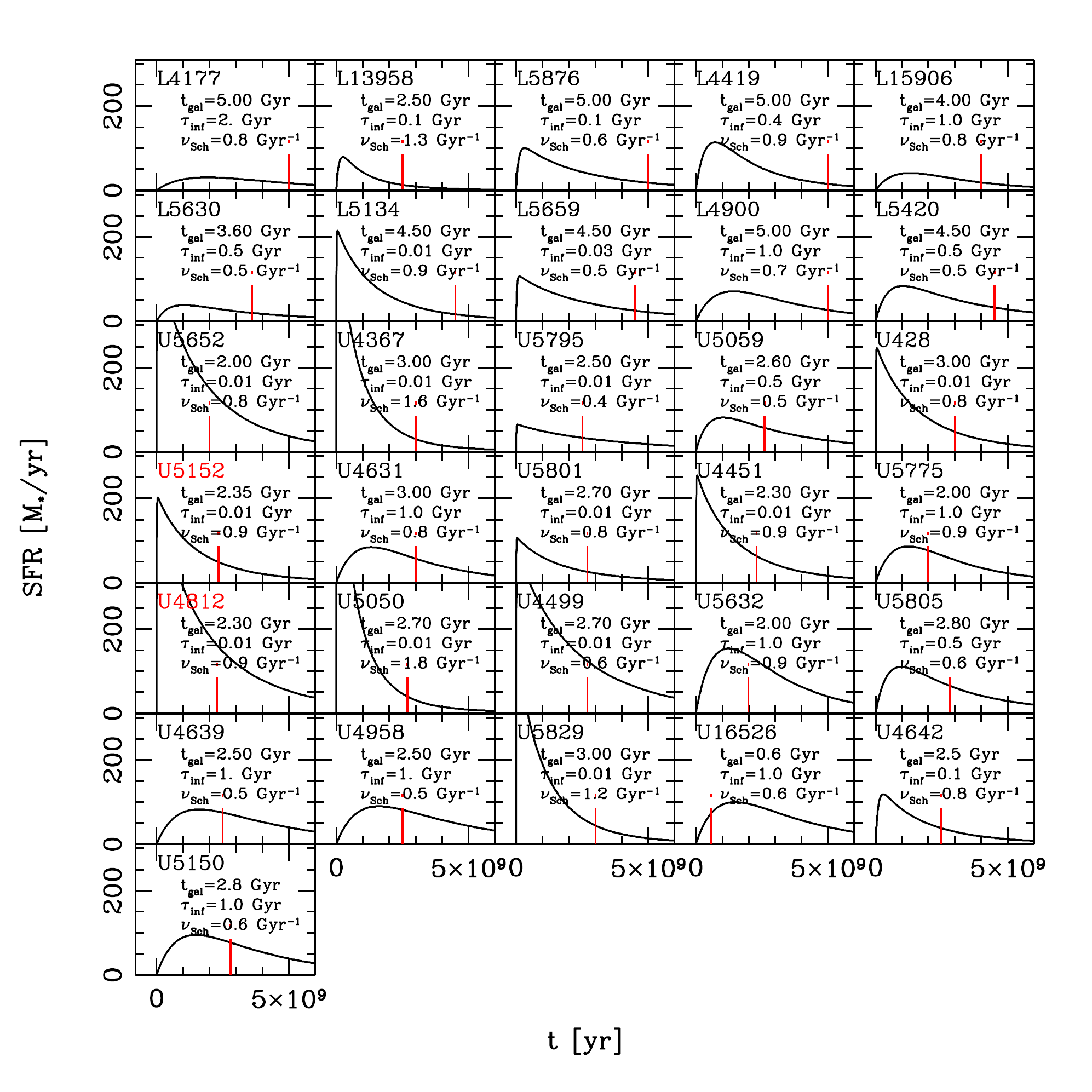}}
\caption{Best-fit SFHs of $z\,\sim$\,1-2 (U)LIRGs. The long dashed red vertical lines indicate the age at which the galaxy is observed. In each panel the two parameters ruling the SFH are also specified, namely the infall timescale and SF efficiency. The two objects labelled in red are those for which the inclusion of radio data substantially changed the SFH (see Fig.~\ref{u4812-u5152-sfh} for details).}
\label{sfhulirgs}
\end{figure*}

The inclusion of radio data into our models thus seems to confirm our physical solutions. In particular, as discussed in BLF13, given the detailed shape of the broadband SED our physical analysis appears to be able to give important hints on the main parameters ruling the source's past SFH, i.e. $\tau_{\mathrm{inf}}$ and $\nu_{\mathrm{Sch}}$ as shown in Fig.~\ref{sfhulirgs}.

In BLF13 we investigated both SF models with and without a starburst on top of the Schmidt-type part of the SF law (see fig. \ref{sfhcheevo}). For the majority of our (U)LIRGs, a suitable calibration of the $\tau_{\mathrm{inf}}$ and $\nu_{\mathrm{Sch}}$ allowed us to obtain good fits to the observed SEDs with the continuous models. Figure~\ref{sfhulirgs} shows the best-fit SFH obtained for each single object in our sample and the relative $\tau_{\mathrm{inf}}$ and $\nu_{\mathrm{Sch}}$. The vertical red line highlights the time, in Gyr, at which the galaxy is observed t$_{gal}$. 

Small values for $\tau_{\mathrm{inf}}$ (in the range 0.01-0.1 Gyr) and high values for $\nu_{\mathrm{Sch}}$ (in the range 0.8-1.4 Gyr$^{-1}$), corresponding to an early fast and efficient SF phase, are required for 16/31 objects, 4 LIRGs and 12 z$\sim$2 (U)LIRGs. These are also the objects showing the strongest stellar bump in the rest-frame near-IR (see for example \textit{L5134} and \textit{U4367} in fig.~\ref{best-fit-radiol1}). Smoother SFHs characterised by longer $\tau_{\mathrm{inf}}$ (ranging between 0.4 and 1.0 Gyr) are instead required for the 6 remaining z$\sim$1 LIRGs and 9/21 z$\sim$2 (U)LIRGs. These galaxies present almost `flat' rest-frame NIR bands and higher UV fluxes.

As already discussed in BLF13 and shown here in Fig.~\ref{sfhulirgs} all our (U)LIRGs appear to include massive populations of old ($>1$\ Gyr) stars with the bulk of stars formed a few Gyr before the observation in essentially all cases. This seems to correspond to a formation redshift of z$\sim$ 5-6. Average estimates can be inferred from Fig~\ref{sfhulirgs}: for the 12 z$\sim$ 2 (U)LIRGs characterised by very peaked SFHs, ($\tau_{\mathrm{inf}} \sim 0.01$), about 66-80\% of the stellar mass has formed within $\sim$ 1 Gyr from the beginning of their star formation activity, proportionally higher at higher SF efficiency. For the z$\sim$2 (U)LIRGs presenting, instead, more regular SFH about 30-43\% of stellar mass is formed within 1 Gyr. Finally for the LIRGs, on average, $\sim$ 28\% of the stellar mass has already been formed within the first 1 Gyr.
%
%
%
%
%
\begin{figure*}
\centerline{
\includegraphics[width=6cm]{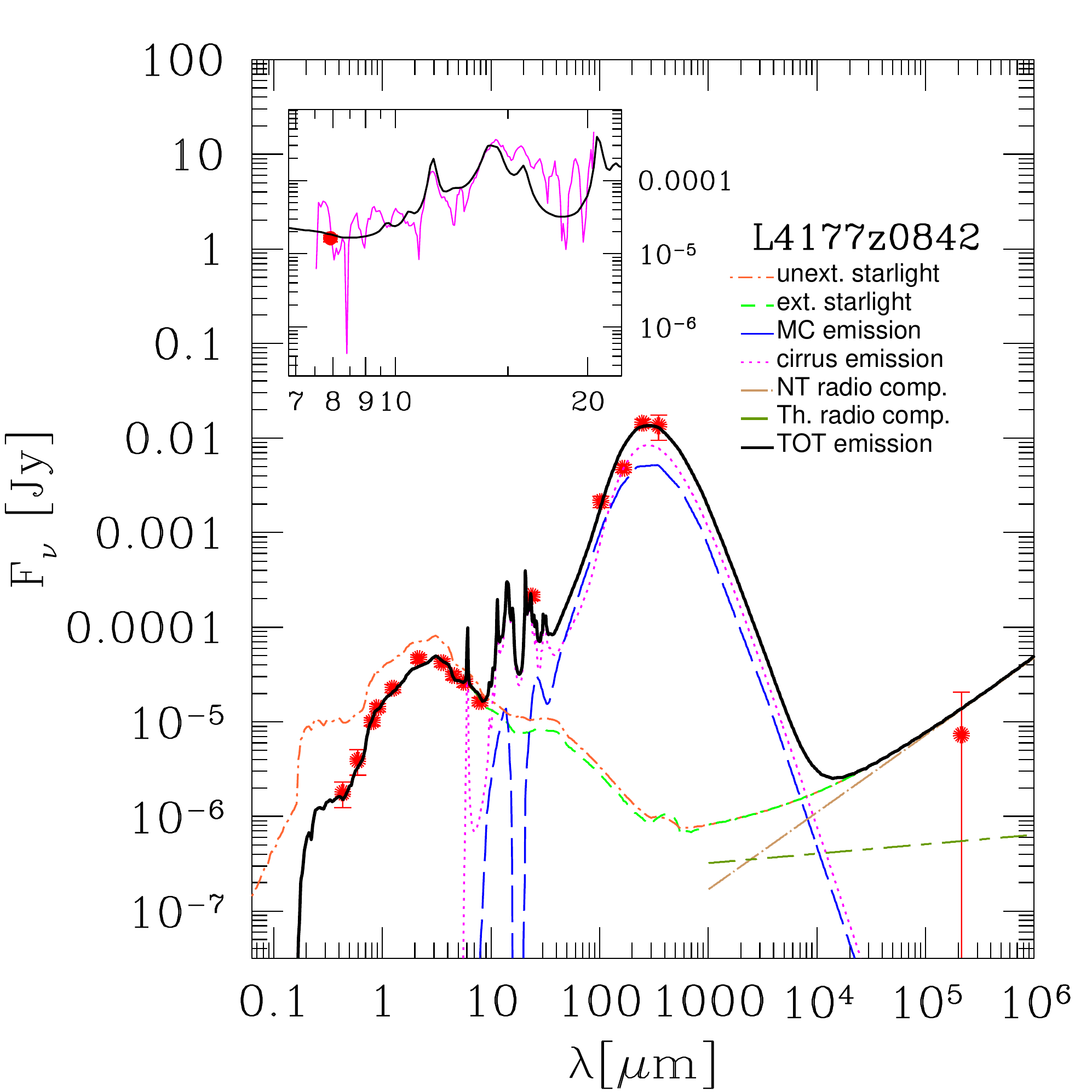}
\includegraphics[width=6cm]{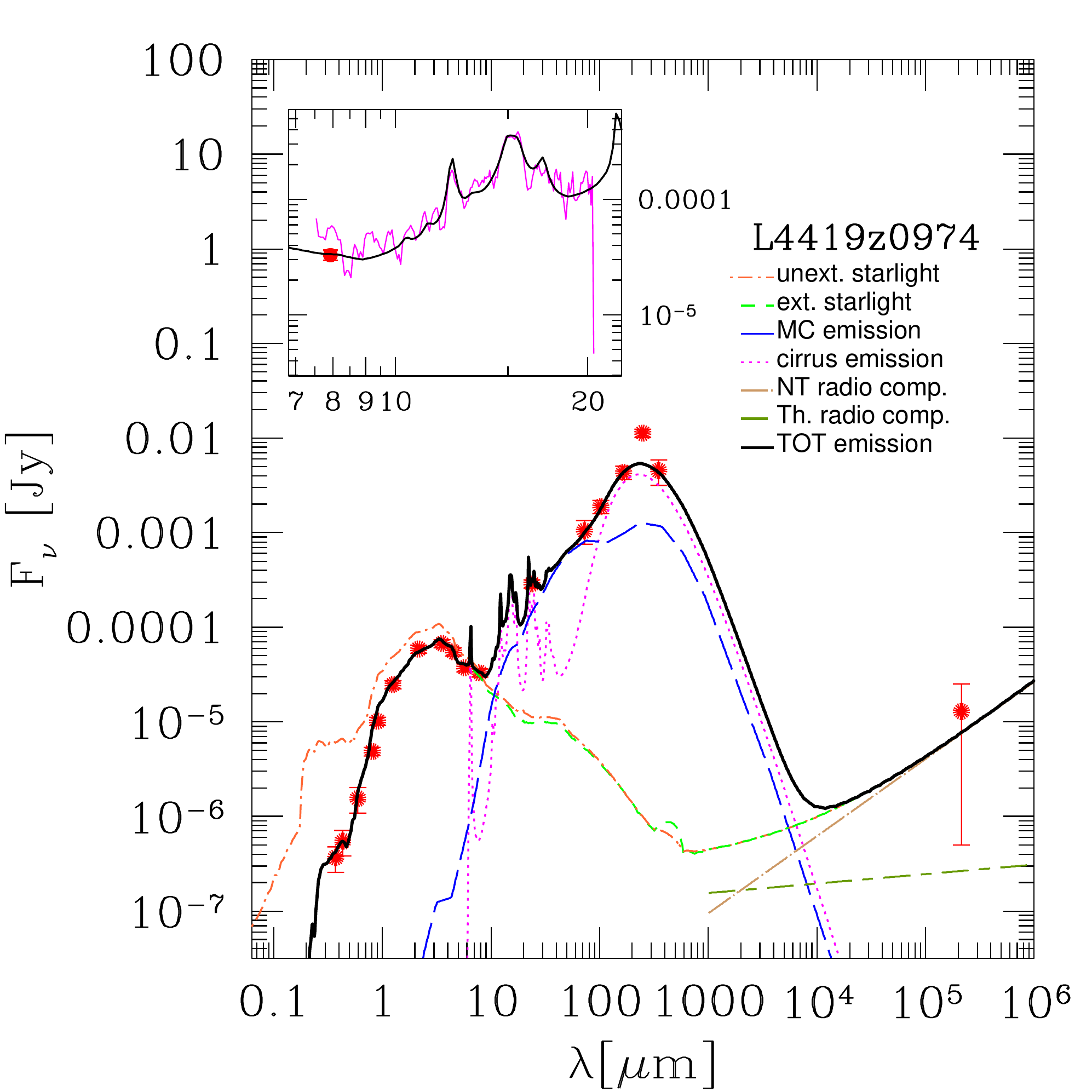}
\includegraphics[width=6cm]{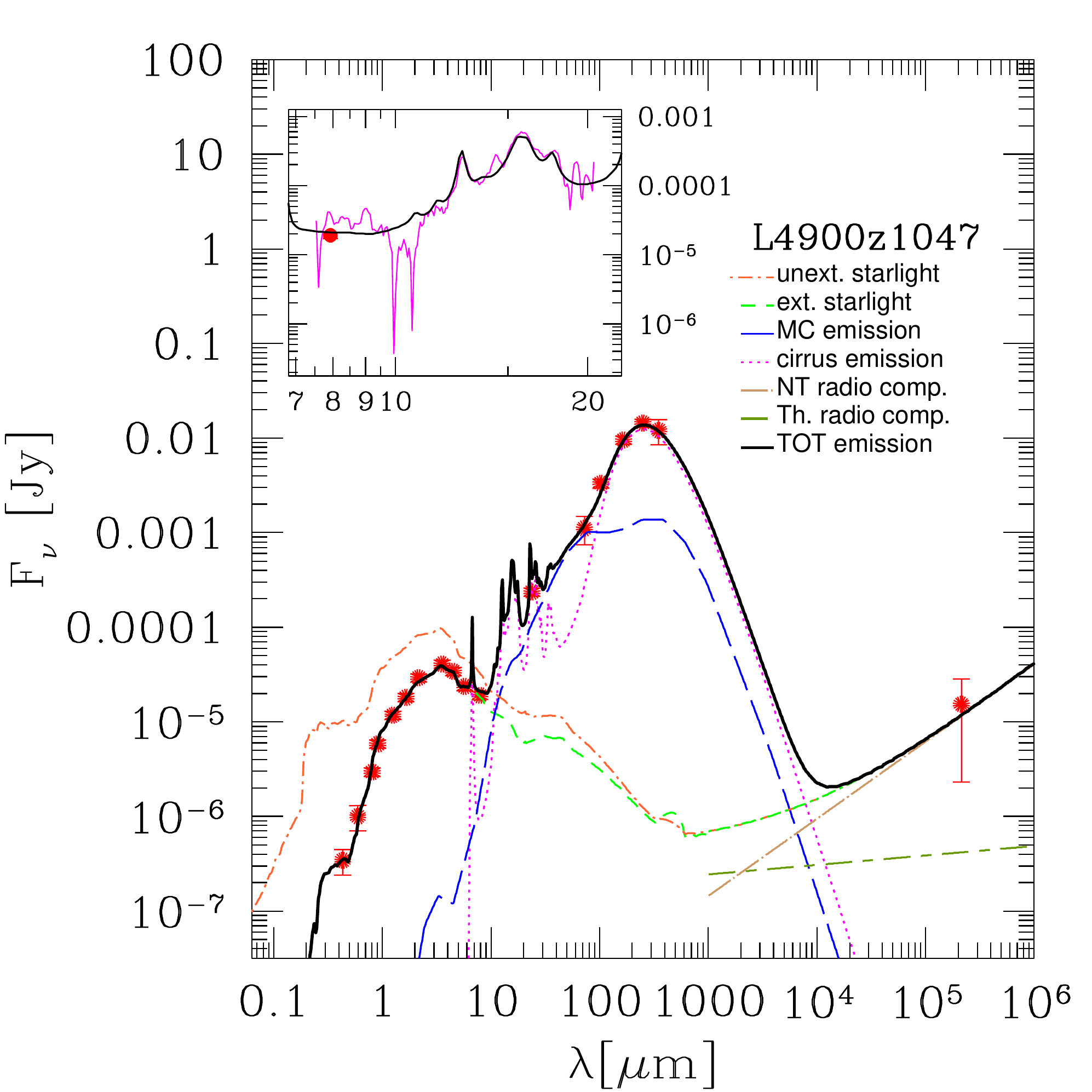}}
\centerline{
\includegraphics[width=6cm]{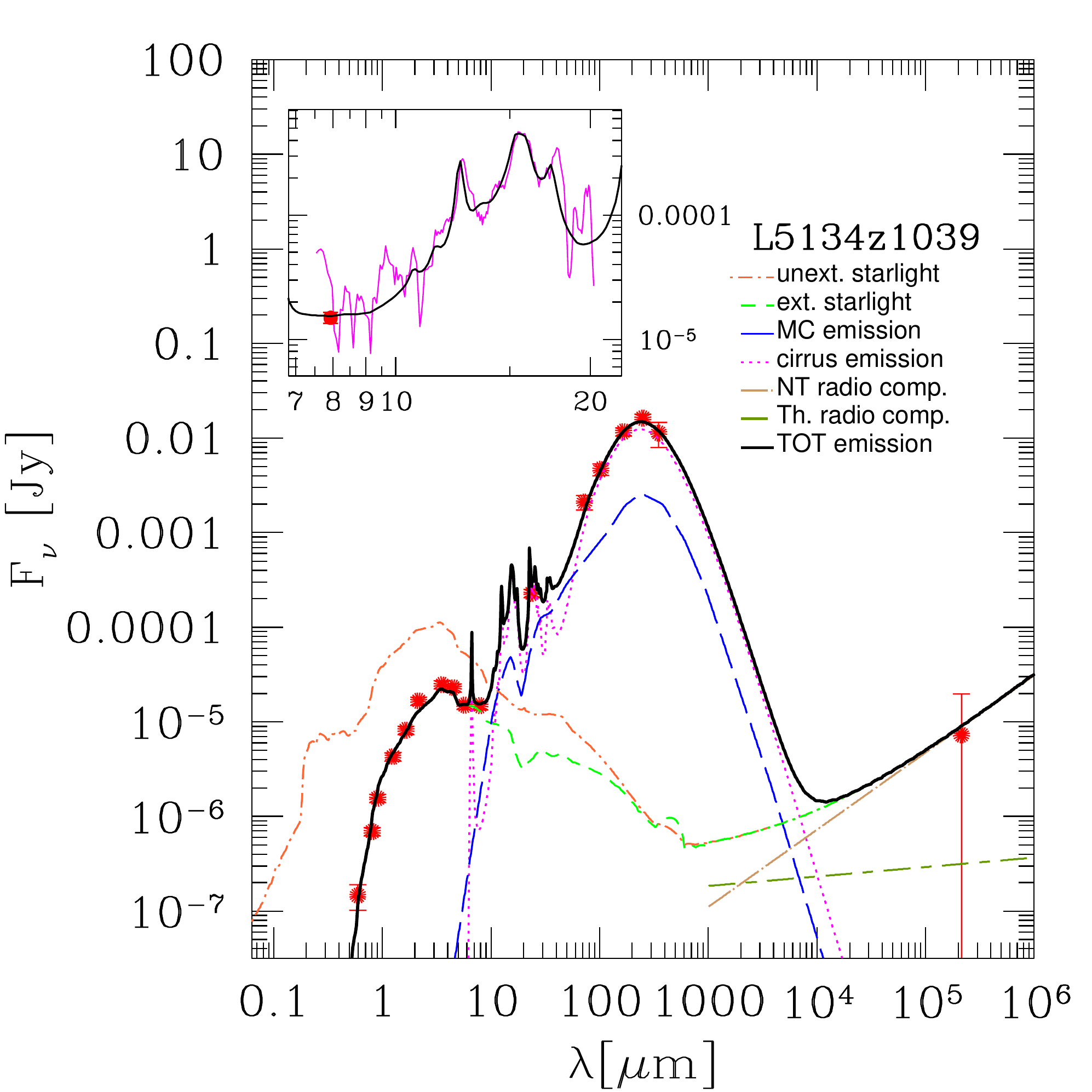}
\includegraphics[width=6cm]{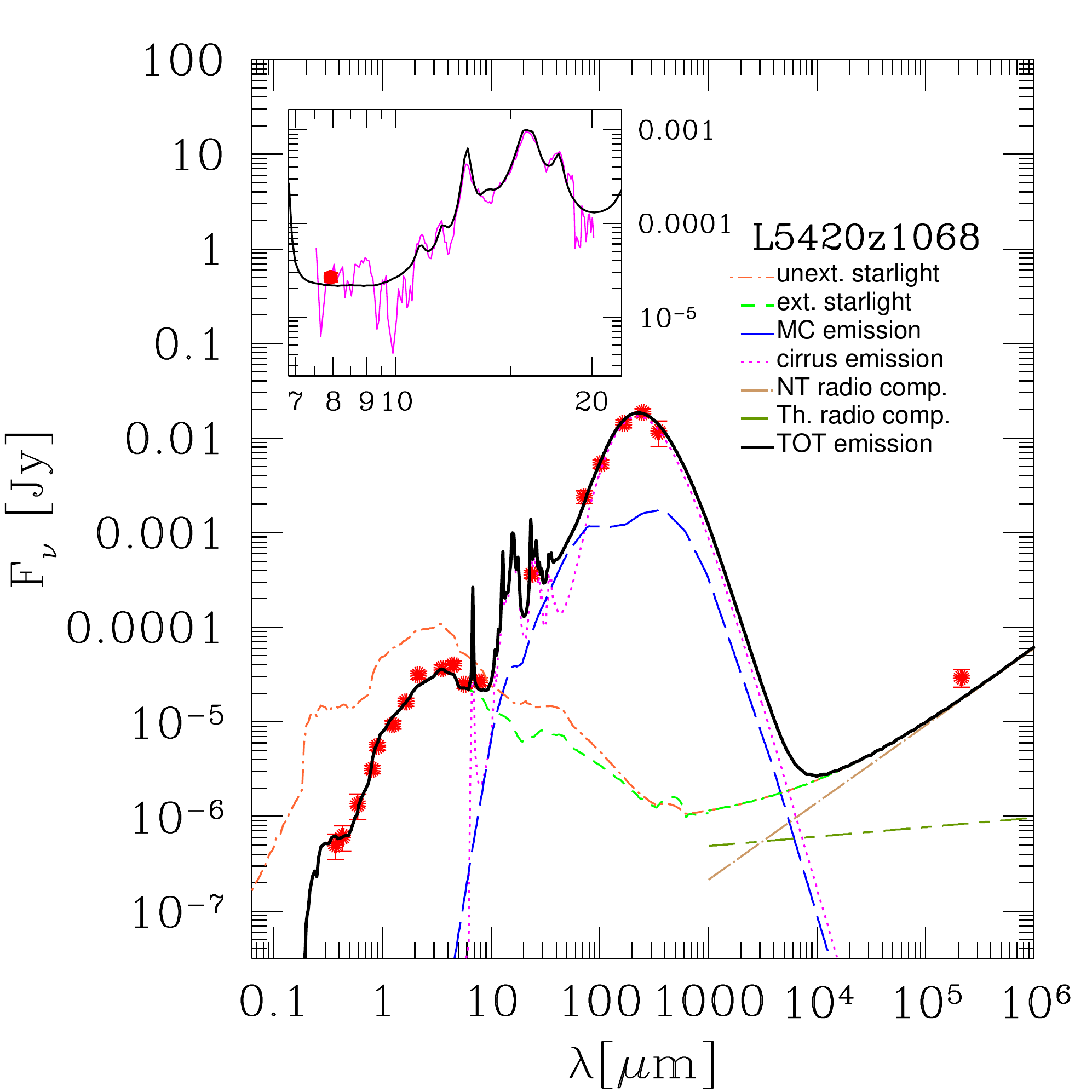}
\includegraphics[width=6cm]{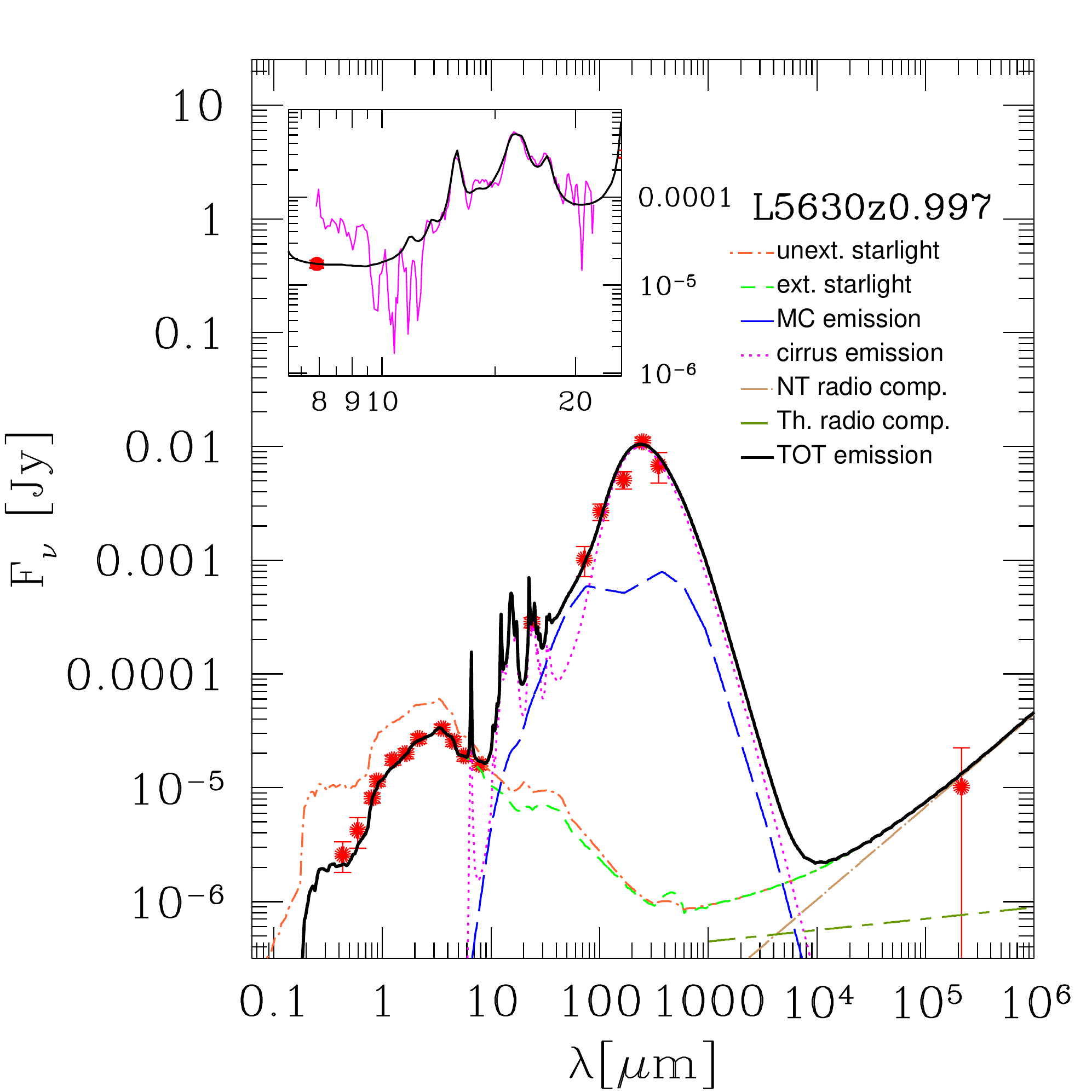}}
\centerline{
\includegraphics[width=6.cm]{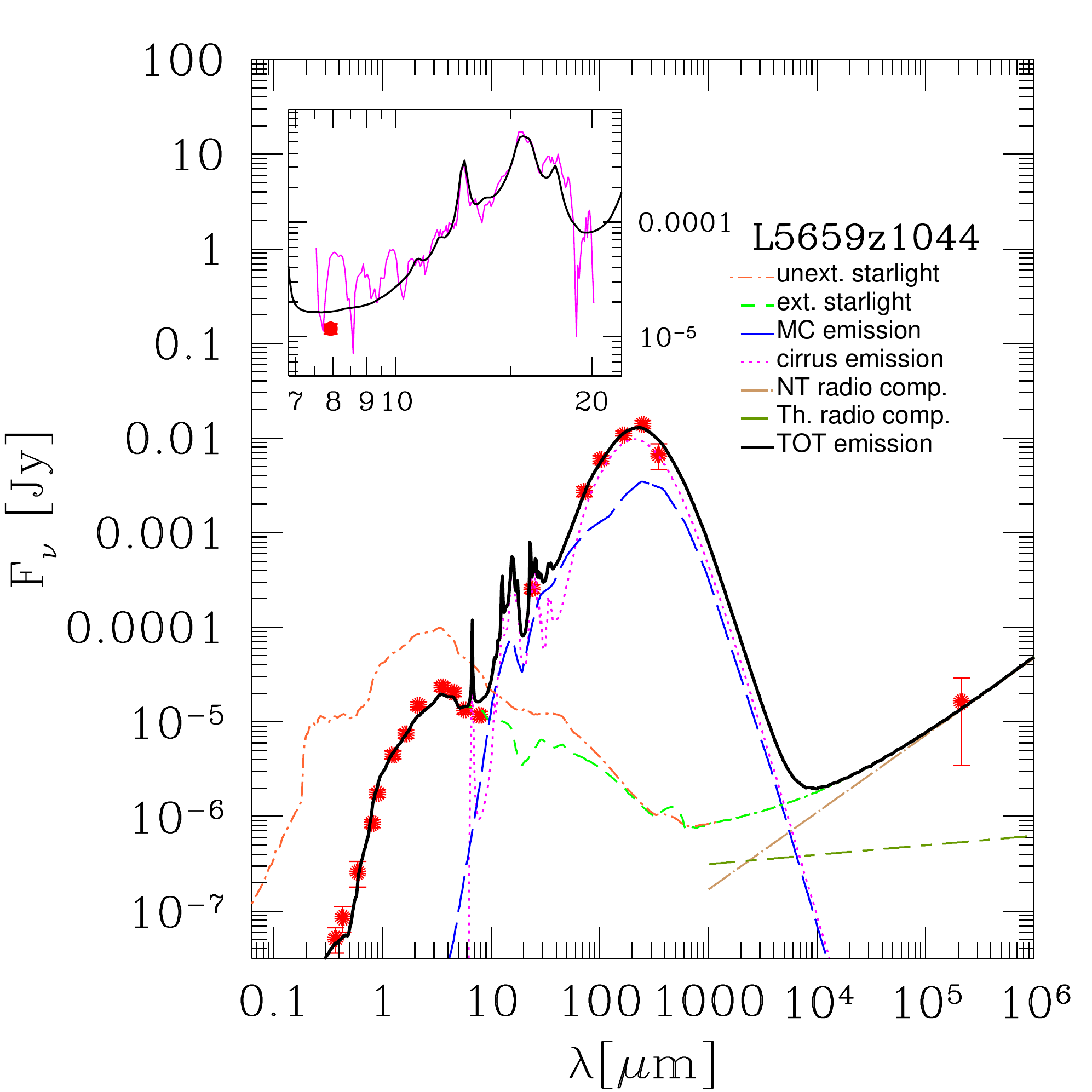}
\includegraphics[width=6.cm]{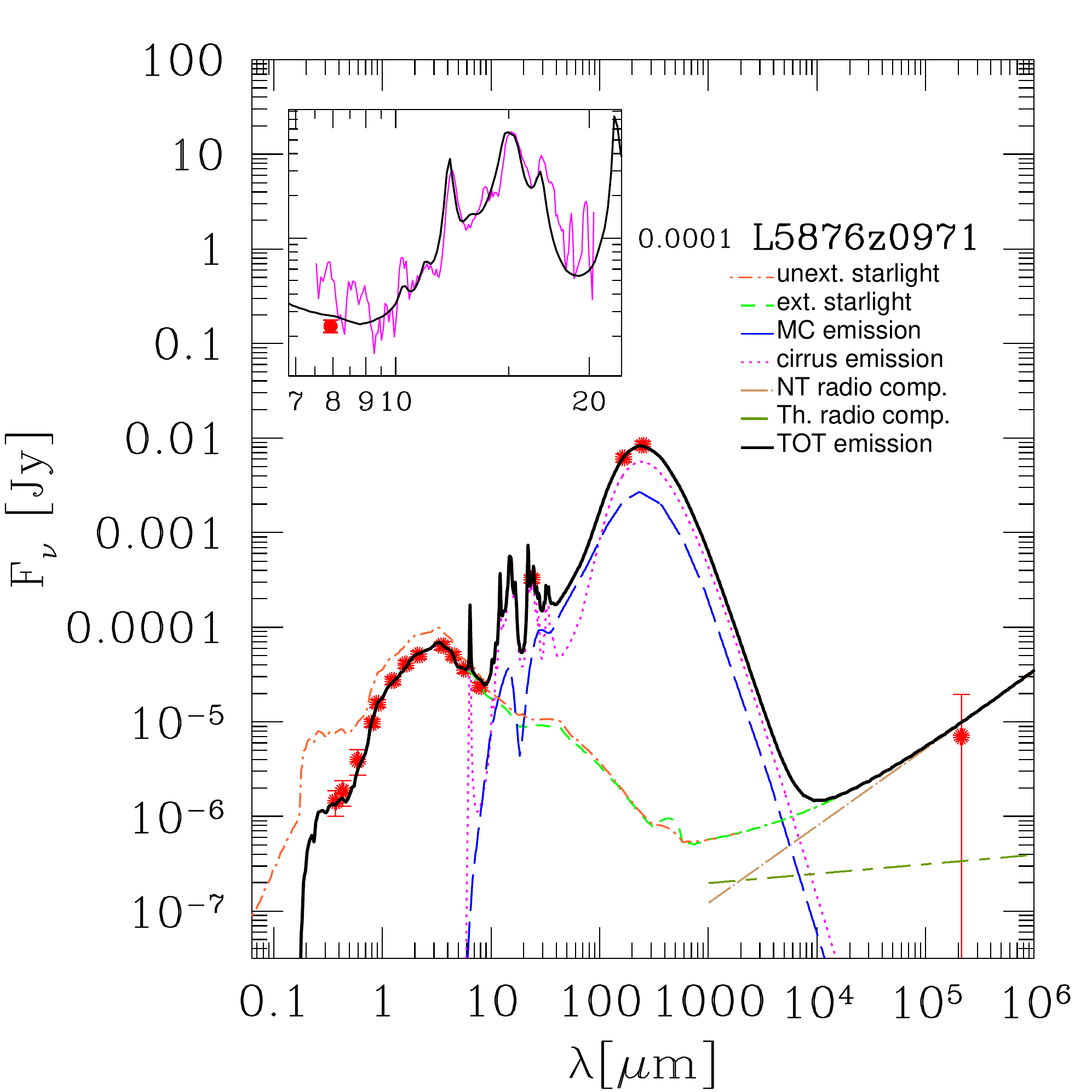}
\includegraphics[width=6.cm]{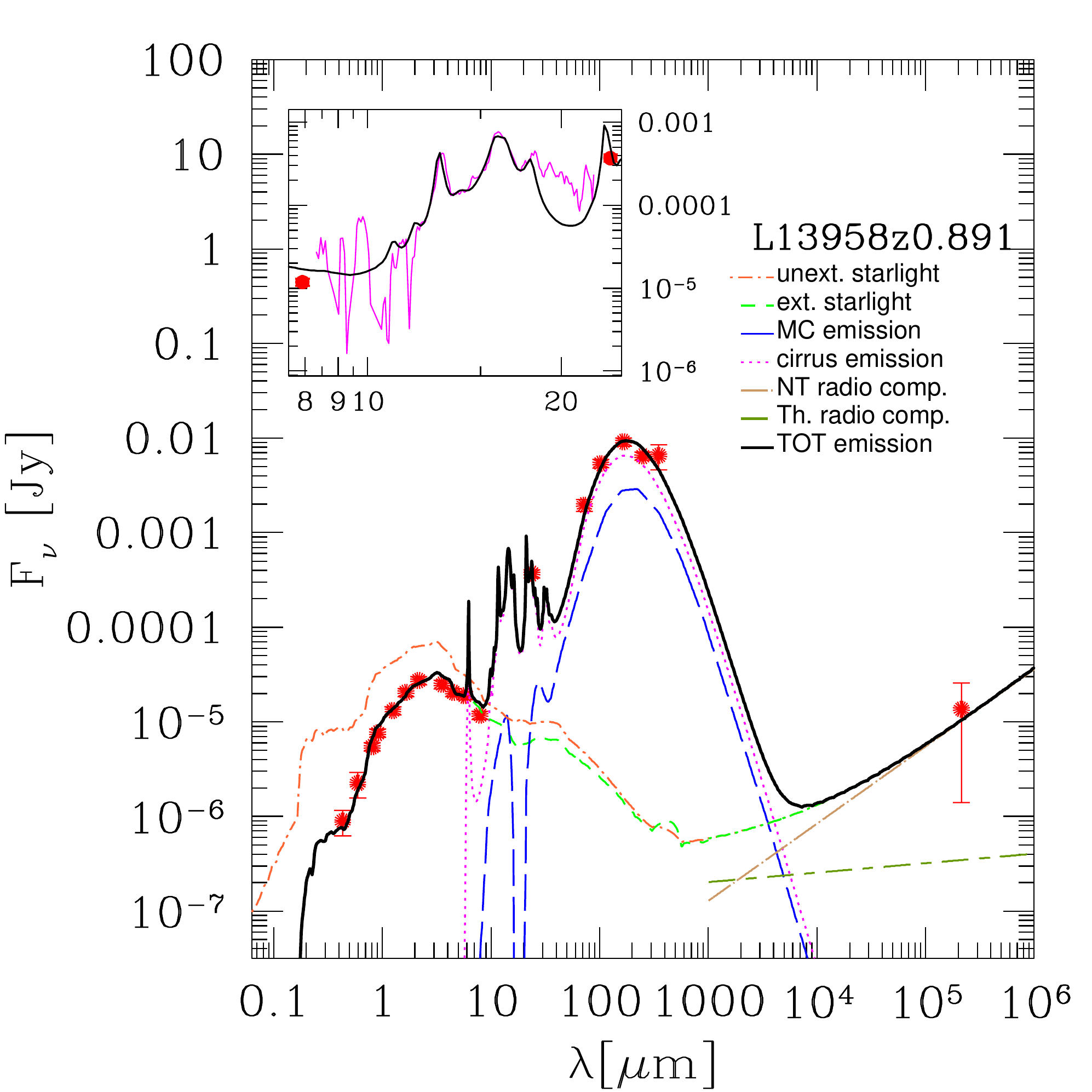}}

\caption{GRASIL best-fits (solid black line) to the observed SED (red circles) of $z\sim1-2$ (U)LIRGs. IRS spectra appear in the inset window (magenta line). The color-coded lines represent the unextinguished starlight (orange dot-dashed), extinguished starlight (green dashed), cirrus emission (magenta dotted), MC emission (blue long dashed) and thermal and non-thermal radio components (salvia long line-dashed and brown dot-long dashed, respectively). The latter are shown in the wavelength range where their contribution is significant. We are able to reproduce, well within a factor of two, the far-UV to radio emission for almost all the (U)LIRGs into our sample. 28/31 (U)LIRGs show modelled radio fluxes within data error-bars.}
\label{best-fit-radiol1}
\end{figure*}

\begin{figure*}
\centerline{
\includegraphics[width=6.cm]{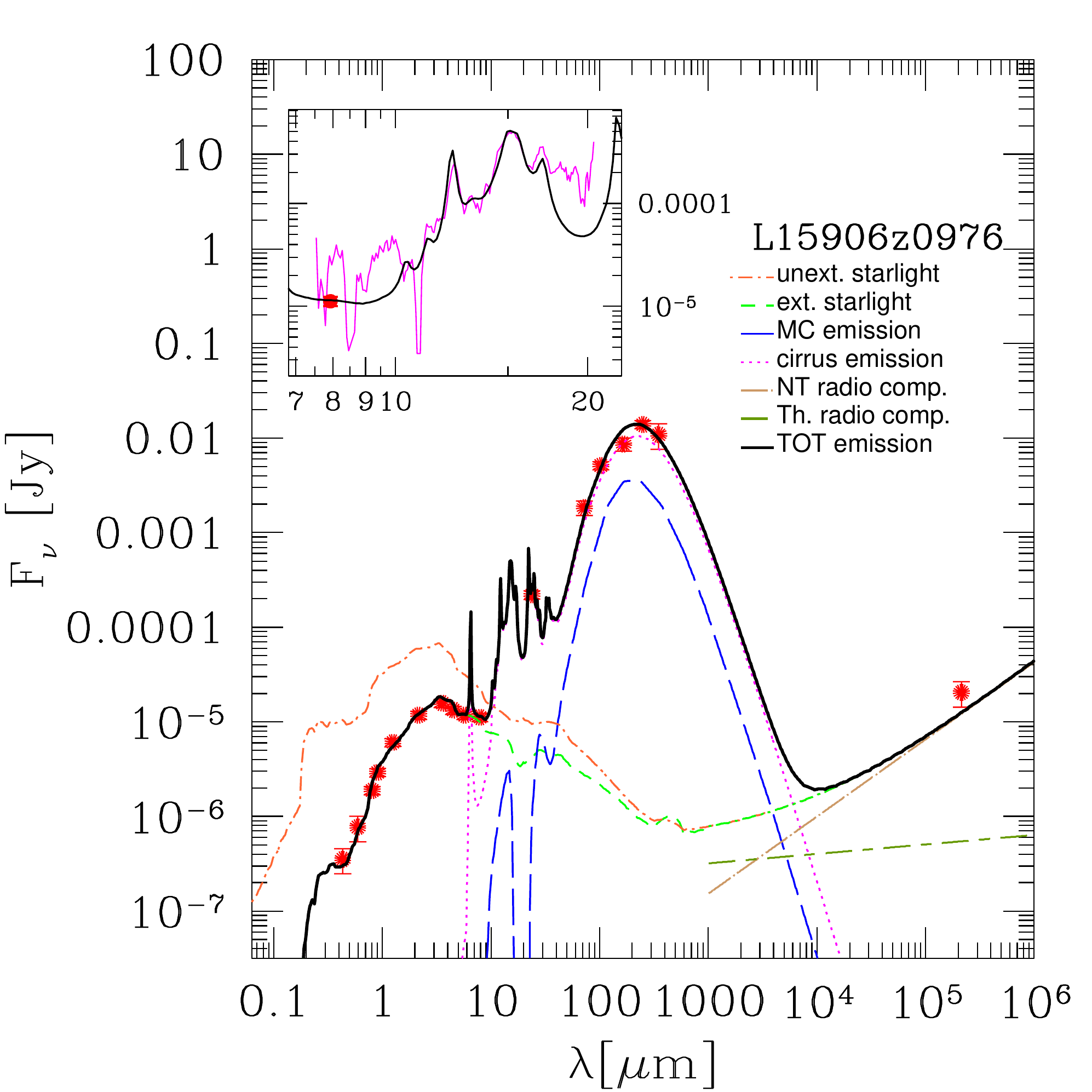}
\includegraphics[width=6.cm]{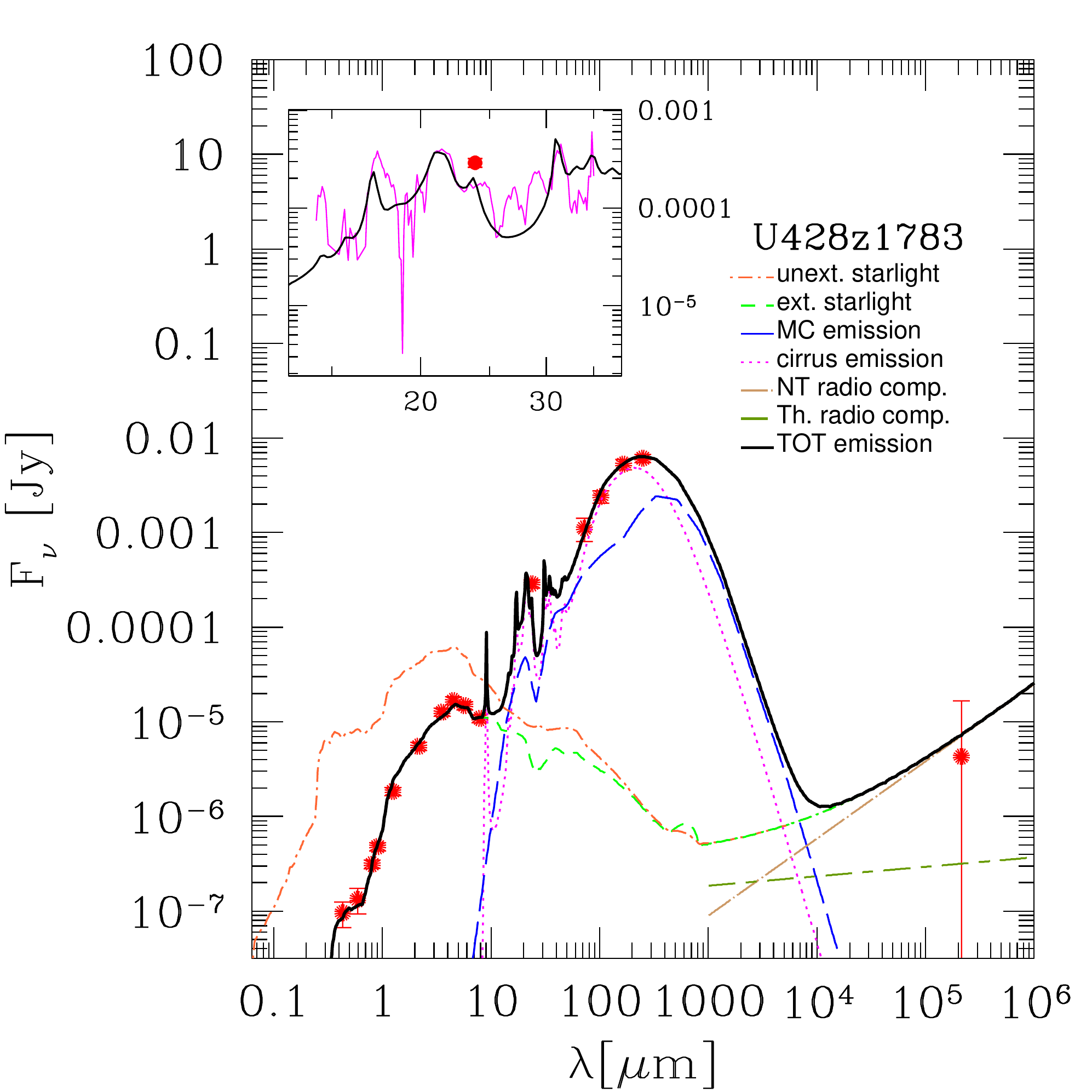}
\includegraphics[width=6.cm]{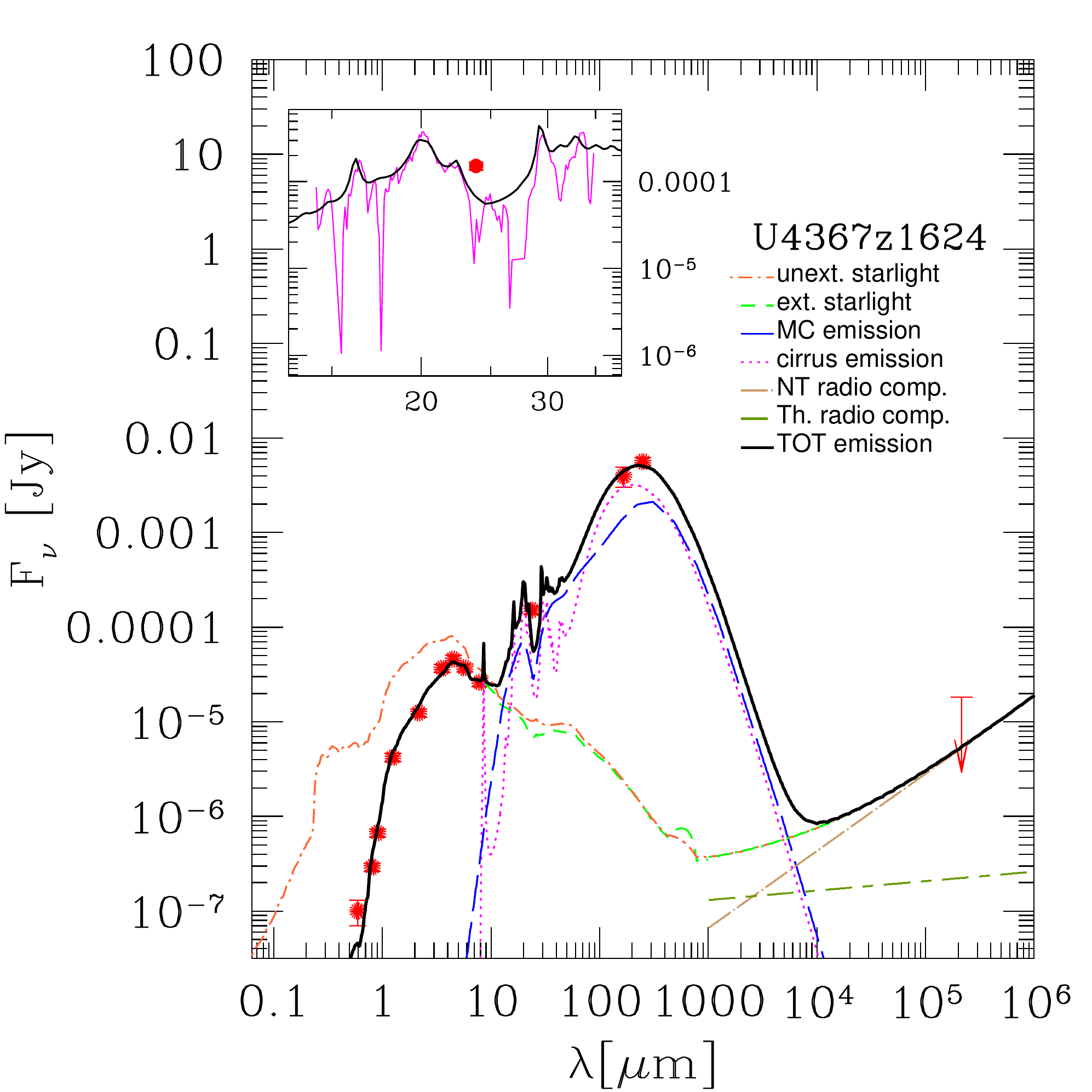}}
\centerline{
\includegraphics[width=6.cm]{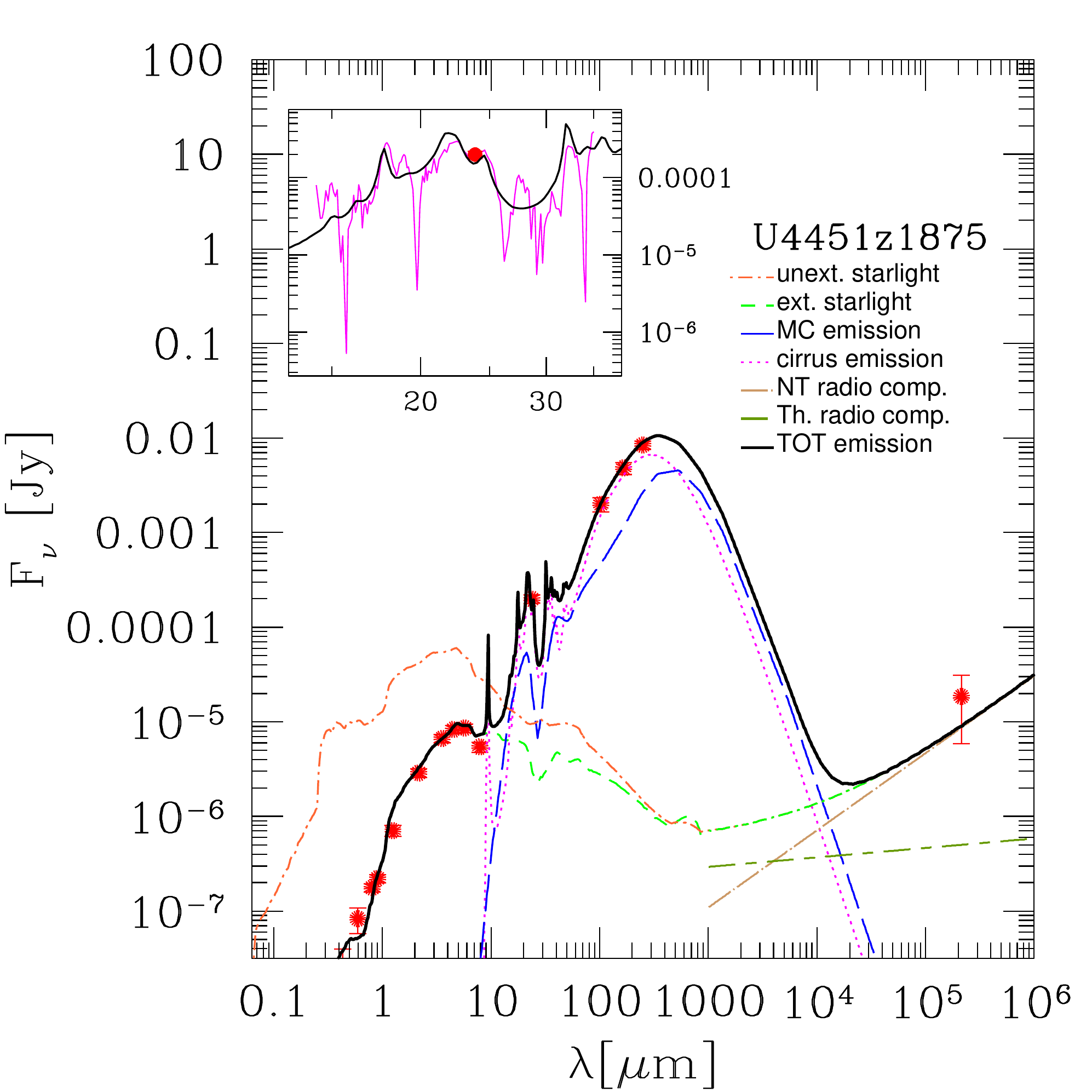}
\includegraphics[width=6.cm]{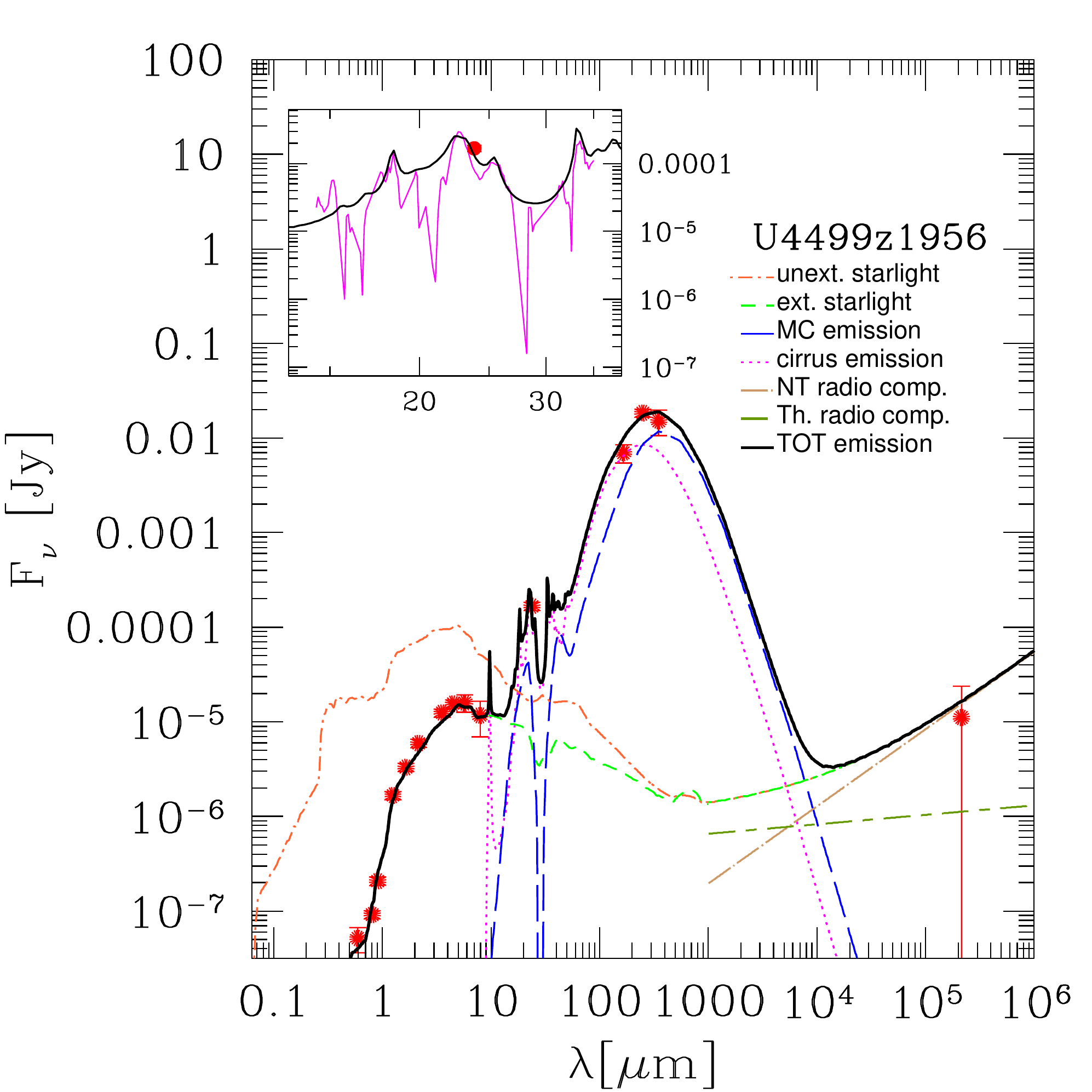}
\includegraphics[width=6.cm]{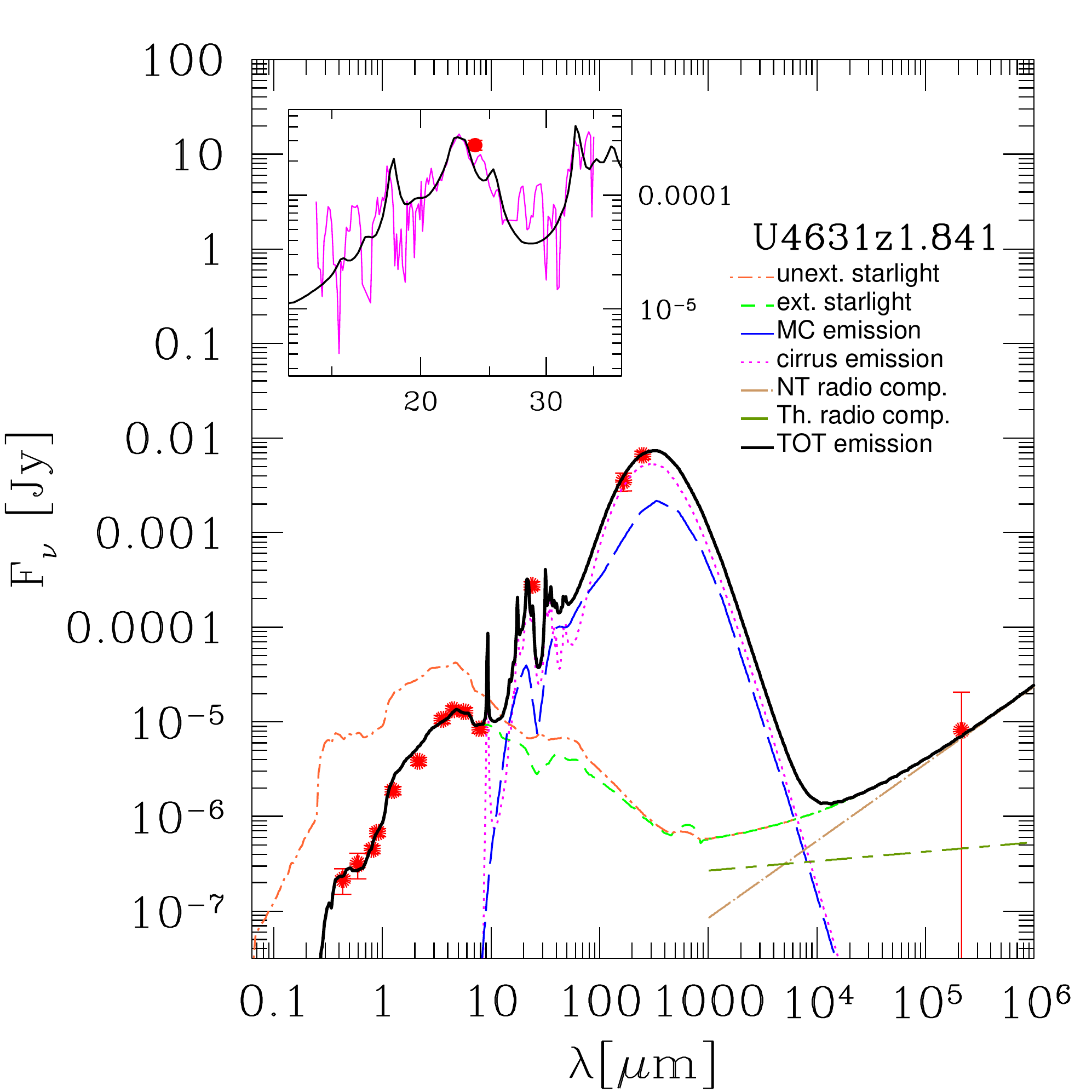}}
\centerline{
\includegraphics[width=6.cm]{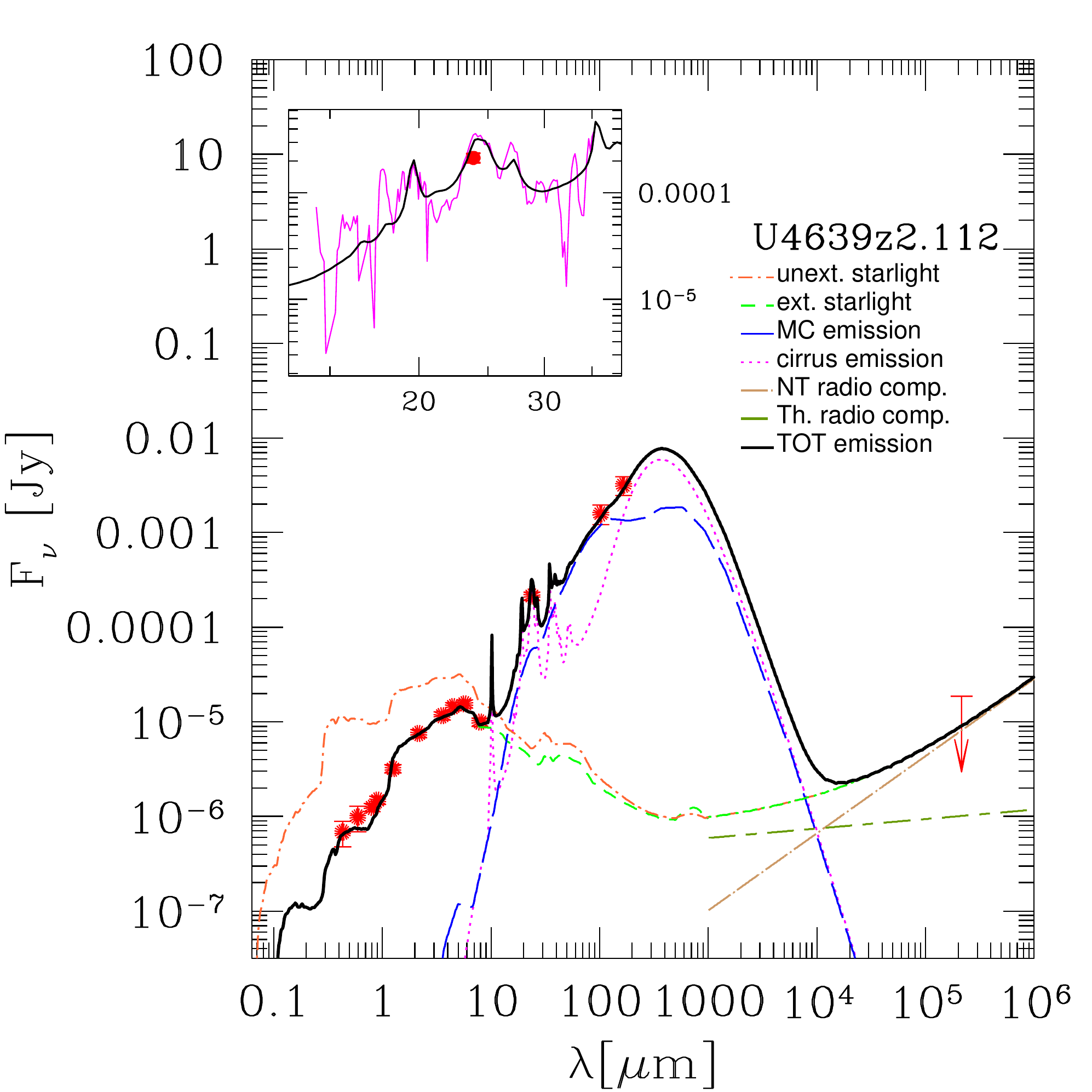}
\includegraphics[width=6.cm]{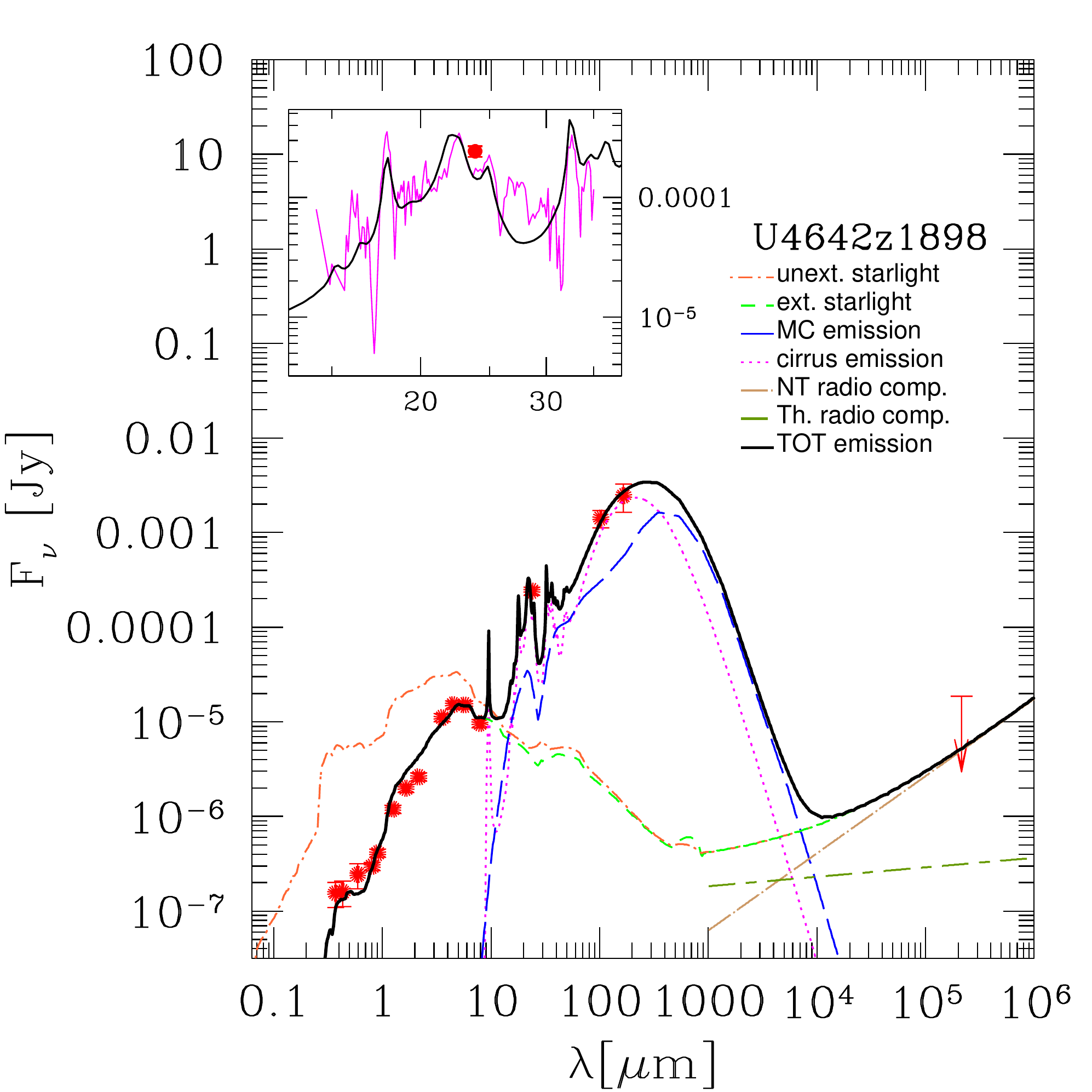}
\includegraphics[width=6.cm]{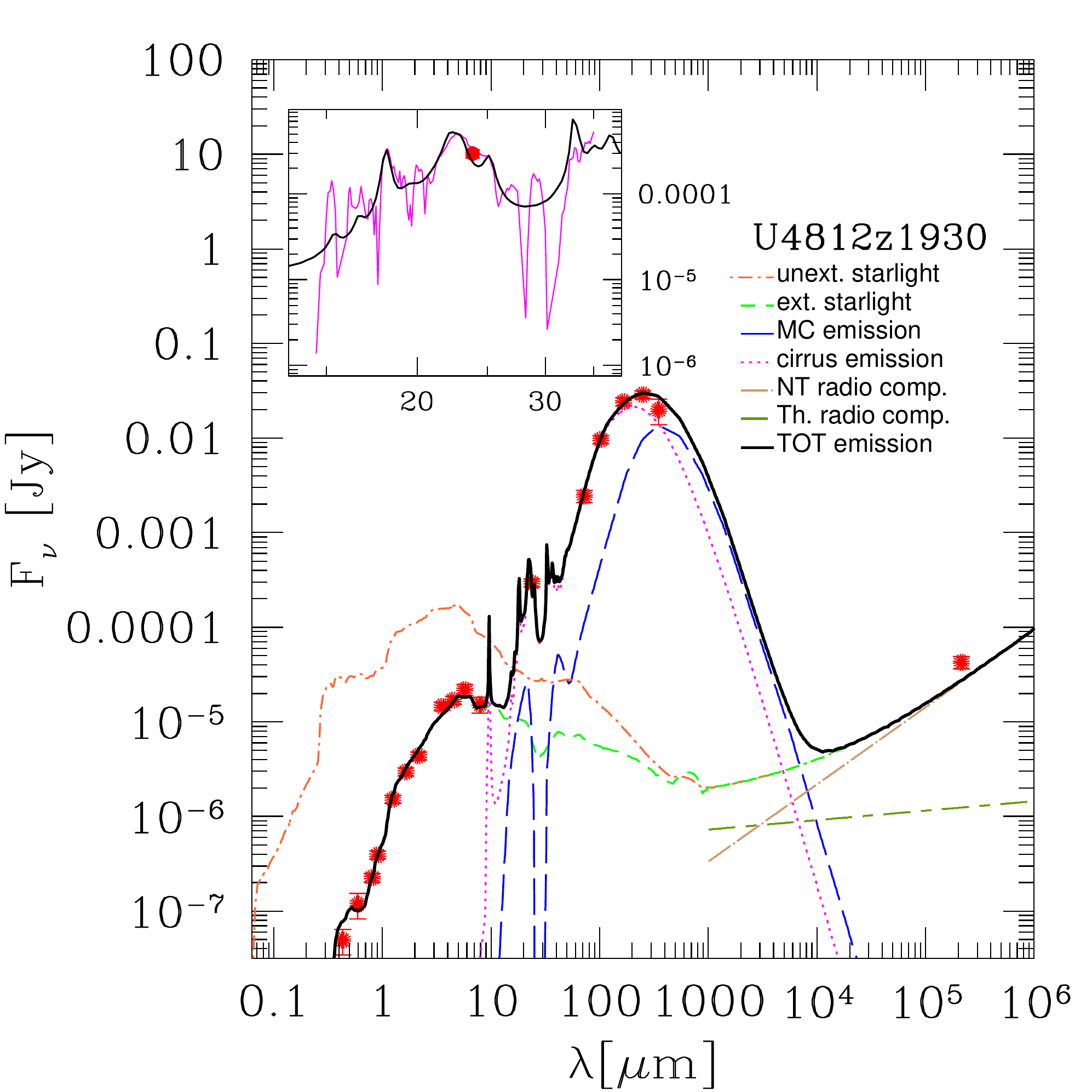}}

\contcaption{}
\label{best-fit-radiol3}
\end{figure*}
\begin{figure*}
\centerline{
\includegraphics[width=6.cm]{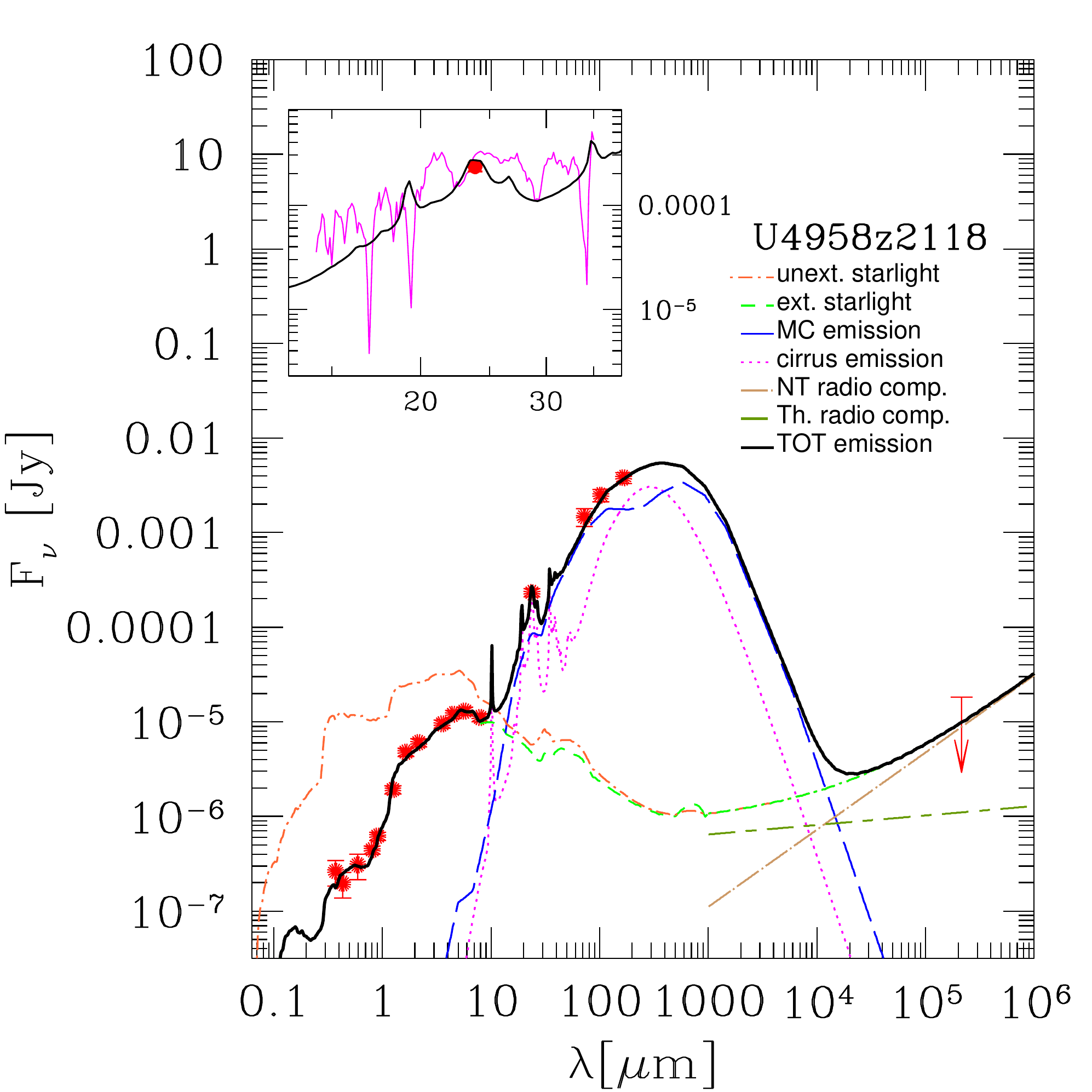}
\includegraphics[width=6.cm]{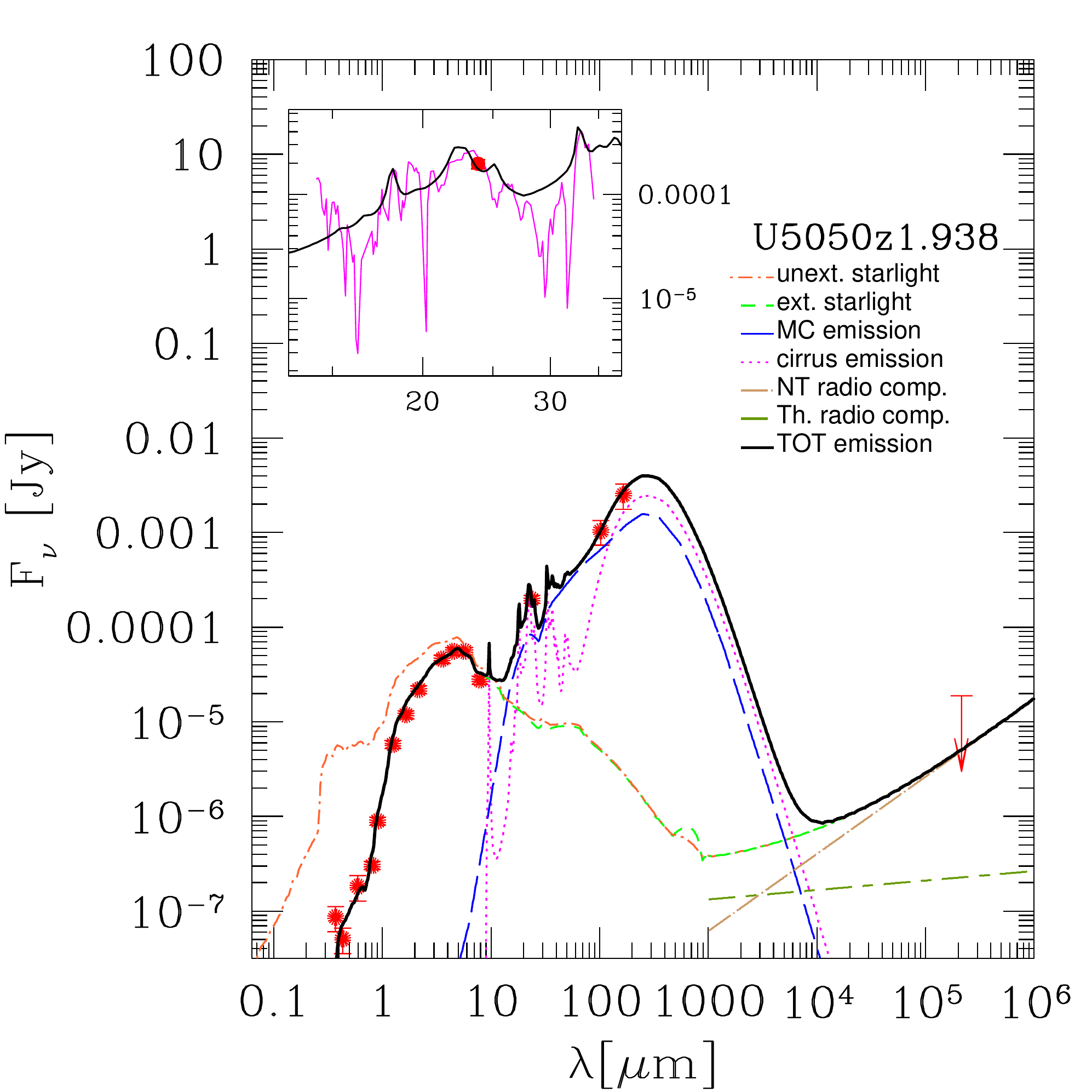}
\includegraphics[width=6.cm]{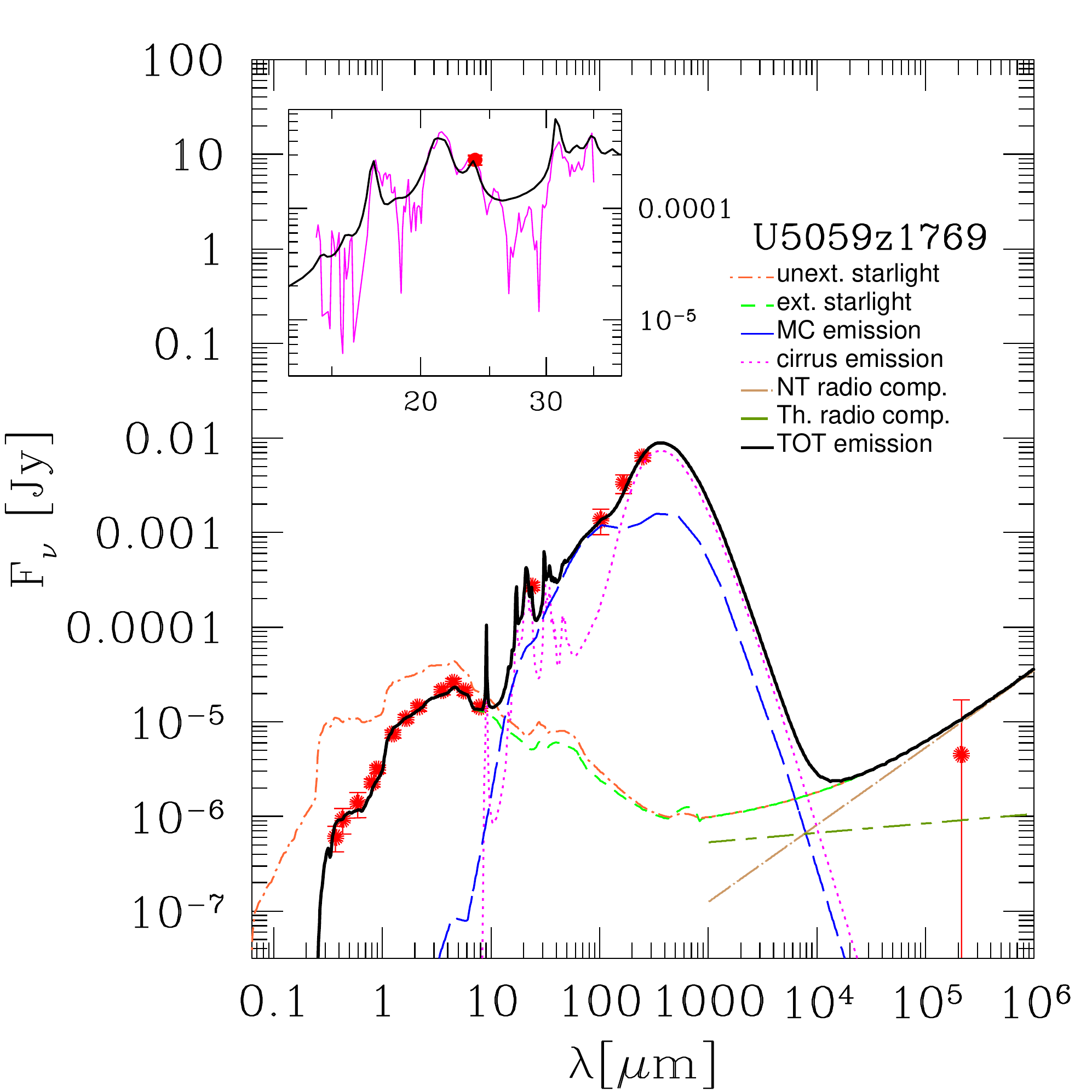}}
\centerline{
\includegraphics[width=6.cm]{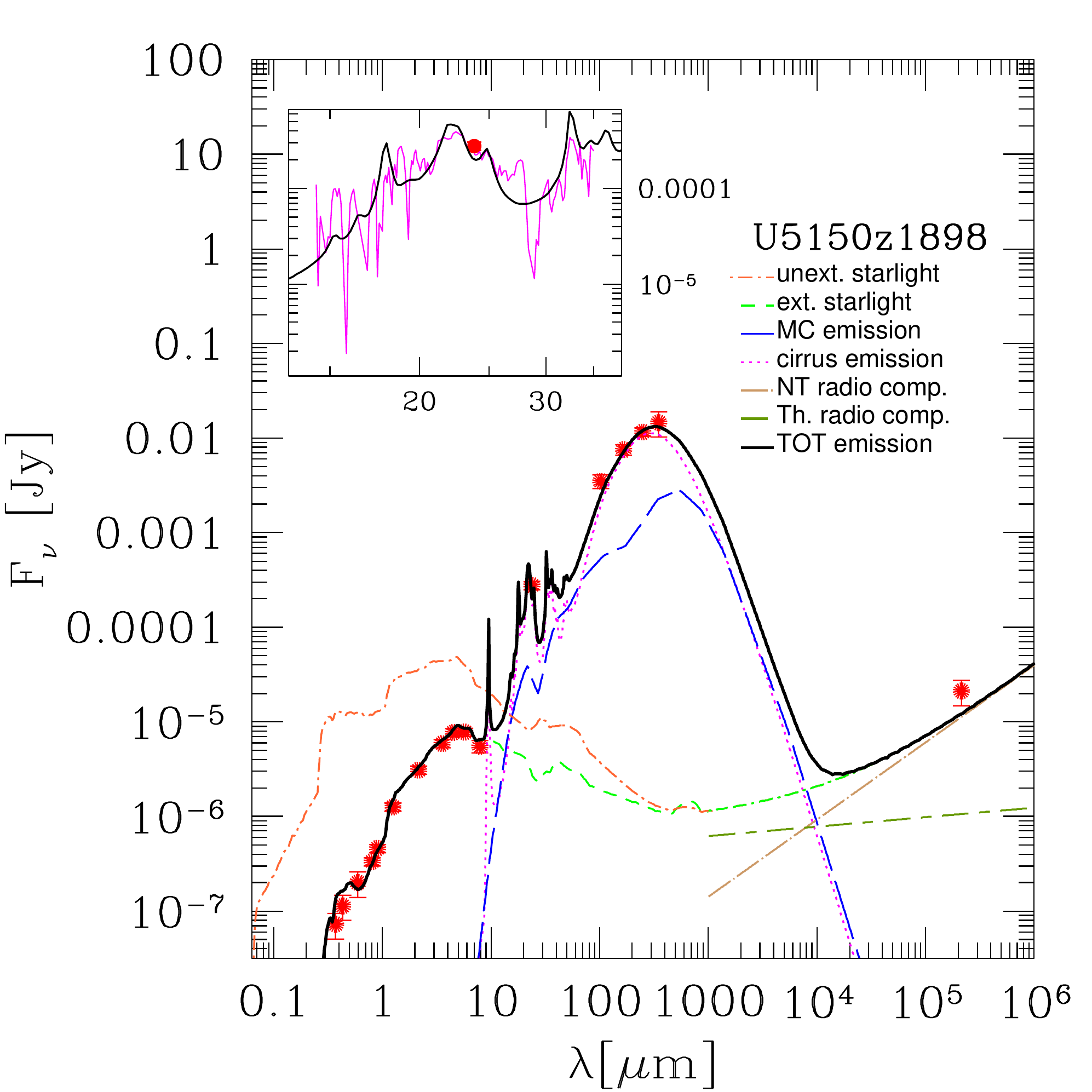}
\includegraphics[width=6.cm]{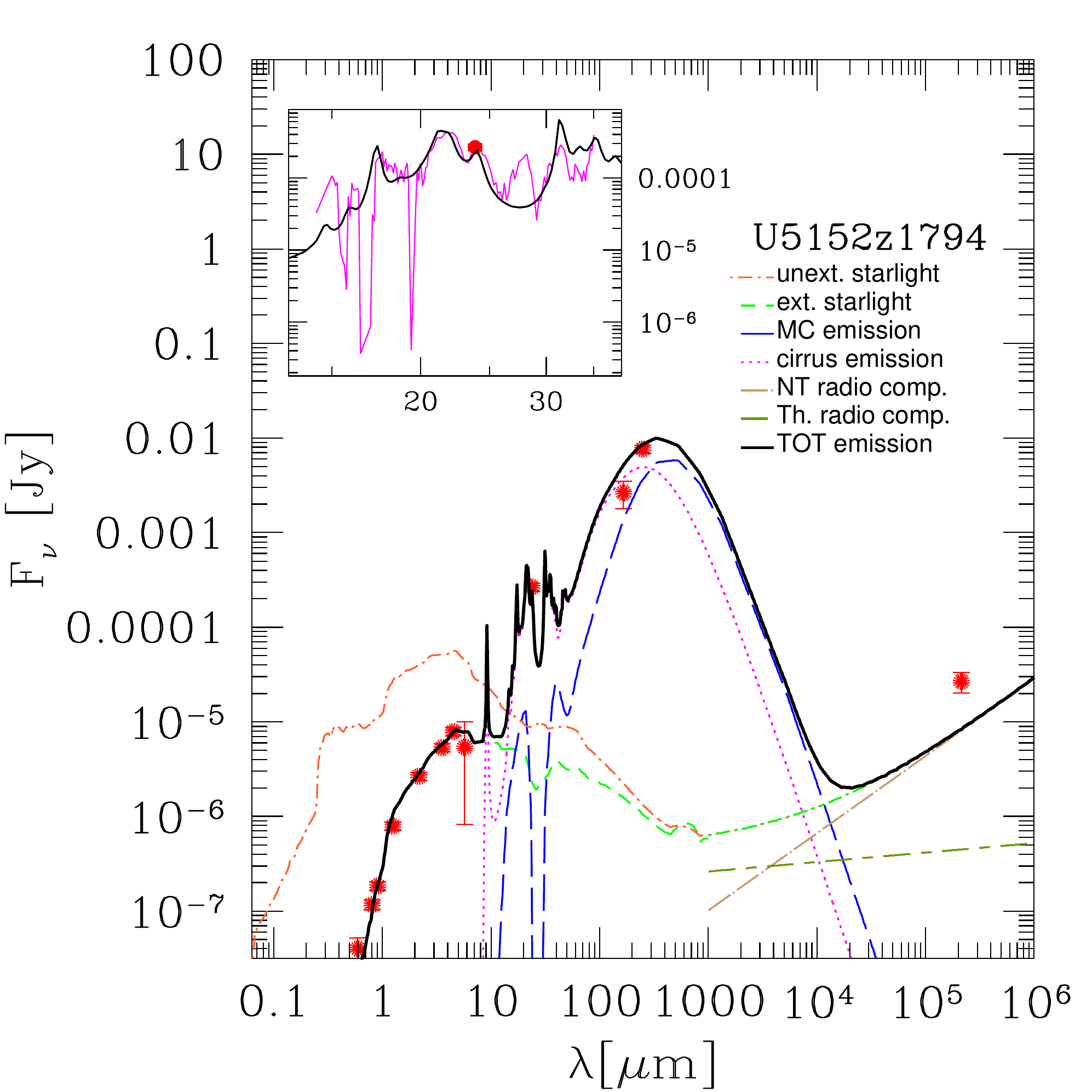}
\includegraphics[width=6.cm]{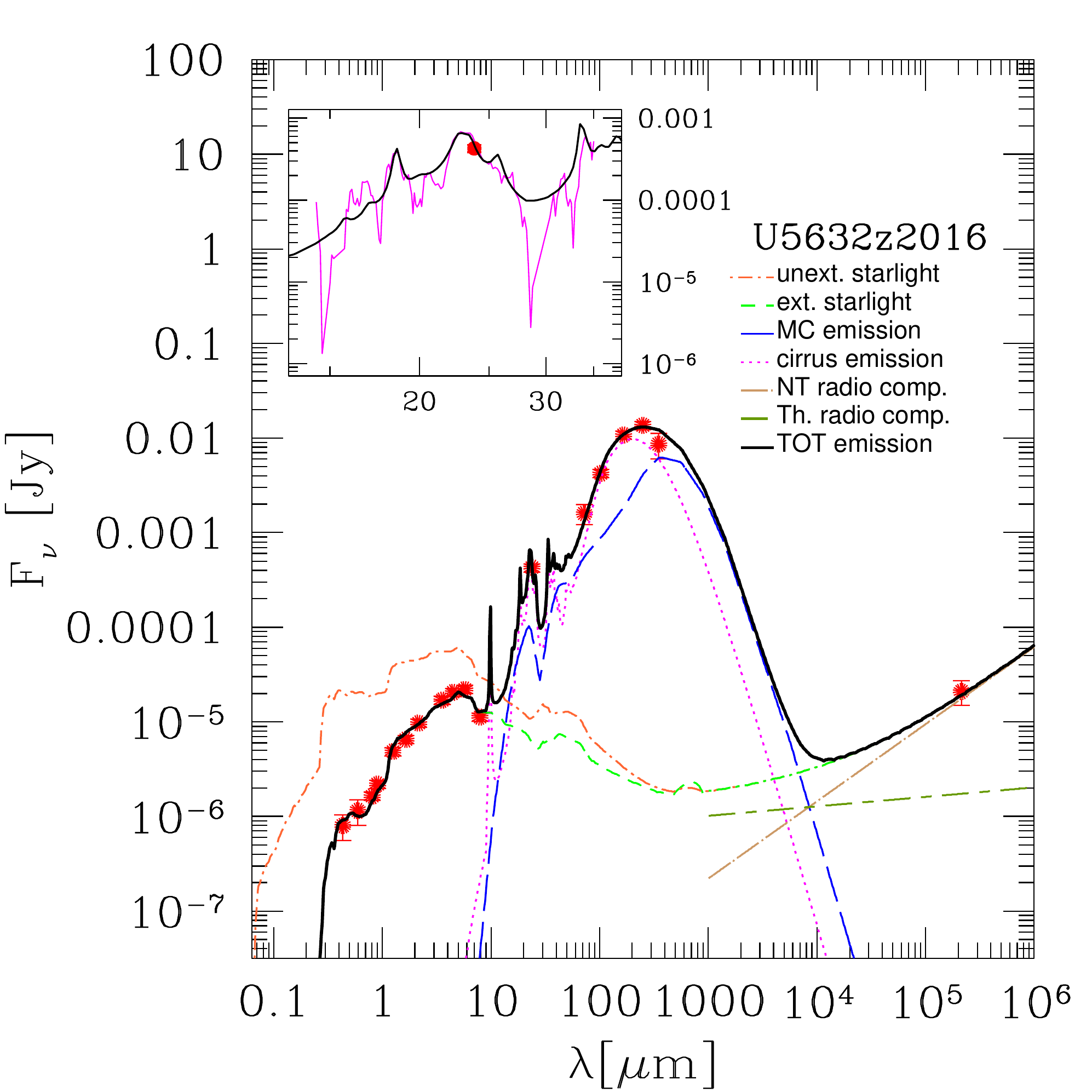}}
\centerline{
\includegraphics[width=6.cm]{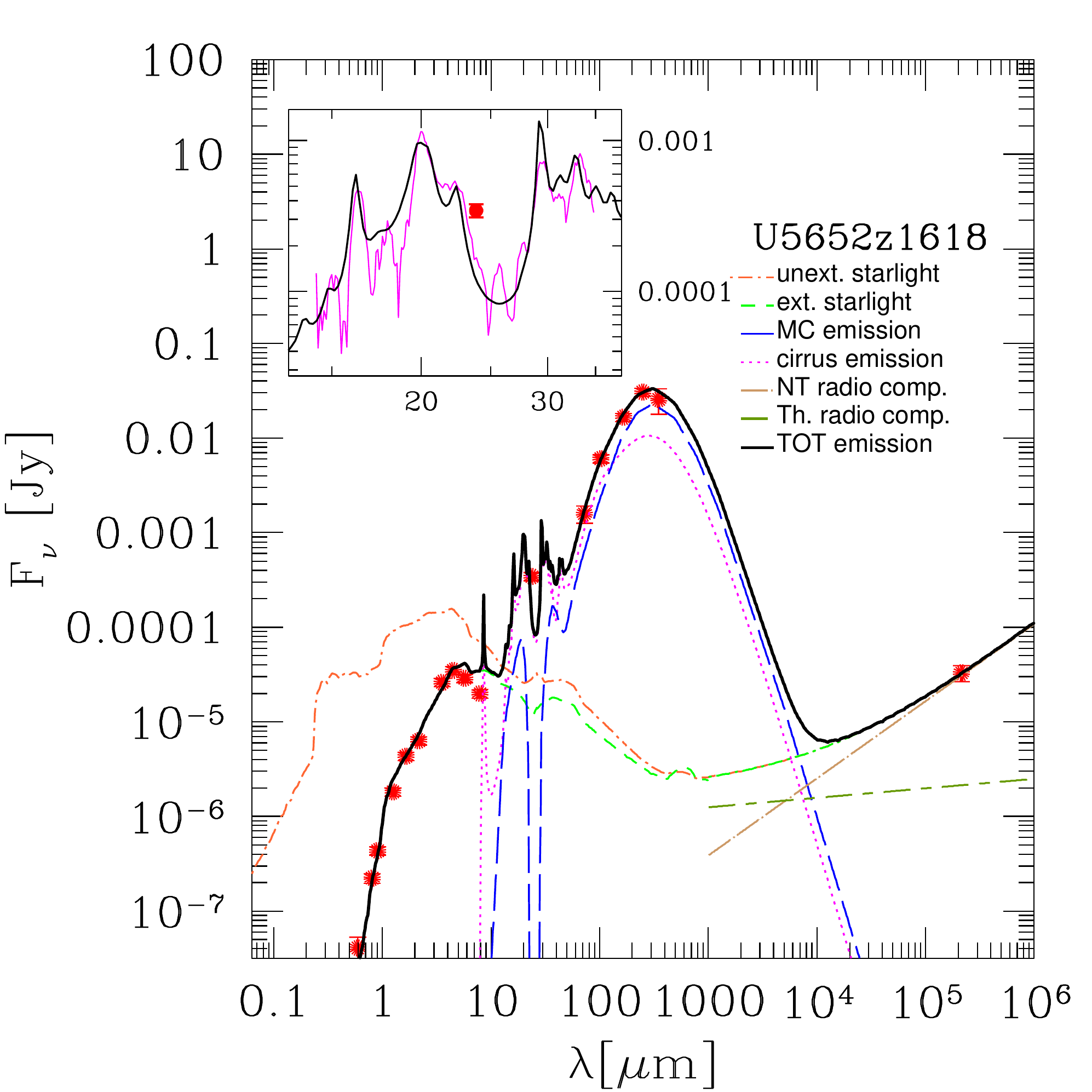}
\includegraphics[width=6.cm]{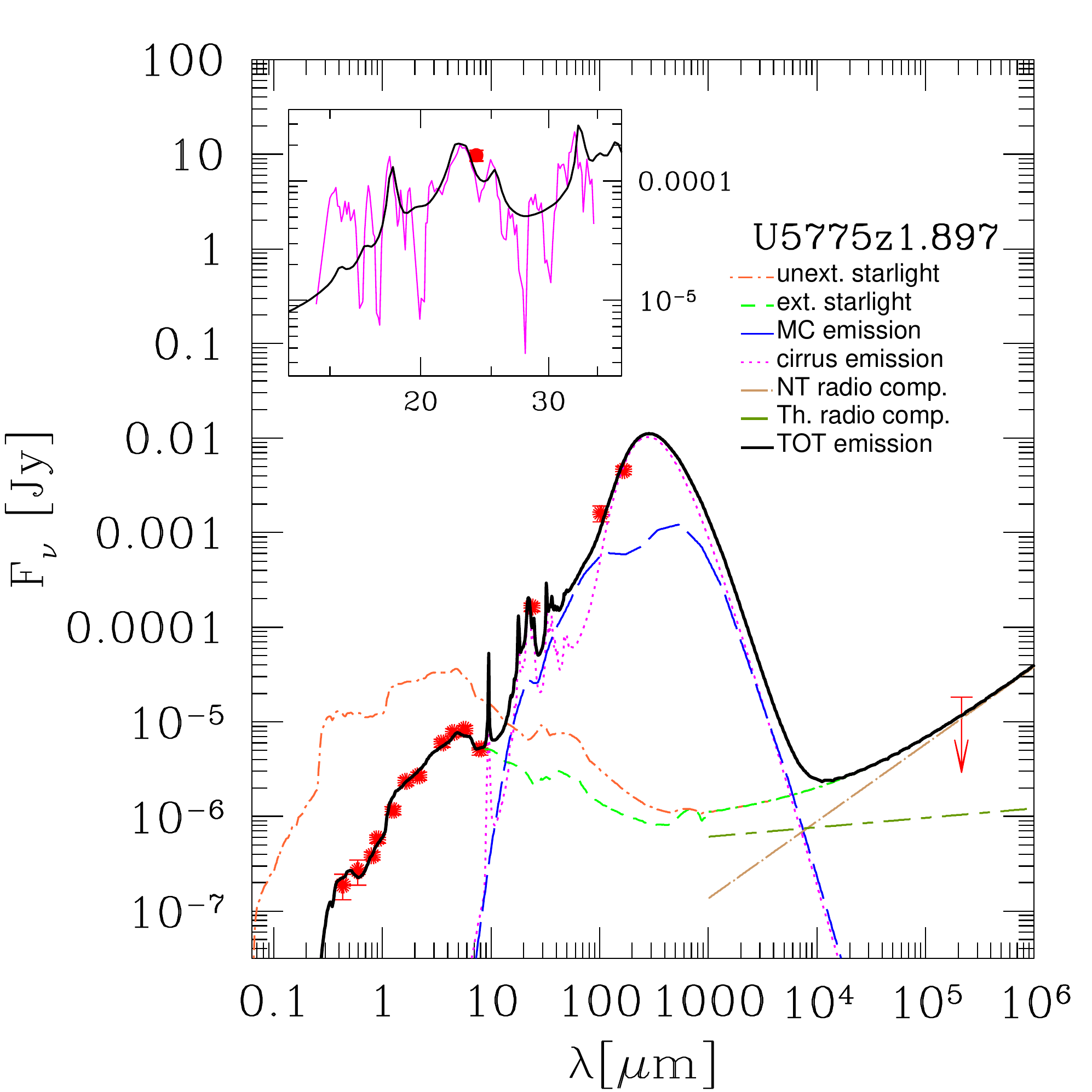}
\includegraphics[width=6.cm]{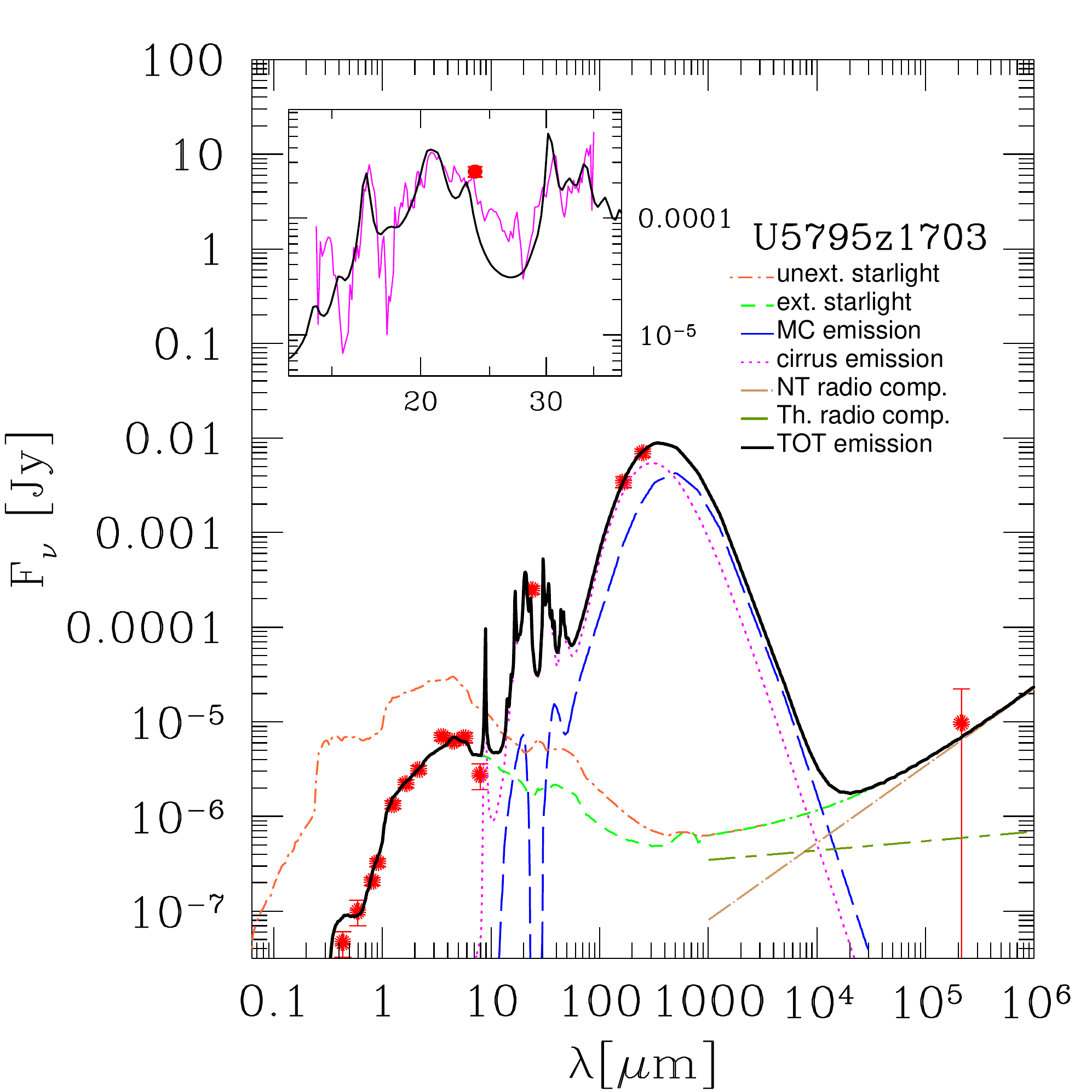}}

\contcaption{}
\label{best-fit-radiou2}
\end{figure*}
\begin{figure*}

\centerline{
\includegraphics[width=6.cm]{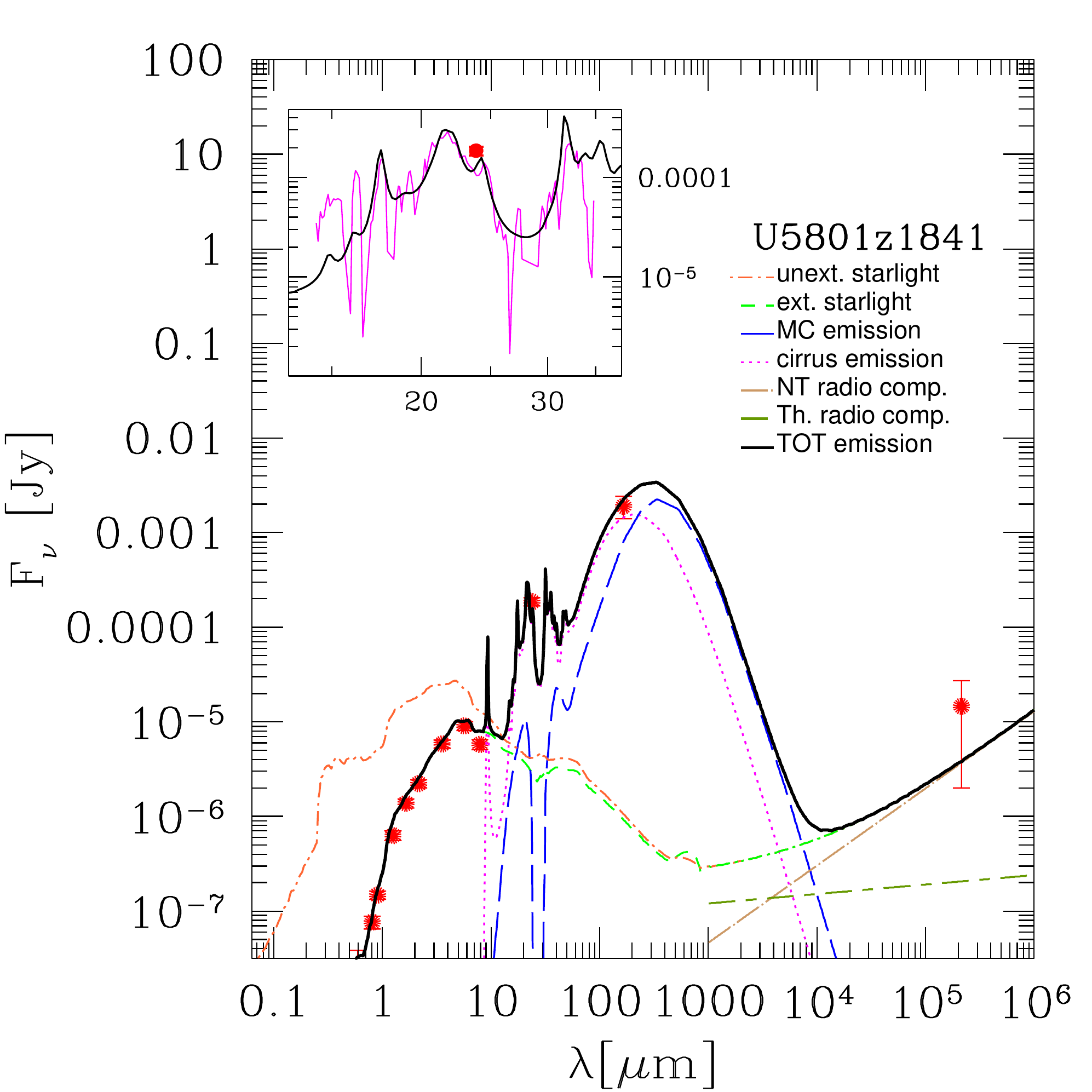}
\includegraphics[width=6.cm]{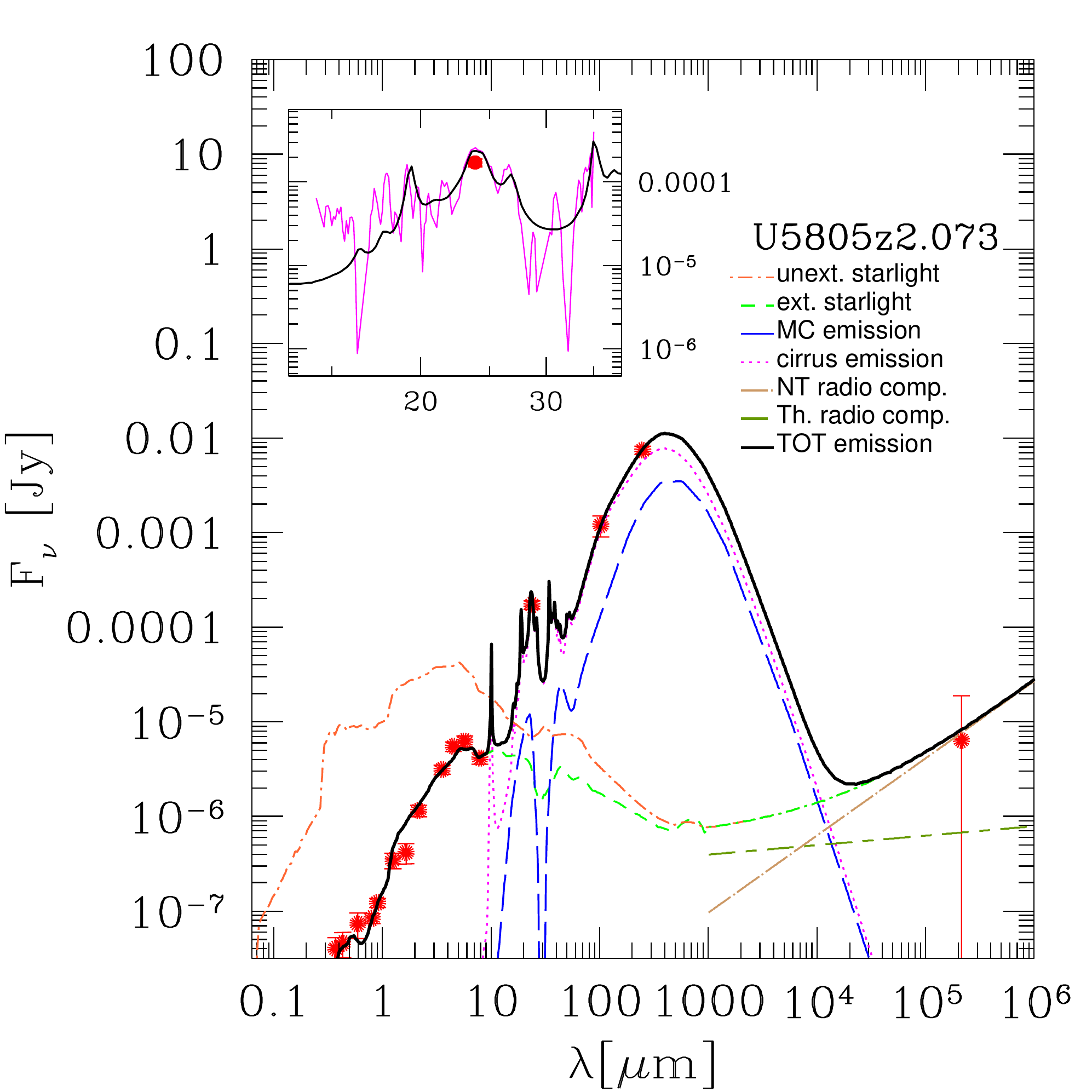}
\includegraphics[width=6.cm]{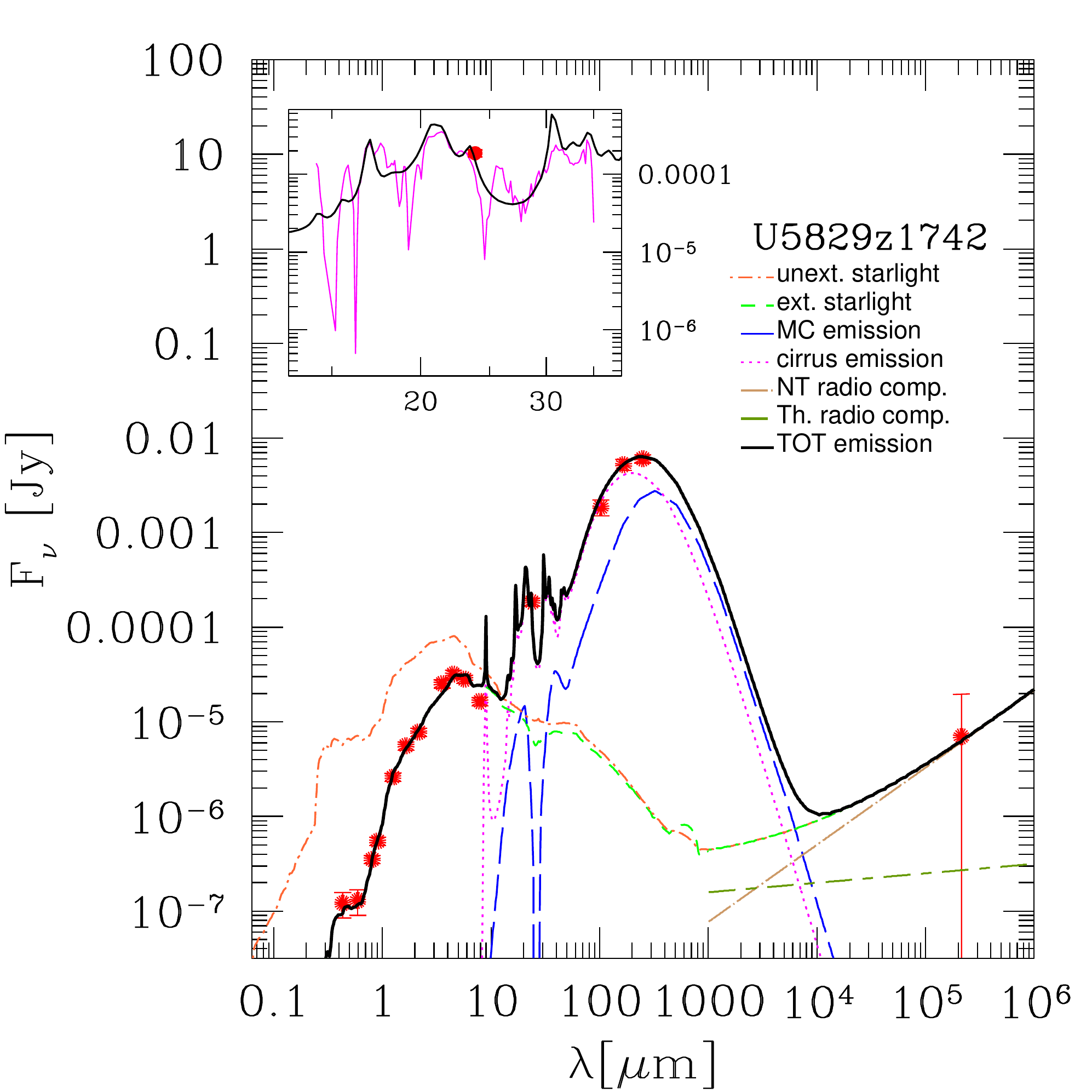}}
\centerline{
\includegraphics[width=6.cm]{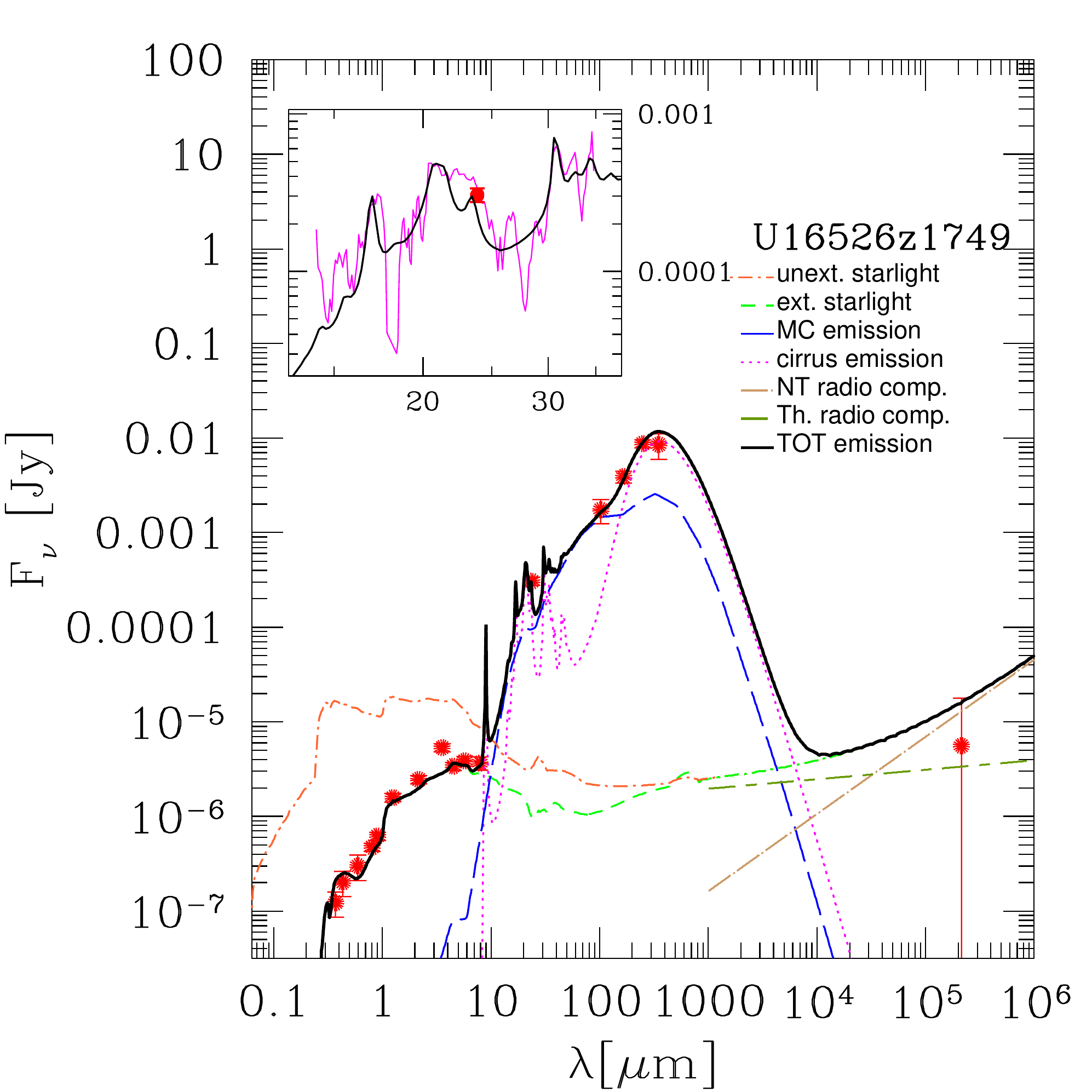}}
\contcaption{}
\label{best-fit-radiou4}
\end{figure*}

\subsubsection{Constraints on physical solutions}
\label{ll}
We recall that all the best-fits shown in
Fig.~\ref{best-fit-radiol1} have been obtained by assuming the
spherically symmetric King profile.

F10 performed a rough analysis of their morphology, based on
HST ACS images in the $B,V,i,z$ filters. Many of these have
been found to be extended sources characterized by many clumps.
Some of them, instead, have been found to be very compact
objects. All of them show red colors indicative of significant
dust contribution. However given their high redshifts it is
difficult to draw a complete picture of their morphology. We
believe our approximation of spheroidal geometry to be a good
choice for these objects, as further confirmed by our results.

We have anyway tested on this sample of high-$z$ (U)LIRGs also a comprehensive library of disk galaxies ($\sim$ 400,000 models described in Section~\ref{sedsfhBzK}), and we have found that for almost all of them a spheroidal geometry is the best choice (see discussion below).

\begin{figure}
\centerline{
\includegraphics[width=9.cm]{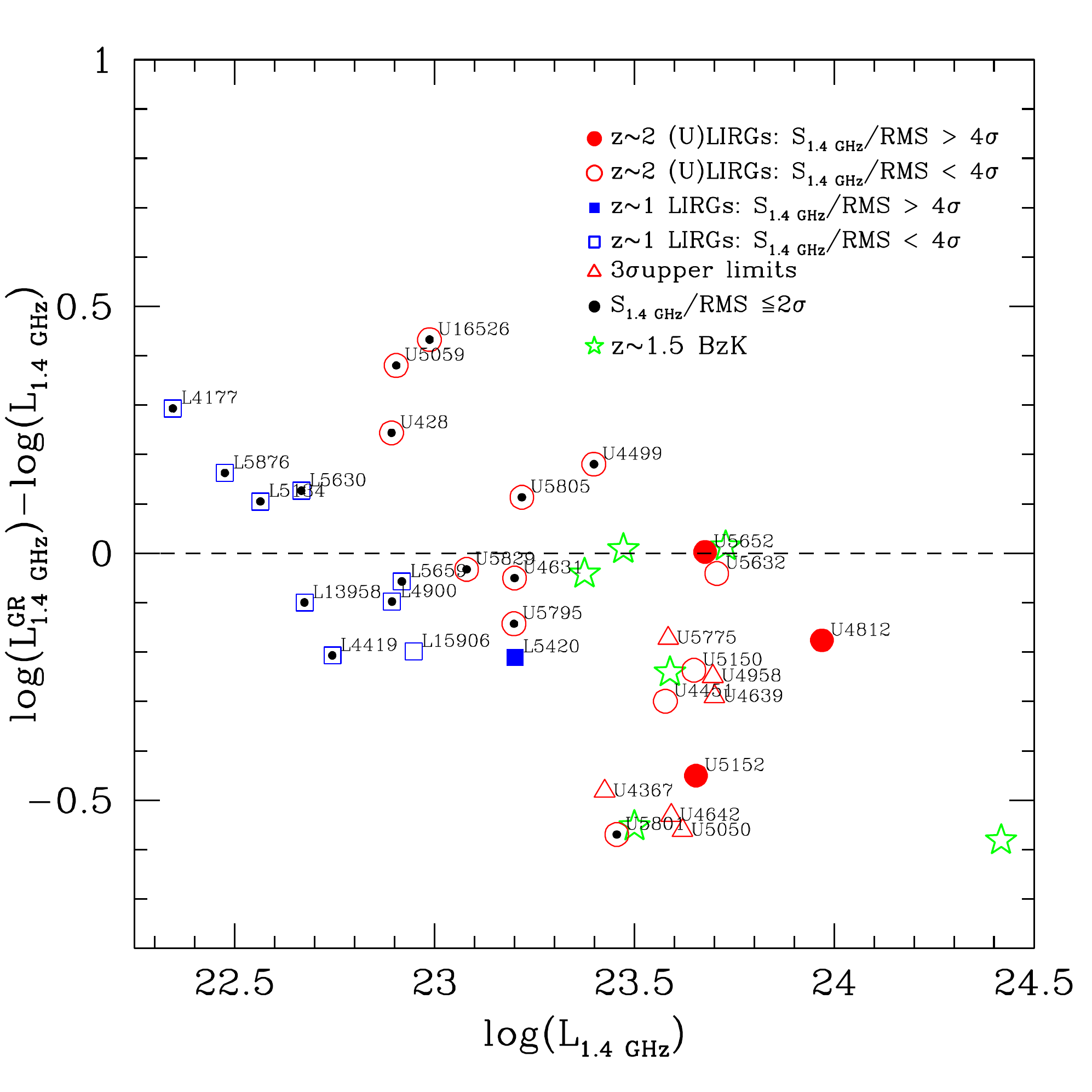}}
\caption{Logarithmic difference between the rest-frame $L^{GR}_{1.4 GHz}$ luminosity computed by GRASIL and that one estimated from the observed flux density at 1.4 GHz using the relation of Eq.~\ref{eqL14}, as a function of the rest-frame observed $L_{1.4 GHz}$. 3$\sigma$ upper limits are highlighted as open red triangles. Errors on luminosities are not reported here in order to avoid crowding in the figure. They are anyway shown in the best-fits. A general trend appears in our scattered solutions with the $\Delta\,L_{1.4 GHz}$ decreasing at increasing $L_{1.4 GHz}$, meaning that at larger rest-frame $L_{1.4 GHz}$ our model tends, on average, to underpredict the observed radio emission. This indeed pertains only to the $\la 2 \sigma$ detections and no trend as a function of luminosity is instead evident when looking to the $> 4\sigma$ detections (filled symbols). We stress here that the values shown in this plot, for the high$z$ (U)LIRGs, refer to the best-fit solutions obtained in our Paper I and as we will see in Section~\ref{newsolutions} the new solutions considered for the objects \textit{U4812} and \textit{U5152} bring these two objects on the $\Delta\,L_{1.4 GHz}=0$ line.}
\label{deltaL}
\end{figure}

Figure~\ref{deltaL} quantitatively summarizes our results. It shows the logarithmic difference between the rest-frame 1.4 GHz luminosity as derived from our best-fit model and the rest-frame $L_{1.4 GHz}$ estimated directly from the observed flux density using the following relation:
\begin{equation} L_{1.4 GHz} [W Hz^{-1}] = \frac{4 \pi D_{L}^{2}(z)}{(1+z)^{1-\alpha}} S_{\nu} (1.4 GHz)
\label{eqL14}\end{equation}
with $S_{\nu}$ measured in units of $\mathrm{Wm}^{-2}\mathrm{Hz}^{-1}$ and assuming a radio spectral index $\alpha\,\sim 0.8$.
Filled symbols in the figure highlight the (U)LIRGs detected at
a high significance, i.e. $S/RMS>4\,\sigma$. Neglecting the upper limits (open triangles), we
have only four cases (\textit{U5152} - \textit{U5801} -
\textit{U16526} - \textit{U5059}) showing a difference between
the rest-frame $L_{1.4 GHz}$ model luminosity and the observed
data larger than a factor of 2.

The first case, \textit{U5152}, is a `genuine' critical case as
it is a 4$\sigma$ detection in the catalogue. For this objects
we measure a $\Delta L_{1.4 GHz}= (L^{GR}_{1.4 GHz}-L_{1.4
GHz})$ of about 0.49 dex.
%
Moreover for this object we also fail in well reproducing the FIR peak of the spectrum. The FIR modelled emission appears to be much broader and hotter than that suggested by the two IR data-points, in particular the PACS 160 $\mu\,m$ flux. The UV-to-MIR part of the spectrum is instead well reproduced. We have thus investigated for this object new solutions including a different geometry and also the possibility for a `late' burst of SF on top of its SFH in order to boost the radio contribution from young stars leaving almost unchanged the rest of the SED. These new solutions are discussed in detail in \S~\ref{newsolutions}.

The three objects \textit{U5801}, \textit{U16526} and
\textit{U5059} are instead faint detections at $\sim$ 2
$\sigma$ so we do not consider these objects as failures of our
SED-fitting solutions. Moreover their overall SED from far-UV
to radio is very well reproduced by our model, with radio
fluxes well within the error-bars. For the object
\textit{U5801}, which shows the largest discrepancy on the
radio flux, the FIR peak is defined by only one data point.
This can bring to some degeneracy in the solutions being the
contribution by young stars less constrained. 

A general trend in our scattered solutions appears in Figure~\ref{deltaL} with the $\Delta\,L_{1.4 GHz}$ decreasing at increasing $L_{1.4 GHz}$, meaning that at larger rest-frame $L_{1.4 GHz}$ our model tends to underpredict the observed radio emission. This holds, however, only for the $\la 2 \sigma$ detections. If we consider, in fact, the high S/N radio detections (filled symbols) together with the BzK galaxies not AGN dominated, there is not evidence for a variation of the $\Delta\,L_{1.4 GHz}$ as a function of luminosity. We emphasize here that all the solutions shown in the plot for the high-$z$ (U)LIRGs refer to the best-fits obtained in our previous paper BLF13 and that, considering the new solutions discussed in Section~\ref{newsolutions}, the two ULIRGs \textit{U4812} and \textit{U5152} are shifted to the $\Delta\,L_{1.4 GHz}=0$ line.
For the low S/N radio detections shown here, the average tendency of our models to underpredict the observed data could be indicative of the need of a little burst of SF on top of our best-fit SFHs, (as discussed above), able to increase the contribution by young stars to the radio emission leaving the rest of the SED unchanged. Another possibility is to consider a different IMF characterized by a larger fraction of massive young stars, with respect to the Salpeter, as the Chabrier for example.  Anyway any strong conclusion in this context is prevented by the high uncertainties in the observed radio flux densities of most of our radio faint sources.

\subsubsection{A test-case: \textit{U4812} and \textit{U5152}}
\label{newsolutions}

\textit{U4812}, together with \textit{U5652} and \textit{U5152}, are the only three objects at $z\,\sim\,2$ to have been detected at $\ga 4\,\sigma$-level and listed in the published radio catalogue by \cite{Miller2013}. These are also among the objects showing the highest fluxes in the FIR \textit{Herschel} bands. For \textit{U5652} our previous physical solution reproduces very well the radio data with a $\Delta\,L_{1.4 GHz}\sim 0$. For the remaining two ULIRGs the modelled radio emission provided by our fits tends to underpredict the observed flux densities by $\sim$ 0.2 and 0.49 dex for \textit{U4812} and \textit{U5152}, respectively. Although the discrepancy between the modelled and observed radio fluxes of \textit{U4812} is well within a factor of two, given that this object is one of the few detected at high significance in the radio band we decided to consider new solutions involving a new SED-fitting for both the objects. 

\begin{figure*}
\centerline{
\includegraphics[width=7.5cm]{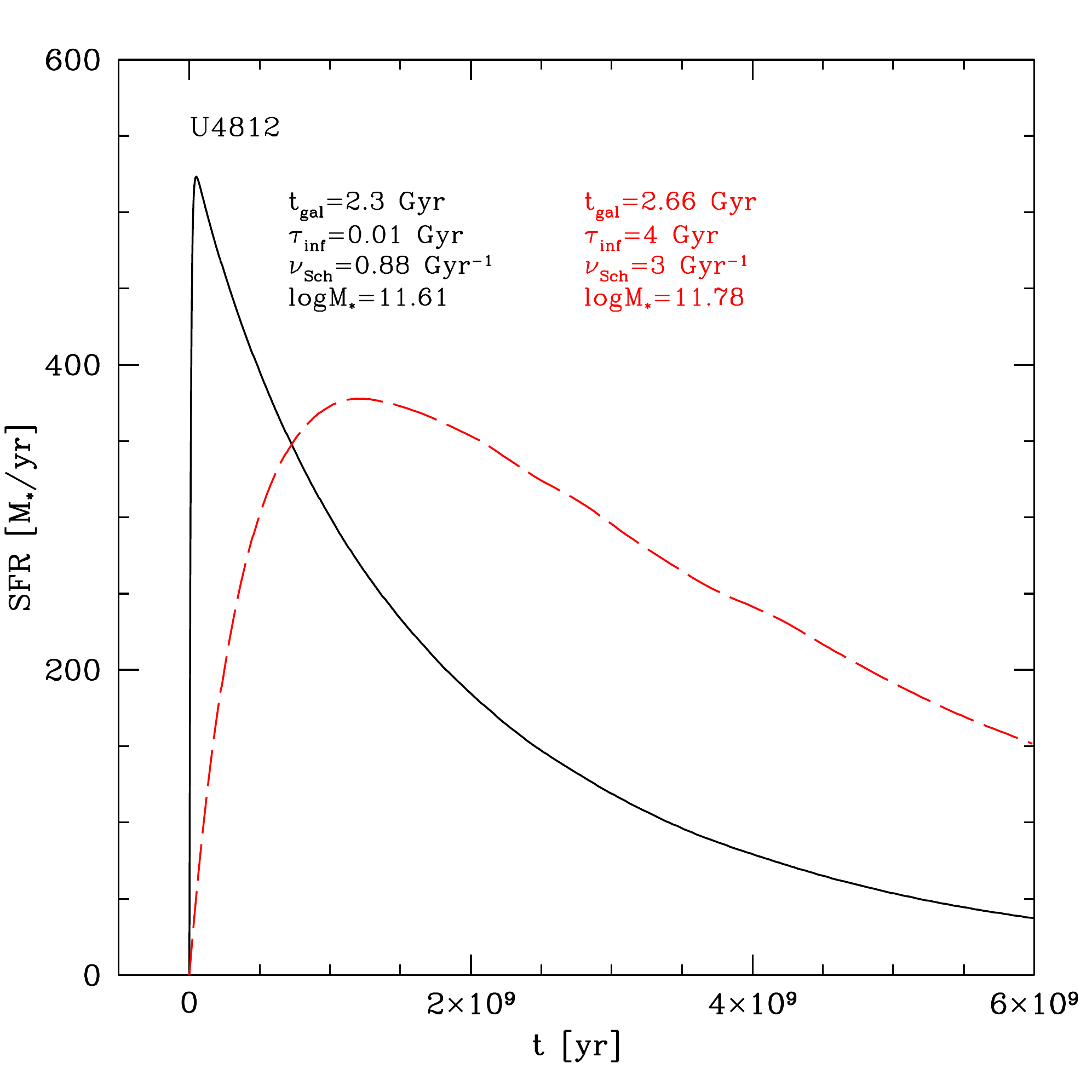}
\includegraphics[width=7.5cm]{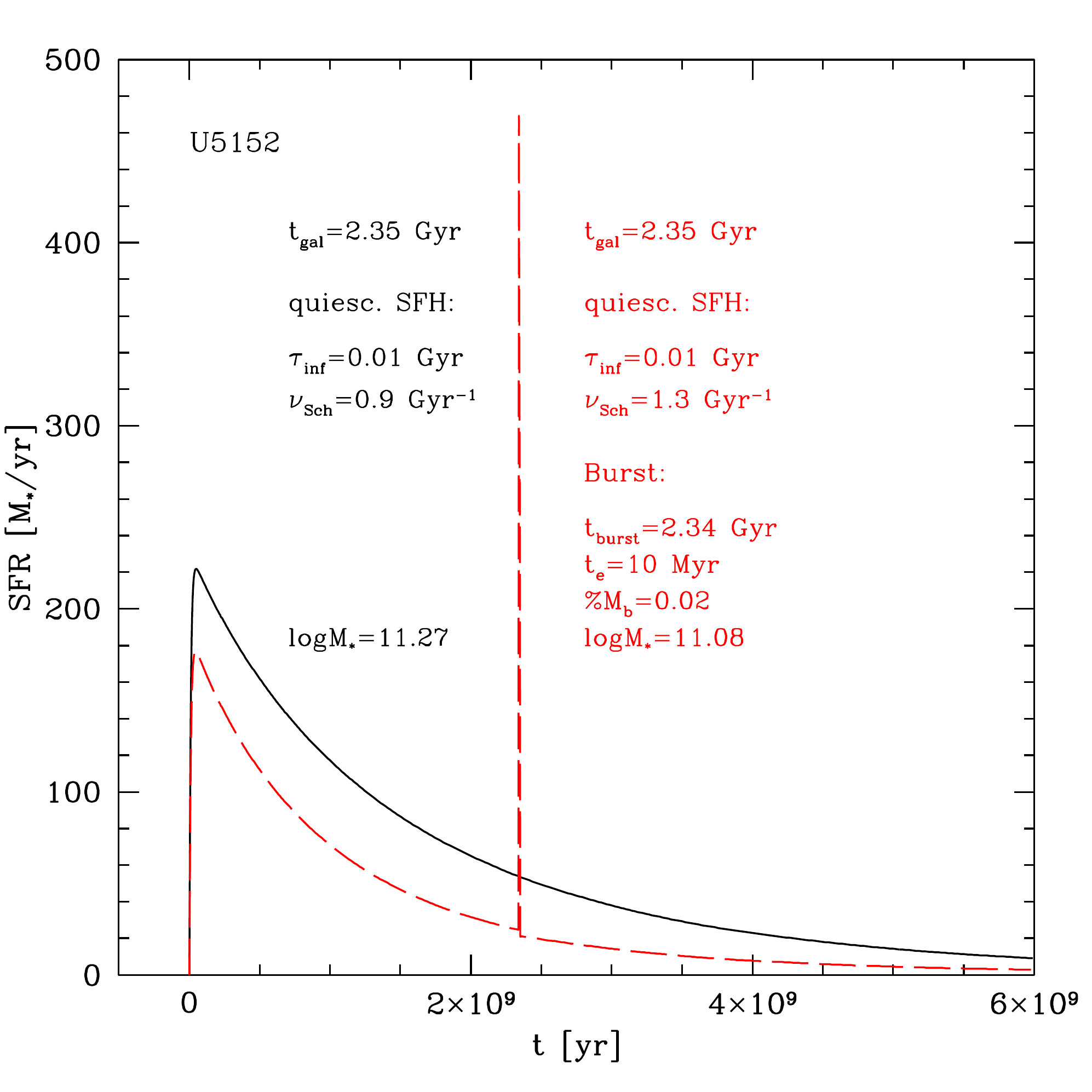}} 
\caption{Comparison between the best-fit SFH obtained with our previous analysis (black solid line) without including the radio data and the new best-fit SFH (red dashed line) resulting from the SED-fitting in which we considered also the radio data. The best-fit galaxy age ($t_{gal}$), infall timescale ($\tau_{inf}$), SF efficiency ($\nu_{Sch}$) and $M_{\star}$ estimates are also highlighted in the figure.}
\label{u4812-u5152-sfh}
\end{figure*}

\begin{figure*}
\centerline{
\includegraphics[width=7.5cm]{Figures/U4812z1930RcfluxOLD}
\includegraphics[width=7.5cm]{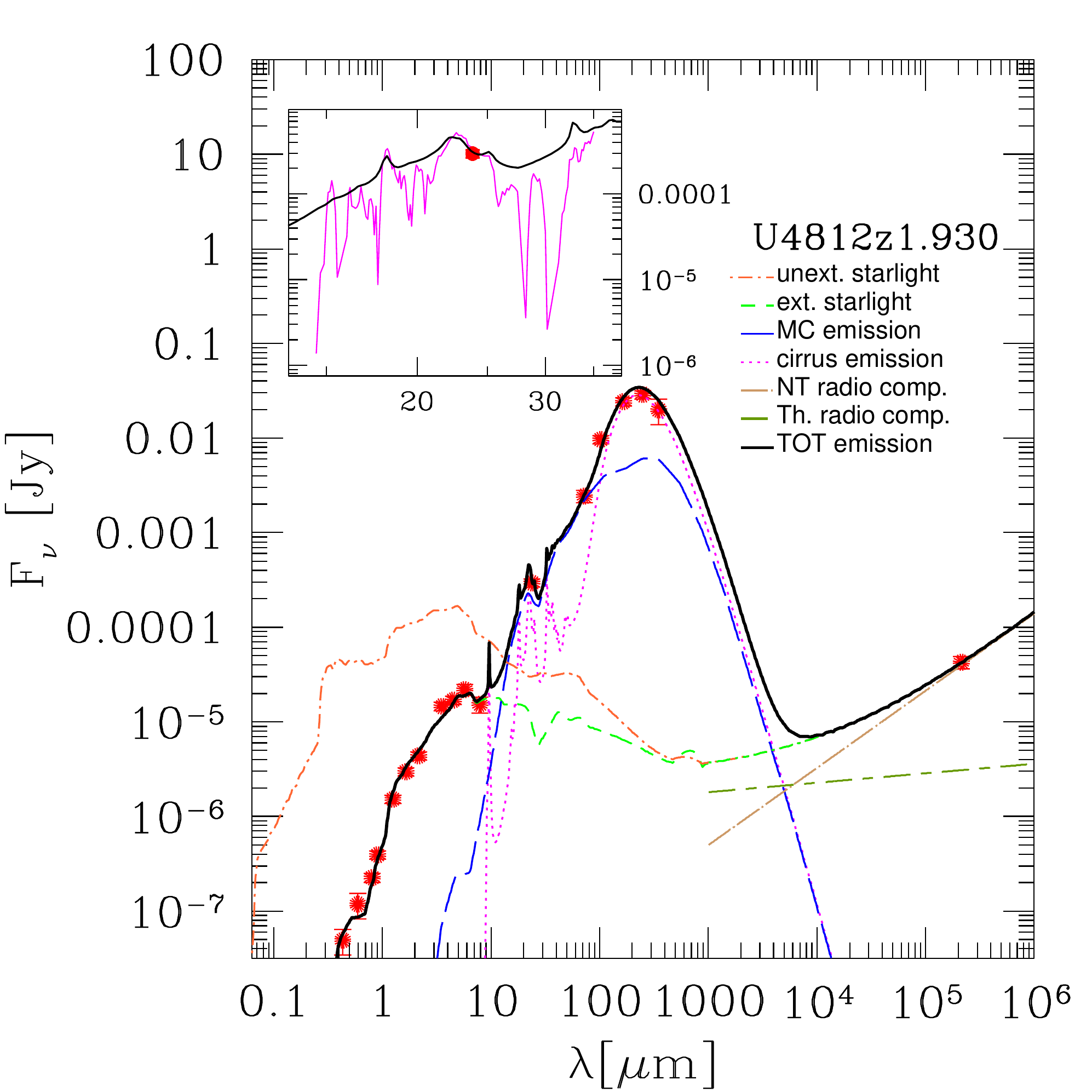}}
\centerline{
\includegraphics[width=7.5cm]{Figures/U5152z1794RcfluxOLD}
\includegraphics[width=7.5cm]{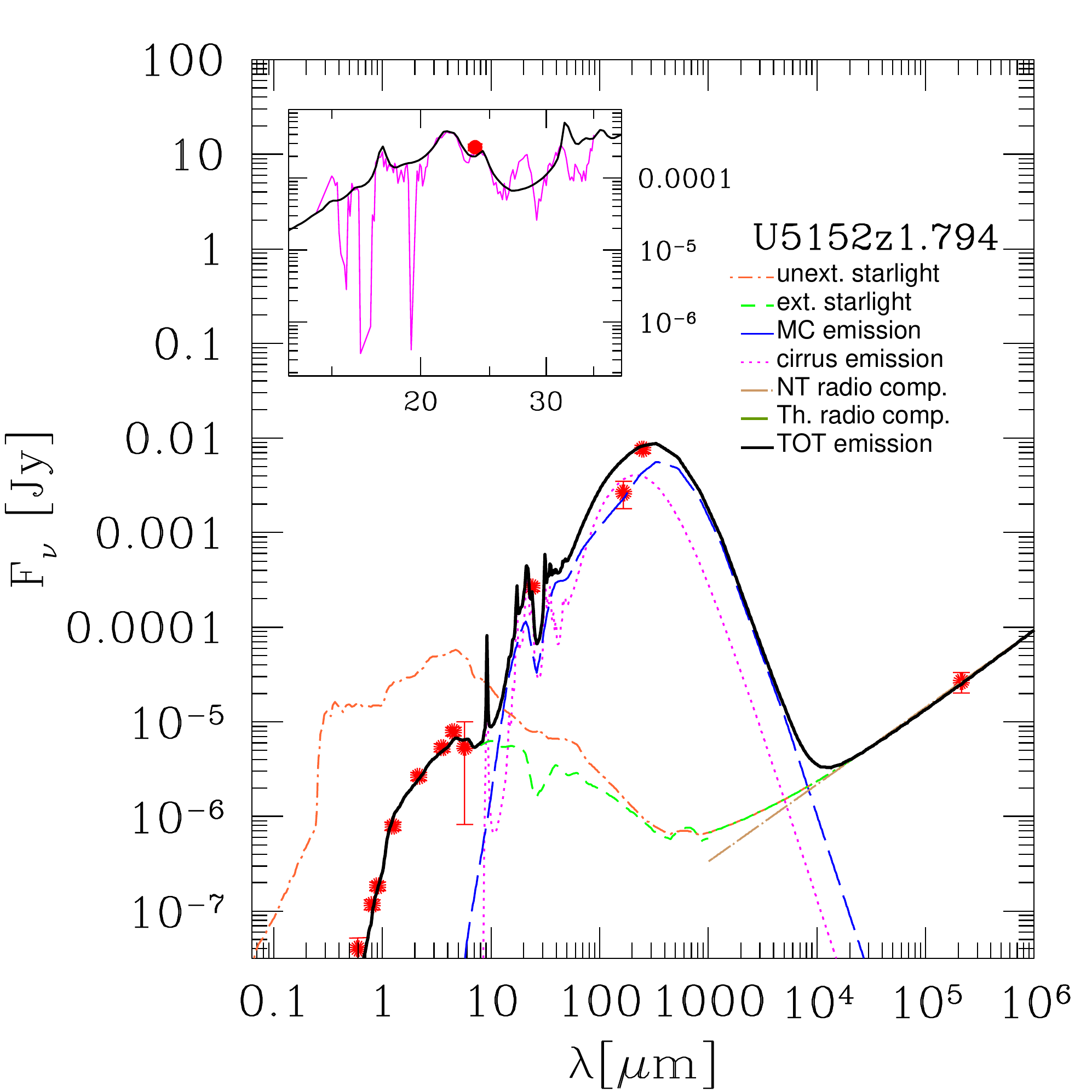}}
\caption{Best-fit SEDs of the ULIRGs \textit{U4812} (top) and \textit{U5152} (bottom) obtained before (left) and after (right) the inclusion of radio data in the fit, the latter including the new solutions considered. For the object \textit{U4812} the two plots shown in top figure refer to a different geometry, spheroidal (left) and disk (right) and different SFHs as reported in Figure~\ref{u4812-u5152-sfh}. For details see text.}
\label{u4812sed}
\end{figure*}

Based on the BLF13 analysis \textit{U4812} has a total IR luminosity of $\sim\,5.0\times10^{12} L/L_{\odot}$ corresponding to a best-fit model SFR$_{10}\,\sim$\,160 M$_{\odot}$/yr. 
It is also the object showing the largest stellar mass discrepancy (about a factor of 4-5), with respect to the estimate based on optical-only SED-fitting procedure, and the largest dust obscuration with an average value of $A_{V}\,\sim\,3.76$. According to our previous analysis its best-fit SFH is characterized by a very short infall timescale and high efficiency of star formation ($\tau_{inf}=0.01$ Gyr, $\nu_{Sch}=0.88$), corresponding to an early fast and intense SF phase with an initial burst followed by a more regular SFR (see Fig.~\ref{u4812-u5152-sfh} (left: black solid line)). The galaxy, however, is then observed a few Gyr after the peak. As shown in Figure~\ref{u4812sed} (top-left) the best-fit SED based on our previous solution works very well in the far-UV-to-sub-mm range but tends to underpredict the radio flux at 1.4 GHz. This is probably due to the significant contribution (much stronger with respect to the other objects in the sample) in our old best-fit by the cirrus component to the rest-frame MIR region.
A good compromise is given by the new solution shown on the
right panel of Fig.~\ref{u4812sed}.This has been obtained by running the SED-fitting procedure described in Sec.~\ref{results} also on an extended library of disc galaxies including more than $\sim$ 400,000 models described in Section~\ref{sedsfhBzK}. In the new solution the contribution of
cirrus emission to the FIR remains unchanged, but its role in
the MIR part of the SED is
overcome by the MC emission whose effect is that of increasing
the young component contributing to the radio continuum. In
this way we are able to reproduce the entire SED of the galaxy,
included the IRS spectrum, although the depth of the Silicate
feature and continuum appear to be better reproduced in the first case.
This new solution corresponds to the best-fit SFH shown in
Fig.~\ref{u4812-u5152-sfh} (left) as a red dashed line. This SFH is clearly different with
respect to the previous one. Here the SFH is characterized by a
longer infall timescale, of 4 Gyr and also a higher efficiency
of SF. Therefore it resembles a more gradually evolving SFH
typical of BzK galaxies at high redshifts. The best galaxy age
is found to be $2.66$ Gyr, very close to our previous estimate
of $2.3$ Gyr. In this new solution the galaxy is observed
closer to the peak of star formation, which contributes to
enhance its SFR to 316 $M_{\odot}$/yr.
Note that this best-fit has been actually obtained using the
disk geometry. This geometry in combination with the SFH
does not seem to affect significantly both the average
extinction ($A^{disk}_{V}=3.92$) and the FIR luminosity
($L_{IR}=4.60\times10^{12} L/L_{\odot}$) of this solution with
respect to the previous one in BLF13. It affects, instead, the
distribution of dust among the dense and diffuse components,
enhancing the contribution from MCs and resulting in a lower
dust mass ($LogM^{disk}_{dust}=9.00$ compared to our previous
estimate of $LogM^{sph}_{dust}=9.39$), i.e. almost a factor $2.5$ difference.
Concerning the stellar mass, the longer infall timescale causes a larger stellar mass by a factor $\la\,1.5$ with respect to our previous best-fit.
Figure~\ref{u4812-u5152-sfh} also reports the best-fit values,
in logarithmic scale, of the $M_{\star}$ relative to the two
best-fits shown in Figure~\ref{u4812sed} (top).

The other case for which we have explored a different fit is
\textit{U5152} whose best-fit SED is shown in
Fig.~\ref{u4812sed} (bottom). For this object our solution
(bottom-left) well reproduces the far-UV to sub-mm SED, (with
some uncertainty on the PACS 160 $\mu\,m$ flux), but
underestimates the observed radio data by a factor of
$\sim$\,3. Based on the BLF13 analysis and similarly to
\textit{U4812}, the best-fit SFH of this objects (shown in
Fig.~\ref{u4812-u5152-sfh} rigth), is also characterized by a
very short infall timescale, of the order of 0.01 Gyr, and high
SF efficiency ($\sim$\,0.9) typical of an early intense SF
phase with the galaxy being observed few Gyr after the peak
($t_{gal}\sim$\,2.35 Gyr).
%

In order to enhance the radio contribution from young stars
leaving almost unchanged the far-UV to FIR SED, we have added a
late starburst on top of the quiescent SFH of this galaxy.
We slightly increased the SF efficiency of the quiescent
SFR in order to limit the amount of gas available for the
burst. Anyway the gas mass involved in the burst accounts to
only a small fraction, of the order of 2\%, of the galactic
mass at that epoch. This burst has an e-folding timescale of
$\sim$ 10 Myr and takes place when the galaxy is 2.34 Gyr old,
with the observational time being 2.35 Gyr (same as in BLF13).
The SFH corresponding to this new solution is shown in
Fig.~\ref{u4812-u5152-sfh} (right) as red dashed line and
compared to the old prescription (black solid line). The red
dashed vertical line in correspondence of $t$=2.34 Gyr
represents the burst.

Under these new prescriptions we are able to reproduce fairly
well the entire SED of the galaxy, including the IRS spectrum
and radio data. The best-fit SED corresponding to the new
solution is shown in Fig.~\ref{u4812sed} (bottom-right). 
In the new fit, the contribution of MCs emission in the MIR
region is enhanced, mainly due to a lower optical depth of MCs as
compared to the previous fit.

This new solution does not seem to affect significantly the main
physical properties of this galaxy. Both the average extinction
($A_{V}=3.20$) and FIR luminosity ($L_{IR}=1.27\times10^12$)
well agree with the previous estimates ($A_{V}=3.27$ and
L$_{IR}=1.09\times10^12$). However given the presence of a
recent burst, its SFR averaged over the last 10 Myr increases
by a factor of $\sim$\,3 with respect to the old solution
(SFR$\sim$\,54 $M_{\odot}$/yr) up to 171 $M_{\odot}$/yr.
Differently from \textit{U4812}, where the effect of a
different geometry (disk vs spheroids) brings to a greater dust
mass by a factor $\sim$\,2.5, here the dust mass corresponding
to the new solution ($Log(M_{dust})$=8.34) is a factor
$\sim$\,3 lower than the BLF13 one, due to the enhancement of
MCs contribution to the MIR (see Fig.~\ref{u4812sed} - bottom)
also dominating at sub-mm wavelengths. For the stellar mass the
new solution provides a $LogM_{\star}$=11.05, a factor 1.5
lower than that one obtained with the old prescription (see
Fig.~\ref{u4812-u5152-sfh} (right)) but still larger compared
to the one based on optical-only SED-fitting procedure.
The lower stellar mass can be explained as due to the
decreased amount of evolved stars.

Our analysis thus demonstrates, at least for these two
test-cases, that radio data are crucial to break model
degeneracies and further constrain the SFH and $M_{\star}$ of
the galaxy.

\begin{figure*}
\centerline{
\includegraphics[width=10.cm]{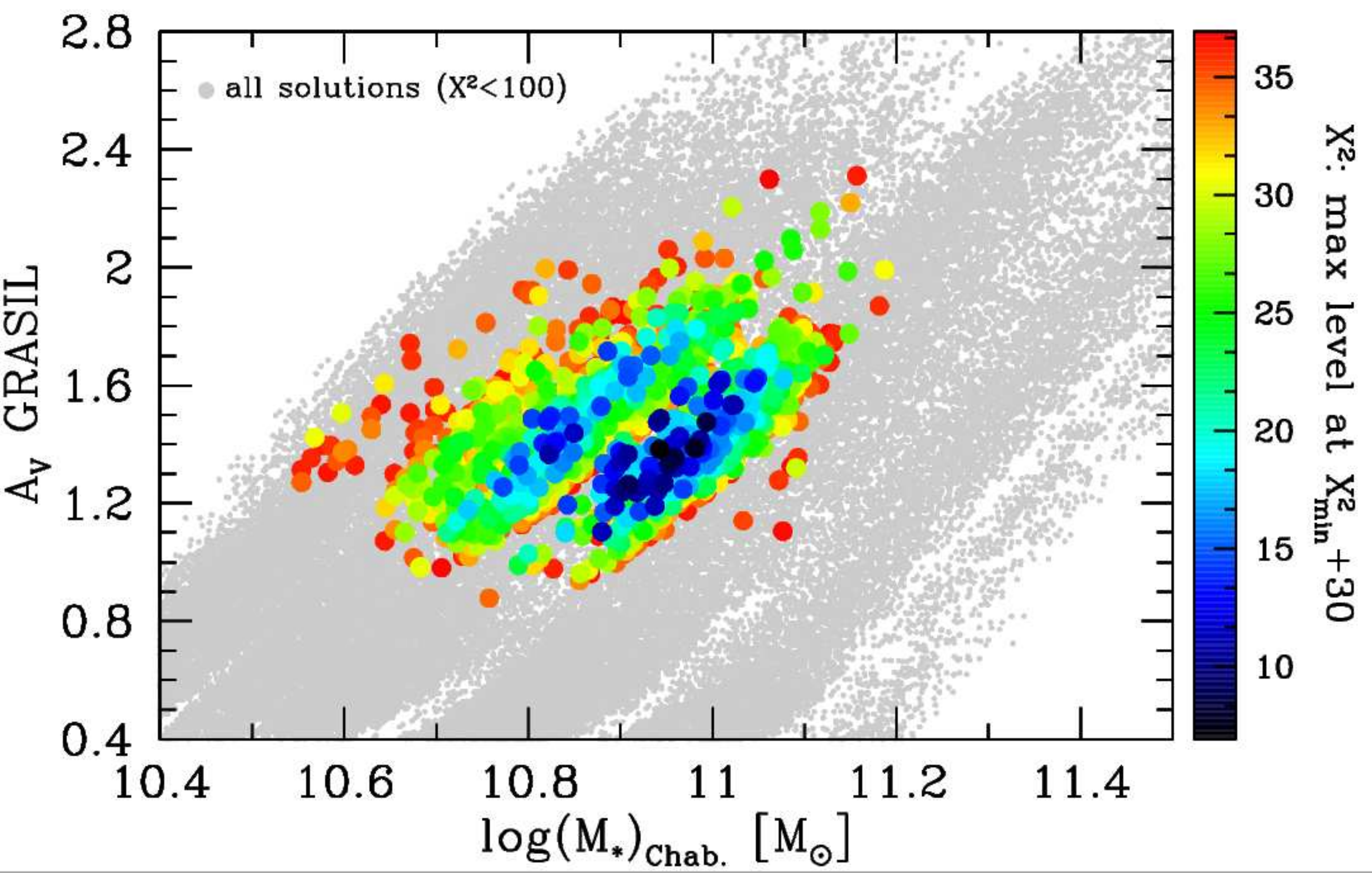}}
\centerline{
\includegraphics[width=10.cm]{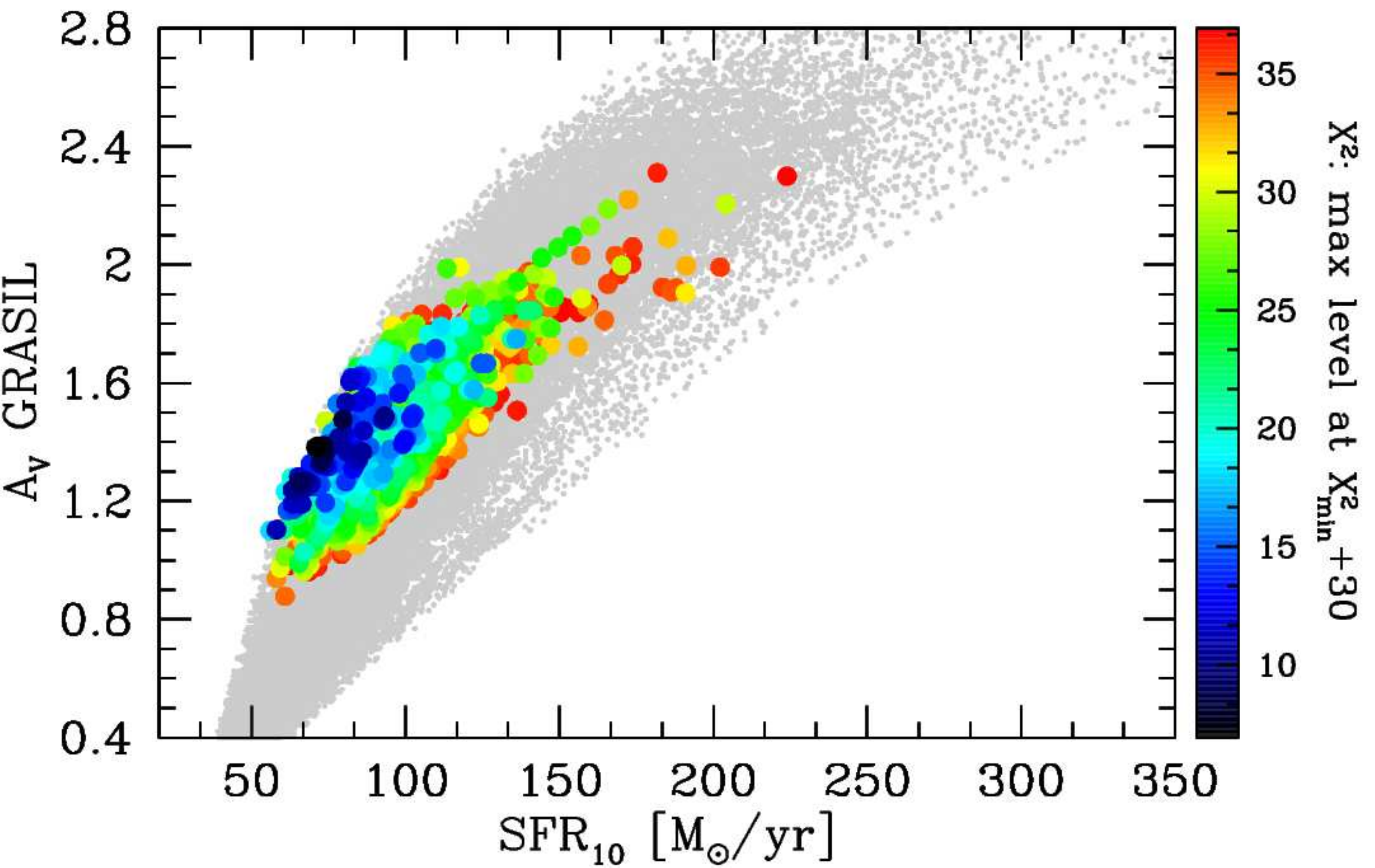}}
\centerline{
\includegraphics[width=10.cm]{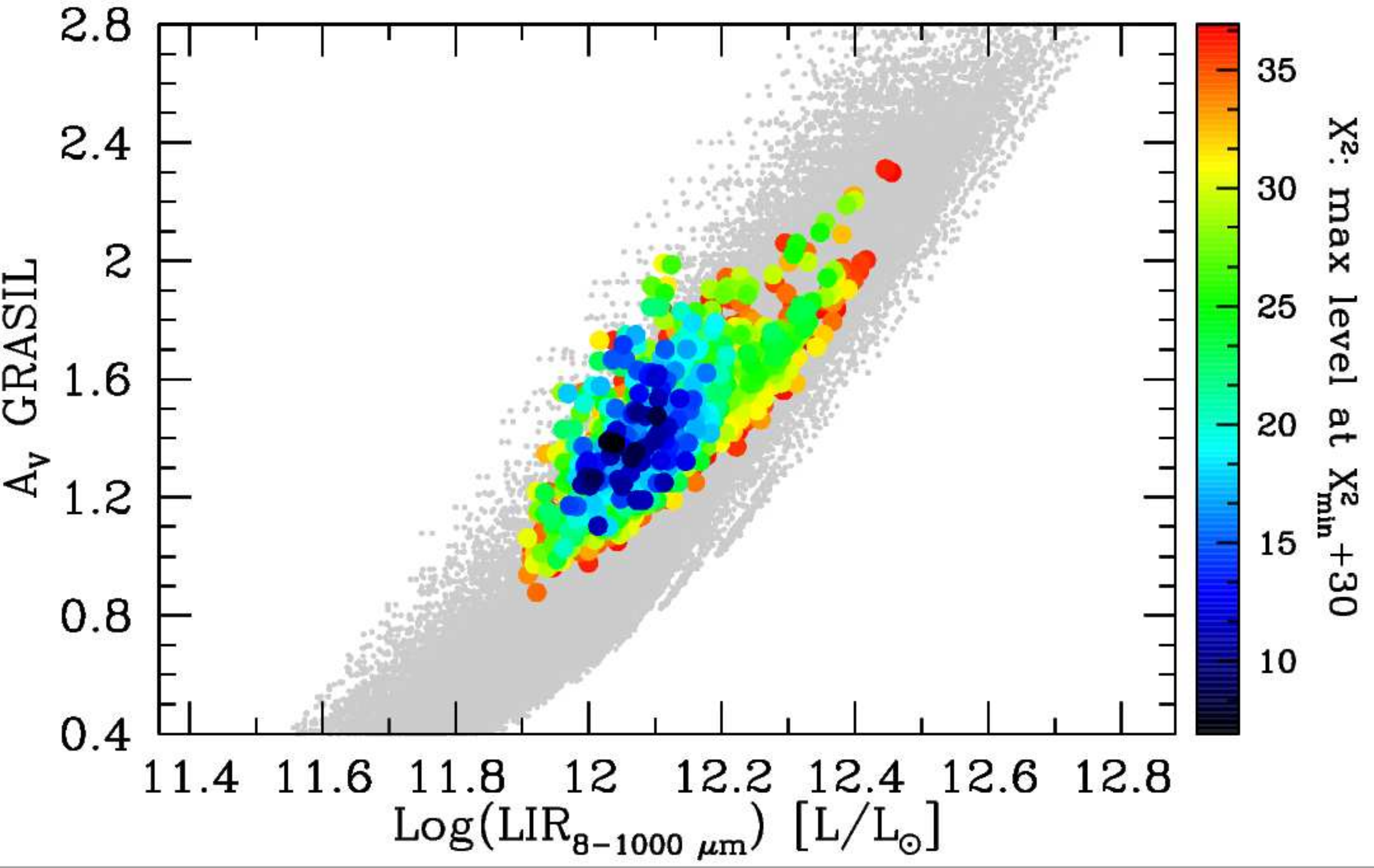}}
\caption{$\chi^{2}$ values for spectral models of a typical z$\sim2$ ULIRG colour-coded ($\chi^{2}$ increasing from dark to light colours) as a function of the average extinction and the stellar mass (TOP), SFR (CENTRE) and total IR Luminosity (BOTTOM). We show here all the solutions with $\chi^{2}$ $\leq$ 100 (gray dots) in order to give an idea of the wide range of parameters (and solutions) explored, at fixed geometry, and those with $\chi^{2}$ within $\chi^{2}_{min}$+30 (colored points). In addition to the variation of the input GRASIL parameters fully discussed in BLF13, we consider here also the variation of the two parameters characterizing the SFH, namely $\nu_{Sch}$ and $\tau_{inf}$. As evident from the figure not much degeneracy is seen among the model parameters. The typical uncertainty on our best-fit M$_{\star}$, A$_{V}$, SFR$_{10}$ and $L_{IR}$ are, respectively, 0.2 dex, 0.3 mag, 0.2 dex and 0.13 dex thus well within the typical uncertainties for this kind of analysis.}
\label{degen}
\end{figure*}

Table~\ref{tabulirgprop} summarizes all the best-fit physical properties
of the high-$z$ (U)LIRGs including the updated solutions for
the two cases discussed in this section. As discussed in our previous paper, when considering in addition to the many combinations of GRASIL parameters also the different combinations of parameters ruling the SFH, the typical uncertainties on our best-fit $L_{IR}$, M$_{\star}$, A$_{V}$ and SFR$_{10}$ are, respectively, 0.13 dex, 0.2 dex, 0.3 mag, and 0.2 dex thus well within the typical uncertainties for this kind of analysis. 
Figure \ref{degen} gives a more specific idea of the degeneracies in our model solutions, between the
average extinction, stellar mass, SFR and IR luminosity color-coded by the value of $\chi^{2}$, for a typical z$\sim$2 ULIRG. As we see, among the many solutions considered, acceptable best-fits, within $\chi^{2}_{min} + 30$, are clearly identified in the parameter space and not much degeneracy is apparent. The gray dots shown in the panels represent all the solutions having $\chi^{2}\la 100$. This large value has been chosen in order to give a better idea of the parameter coverage of our libraries. The degeneracy shown in this figure refers to the case of fixed geometry (King's profile here). Except for the ULIRG \textit{U4812} discussed above for which we have been able to obtain a better fit to the entire SED (the radio data here has been crucial) by considering disc geometries for all the other ULIRGs in the sample we did not get any physical solution when exploring a geometry different from a spheroidal one. We cannot therefore provide in this case a reliable measurement of model degeneracies including the assumption of different geometries. Based on the unique case investigated here we can indeed notice that the use of a disc geometry results in a more `classical' and less `bursty' SFH more typical of normal SF galaxies (see discussion below in Sec.~\ref{sedsfhBzK}). The typical uncertainties on the main physical quantities listed above settle, in this case, around a factor of 2.
%
\begin{table*}
\centering
\scriptsize
\begin{tabular}{|l|l|l|l|l|l|l|l|l|l|l|l|l|l|l|l|}
\hline
\hline
  \multicolumn{1}{|c|}{ID} &
  \multicolumn{1}{c|}{z} &
  \multicolumn{1}{c|}{$\chi^{2}$} &
  \multicolumn{1}{c|}{L$_{IR}$} &
  \multicolumn{1}{c|}{L$_{IR,cirr}$} &
  \multicolumn{1}{c|}{L$_{IR,MC}$} &
  \multicolumn{1}{c|}{SFR$_{10}$} &
  \multicolumn{1}{c|}{SFR$_{K}$} &
  \multicolumn{1}{c|}{M$_{\star}$} &
  \multicolumn{1}{c|}{M$_{gas}$} &
  \multicolumn{1}{c|}{A$_{V}$} &
  \multicolumn{1}{c|}{A$_{FUV}$} &
  \multicolumn{1}{c|}{q$_{TIR}$} &
  \multicolumn{1}{c|}{L$_{1.4 GHz}^{GR}$} &
  \multicolumn{1}{c|}{L$_{1.4 GHz}$} &
  \multicolumn{1}{l|}{S/N}\\
      &        &   & (L/L$_{\odot}$)   & (L/L$_{\odot}$) & (L/L$_{\odot}$) & (M$_{\odot}$/yr) & (M$_{\odot}$/yr) & (M$_{\odot}$) & (M$_{\odot}$) & & & & (W Hz$^{-1}$) & (W Hz$^{-1}$) & radio det.\\
\hline
\hline
  L4177 & 0.842 & 4.69 &  2.11E11 & 1.25E11 & 8.54E10 & 15 & 21 & 8.38E10 & 1.84E10 & 0.94 & 2.18 & 2.70 & 4.35E22 & 2.21E22 & $\la 2 \sigma$\\
  L4419 & 0.974 & 4.81 & 2.06E11 & 1.11E11 & 9.27E10 & 12 & 20 & 1.86E11 & 1.39E10 & 0.75 & 3.06 & 2.79 & 3.45E22 & 5.55E22 & $\sim 2 \sigma$\\
  L4900 & 1.047 & 1.89 & 4.46E11 & 3.27E11 & 1.17E11 & 22 & 45 & 1.70E11 & 3.32E10 & 1.58 & 3.80 & 2.87 & 6.25E22 & 7.83E22 & $\sim 2 \sigma$\\
  L5134 & 1.039 & 2.59 & 4.95E11 & 3.93E11 & 1.00E11 & 16 & 49 & 2.08E11 & 1.76E10 & 2.79 & 7.22 & 3.04 & 4.67E22 & 3.67E22 & $\la 2 \sigma$\\
  L5420 & 1.068 & 5.00 & 6.98E11 & 5.66E11 & 1.31E11 & 32 & 70 & 1.73E11 & 6.50E10 & 1.90 & 3.46 & 2.87 & 9.79E22 & 1.59E23 & $\sim 5 \sigma$\\
  L5630 & 0.997 & 1.09 & 3.10E11 & 2.51E11 & 5.76E10 & 19 & 31 & 7.06E10 & 3.90E10 & 1.04 & 1.75 & 2.71 & 6.21E22 & 4.64E22 & $\la 2 \sigma$\\
  L5659 & 1.044 & 3.80 & 5.30E11 & 3.68E11 & 1.60E11 & 25 & 53 & 1.63E11 & 5.00E10 & 2.78 & 6.09 & 2.87 & 7.26E22 & 8.28E22 & $2-3 \sigma$\\
  L5876 & 0.971 & 0.44 & 2.24E11 & 1.41E11 & 8.10E10 & 15 & 22 & 1.63E11 & 2.63E10 & 0.73 & 1.97 & 2.72 & 4.35E22 & 3.00E22 & $\la 2 \sigma$\\
  L13958 & 0.891 & 2.58 & 2.49E11 & 1.82E11 & 6.57E10 & 13 & 245 & 6.68E10 & 1.01E10 & 1.34 & 2.76 & 2.83 & 3.76E22 & 4.72E22 & $\sim 2 \sigma$\\
  L15906 & 0.976 & 1.75 & 4.01E11 & 3.11E11 & 8.94E10 & 19 & 40 & 8.22E10 & 2.35E10 & 2.27 & 3.81 & 2.87 & 5.610E22 & 8.86E22 & $3-4 \sigma$\\	
  U428 & 1.783 & 1.79 & 9.60E11 & 6.70E11 & 2.84E11 & 48 & 96 & 2.38E11 & 5.77E10 & 2.45 & 4.73 & 2.86 & 1.37E23 & 7.80E22 & $\la 2 \sigma$\\
  U4367 & 1.624 & 4.53 & 6.78E11 & 3.93E11 & 2.81E11 & 31 & 68 & 3.18E11 & 1.92E10 & 1.84 & 6.08 & 2.9 & 8.79E22 & 2.66E23 & $3 \sigma$ upp. lim.\\
  U4451 & 1.875 & 3.47 & 1.27E12 & 8.48E11 & 4.23E11 & 63 & 127 & 2.14E11 & 7.02E10 & 3.14 & 5.80 & 2.84 & 1.90E23 & 3.78E23 & $2-3 \sigma$\\
  U4499 & 1.956 & 2.26 & 2.31E12 & 1.26E12 & 1.05E12 & 122 & 231 & 4.24E11 & 1.96E11 & 3.31 & 7.03 & 2.80 & 3.79E23 & 2.50E23 & $\la 2 \sigma$\\
  U4631 & 1.841 & 4.18 & 8.18E11 & 5.78E11 & 2.37E11 & 46 & 82 & 1.59E11 & 5.85E10 & 2.07 & 3.58 & 2.78 & 1.41E23 & 1.58E23 & $\la 2 \sigma$\\
  U4639 & 2.112 & 4.74 & 1.34E12 & 7.10E11 & 6.28E11 & 75 & 134 & 1.20E11 & 1.50E11 & 1.40 & 2.92 & 2.73 & 2.57E23 & 5.02E23 & $3 \sigma$ upp. lim.\\
  U4642 & 1.898 & 5.00 & 6.33E11 & 4.15E11 & 2.15E11 & 38 & 63 & 1.27E11 & 4.72E10 & 2.15 & 3.89 & 2.75 & 1.15E23 & 3.91E23 & $3 \sigma$ upp. lim.\\
  U4812 & 1.93 & 1.27 & 4.55E12 & 2.93E12 & 1.61E12 & 316 & 455 & 6.04E11 & 1.05E11 & 3.92 & 6.98 & 2.67 & 9.98E23 & 9.31E23 & $> 5 \sigma$ \\
  U4958 & 2.118 &  1.58 & 1.43E12 & 5.46E11 & 8.80E11 & 82 & 143 & 1.31E11 & 1.63E11 & 1.92 & 4.01 & 2.72 & 2.80E23 & 4.97E23 & $3 \sigma$ upp. lim.\\
  U5050 & 1.938 & 3.17 & 6.90E11 & 3.25E11 & 3.58E11 & 40 & 69 & 3.76E11 & 2.25E10 & 1.00 & 4.43 & 2.79 & 1.15E23 & 4.17E23 & $3 \sigma$ upp. lim.\\
  U5059 & 1.769 & 1.04 & 8.91E11 & 5.05E11 & 3.83E11 & 57 & 89 & 1.22E11 & 1.14E11 & 1.13 & 2.53 & 2.68 & 1.93E23 & 8.02E22 & $\la 2 \sigma$\\
  U5150 & 1.898 & 1.91 &  1.64E12 & 1.32E12 & 3.13E11 & 77 & 164 & 1.53E11 & 1.28E11 & 2.89 & 4.55 & 2.81 & 2.58E23 & 4.46E23 & $3-4 \sigma$\\
  U5152 & 1.794 & 2.74 &  1.27E12 & 6.78E11 & 5.94E11 & 171 & 127 & 1.22E11 & 1.60E10 & 3.20 & 6.78 & 2.46 & 5.48E23 & 4.51E23 & $\sim 4 \sigma$ \\
  U5632 & 2.016 & 1.21 &  2.45E12 & 1.66E12 & 7.86E11 & 137 & 245 & 1.82E11 & 1.52E11 & 1.90 & 3.25 & 2.74 & 4.63E23 & 5.08E23 & $3-4 \sigma$\\
  U5652 & 1.618 & 2.89 &  2.60E12 & 1.10E12 & 1.49E12 & 151 & 260 & 3.81E11 & 1.89E11 & 3.72 & 8.02 & 2.75 & 4.78E23 & 3.64E23 & $\sim 4 \sigma$ \\
  U5775 & 1.897 & 3.76 &  1.23E12 & 9.91E11 & 2.40E11 & 76 & 123 & 1.01E11 & 8.48E10 & 2.69 & 4.21 & 2.69 & 2.59E23 & 3.84E23 & $3 \sigma$ upp. lim.\\
  U5795 & 1.703 & 5.00 & 6.98E11 & 4.66E11 & 2.31E11 & 34 & 70 & 8.21E10 & 8.82E10 & 2.47 & 4.70 & 2.8 & 1.14E23 & 1.58E23 & $\la 2 \sigma$\\
  U5801 & 1.841 & 4.0 &  4.58E11 & 2.70E11 & 1.86E11 & 26 & 46 & 1.02E11 & 3.32E10 & 2.57 & 5.25 & 2.79 & 7.70E22 & 2.86E23 & $2-3 \sigma$\\
  U5805 & 2.073 & 3.04 &  1.34E12 & 1.01E12 & 3.31E11 & 66 & 134 & 1.70E11 & 1.10E11 & 3.69 & 5.60 & 2.81 & 2.15E23 & 1.65E23 & $\la 2 \sigma$\\
  U5829 & 1.742 & 2.48 &  8.38E11 & 6.17E11 & 2.17E11 & 39 & 84 & 3.34E11 & 3.17E10 & 2.36 & 4.44 & 2.90 & 1.12E23 & 1.20E23 & $\la 2 \sigma$\\
  U16526 & 1.749 & 3.10 &  1.10E12 & 6.06E11 & 4.99E11 & 74 & 110 & 2.08E10 & 1.24E11 & 2.62 & 4.45 & 2.64 & 2.63E23 & 9.71E22 & $\la 2 \sigma$\\
\hline
\hline
\end{tabular}
\caption[Best-fit physical parameters of high-$z$ (U)LIRGs including radio data]{Best-fit physical parameters of high-$z$ (U)LIRGs including radio data.}
\label{tabulirgprop}
\end{table*}
%

\subsection{Best-fit SEDs and SFHs of $z\,\sim\,1.5$ BzK star forming galaxies}
\label{sedsfhBzK}

We discuss here the results relative to the 6 $z\,\sim\,1.5$ BzK SF galaxies presented in \S~\ref{bzk}. 

In addition to the library of star forming spheroids discussed above, it was particularly important to consider for these objects also the model libraries with disc geometry. The scale lengths $R_{d}$ and $z_{d}$ of the double exponential profile have been assumed to be equal for stars and dust in order to limit the number of free parameters of the model. Several values for these scale lengths have been
considered in combination with the other model parameters
widely discussed in BLF13. Concerning the SFH, we have
considered both SFHs typical of normal SF galaxies, namely
long infall timescales (1-5 Gyr) and moderate to high star
formation efficiencies (0.5-3.0 Gyr$^{-1}$), and more `bursty'
SFHs characterized by shorter infall timescales. We have built
in this way a library including more than $400,000$ disk spectra.

For each BzK galaxy of the sample, we have run the
SED-fitting procedure on the full model library.
Based on the $\chi^{2}_{\nu}$ minimization procedure
and physical parameter analysis, we have obtained the best-fits
shown in Figure~\ref{bestfitbzk1} all having a disk geometry,
in agreement with D10 morphological analysis.

For all the six BzK galaxies in the sample, we are able to reproduce the observed SEDs from far-UV-to-sub-mm very well.
Among these galaxies, 3/6 (\textit{BzK-16000, BzK-21000, BzK-17999}) have also modelled radio emission in perfect agreement with radio data,
1/6 (\textit{BzK-4171}) has modelled radio fluxes in agreement with radio data within a factor of $\la$\,1.5 and for 2 out to 6 (\textit{BzK-12591, BzK-25536})
objects our solutions appear to underpredict the radio data by a factor larger than $\sim$\,2. All these results are quantitatively summarized in Fig.~\ref{deltaL}
where the $z\sim1.5$ BzKs are represented by the starred symbols.

The two objects showing a `deficit' of the models with respect to the radio data, namely \textit{BzK-12591} and \textit{BzK-25536}, also show a low value of the observed
FIR-Radio Luminosity ratio $q_{\mathrm{TIR}}$, both $\sim$\,2.06 (see \S~\ref{qtirsection} for details).
The galaxy \textit{BzK-12591}, as discussed in \S~\ref{bzk},
shows a strong bulge in HST imaging and has a possible
detection of [NeV]$\lambda$\,3426 $\AA$ emission line,
indicative of the presence of an AGN. As our model does not
include an AGN component, we tend to underpredict the radio
emission for this object. The other galaxy, \textit{BzK-25536},
has no explicit indication for a presence of an AGN but shows
the same low value of $q_{\mathrm{TIR}}$. Apart from a hidden AGN, it has been argued that also star forming galaxies in dense environments or particularly star forming phase may show such low q values. Miller \& Owen~(2001) observed cluster galaxies with no sign of AGN and low-q values, which they ascribed to thermal pressure of the ICM causing a compression of the galaxy magnetic field and therefore a radio excess. On the other hand, B02 proposed for these same cases that an environment induced fast damping of the SF could
give rise to an apparent radio excess since radio emission fades less rapidly than the IR.    

However given the nature of these galaxies and their selection we suggest that an optically-obscured AGN more probably contributes to the radio emission of this object. 
A `radio excess' for these two galaxies has been also independently claimed  by Magdis and collaborators (private communication).
The predicted and observed $q_{\mathrm{TIR}}$ values of each BzK galaxy in the sample, estimated according to the Eq.~\ref{qtir_eq}, are listed in Table~\ref{BzKpar1} and discussed in \S~\ref{qtirsection}.

\begin{figure*}
\centerline{
\includegraphics[width=12.cm,height=10.5cm]{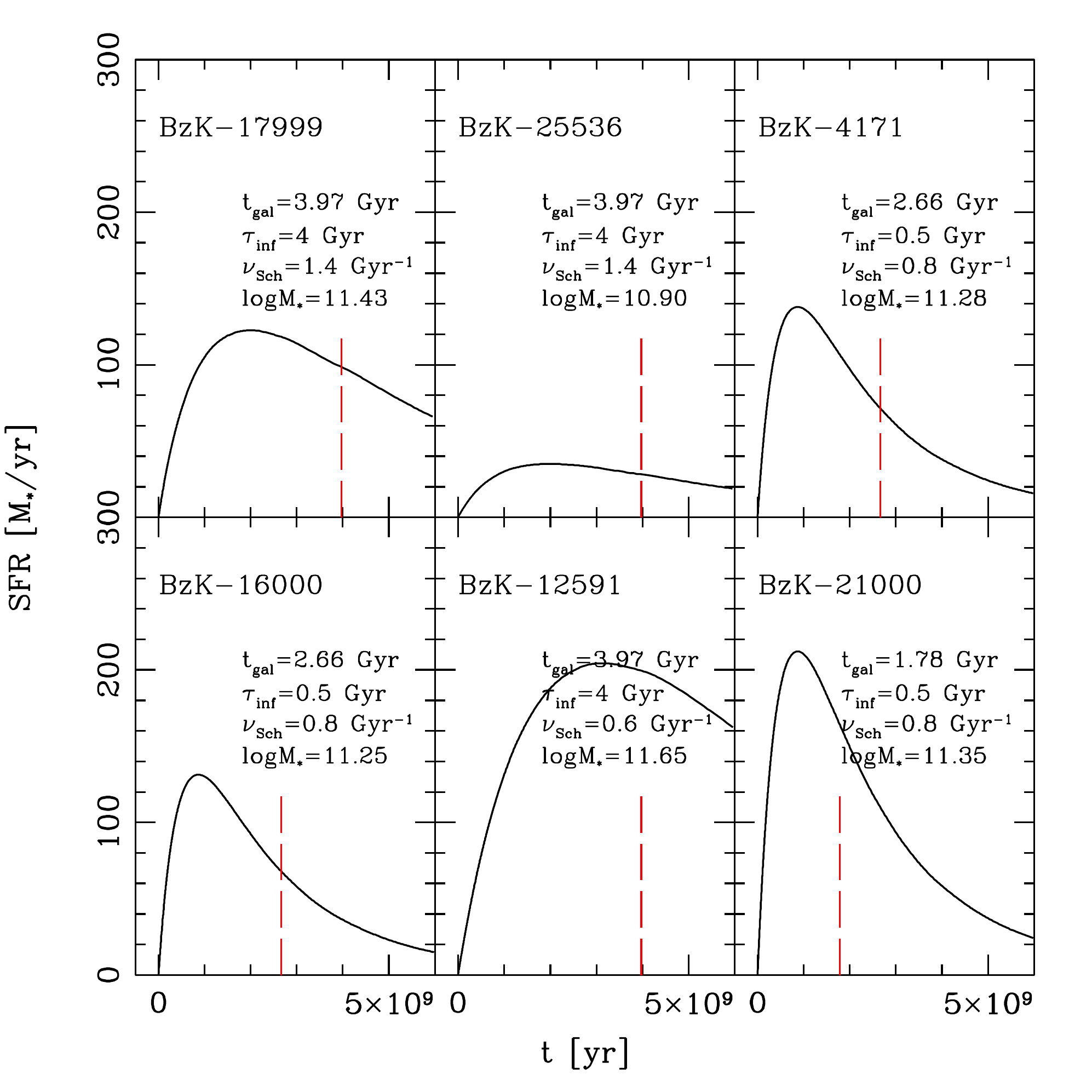}}SFHpaper1.ps
\caption[Best-fit SFHs of z\,$\sim$\,1.5 SF BzK galaxies]{Best-fit SFHs of $z\,\sim$\,1.5 SF BzK galaxies. The long dashed red vertical lines indicate the age at which the galaxy is observed.}
\label{sfhBzK}
\end{figure*}

\begin{figure*}
\centerline{
\includegraphics[width=6.cm]{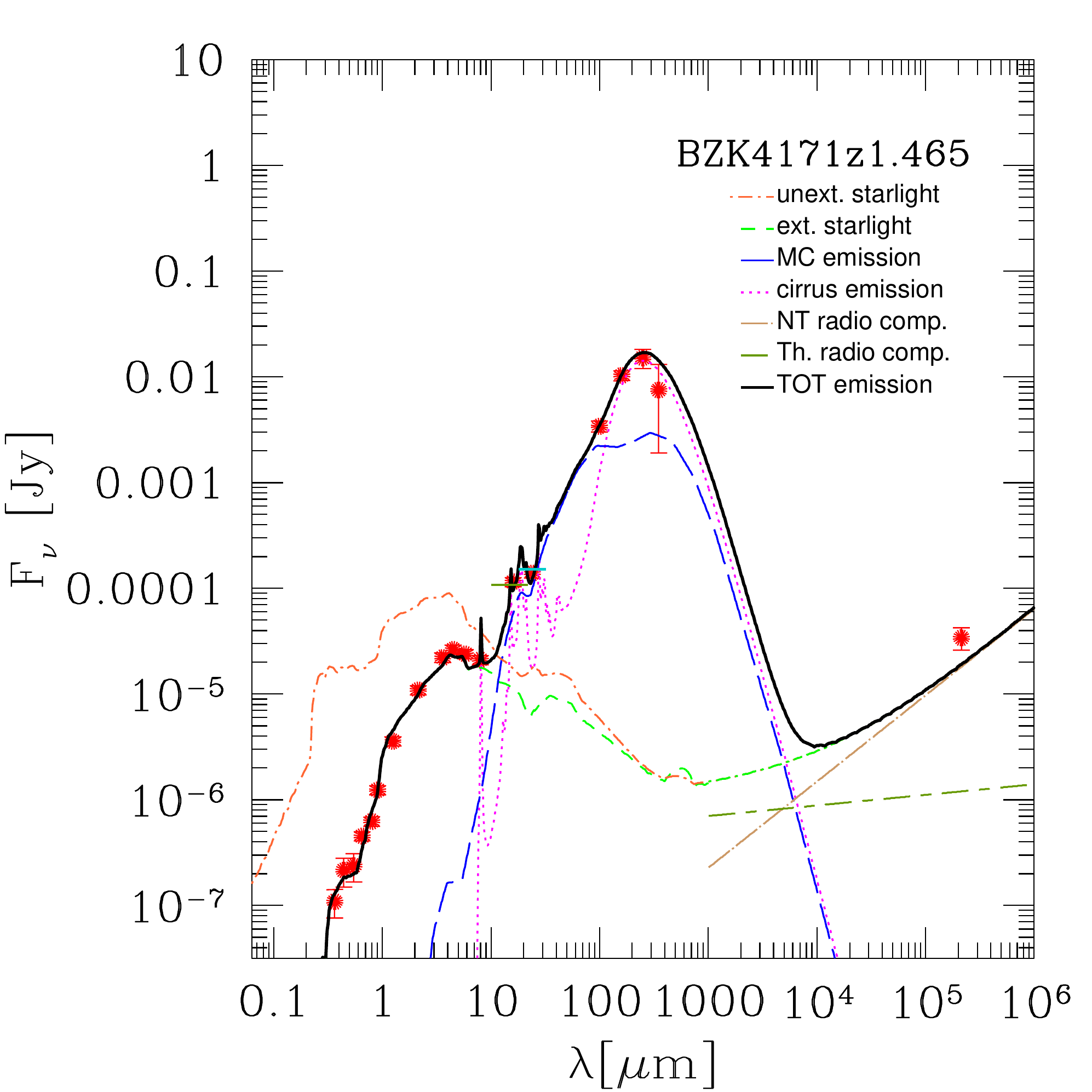}
\includegraphics[width=6.cm]{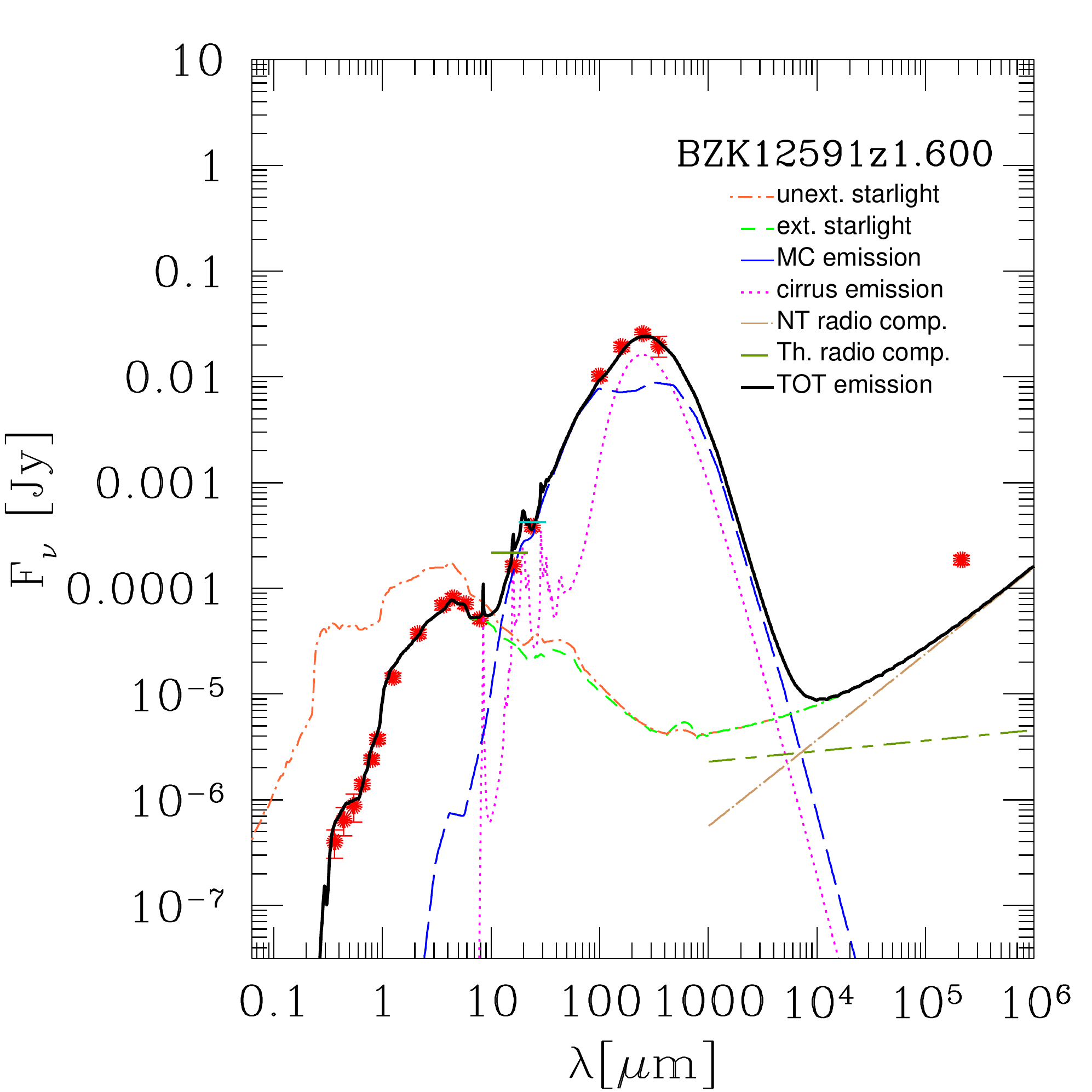}
\includegraphics[width=6.cm]{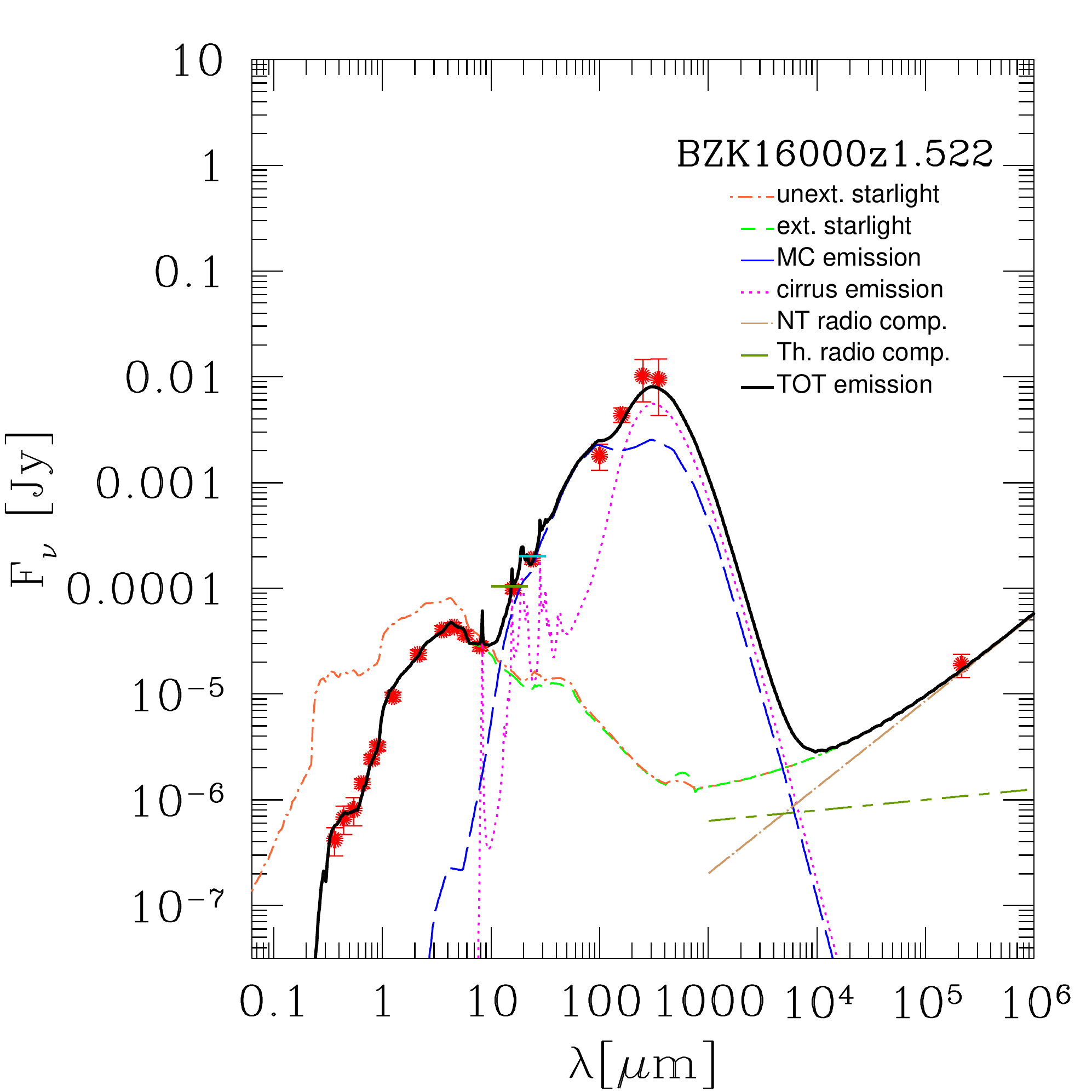}}
\centerline{
\includegraphics[width=6.cm]{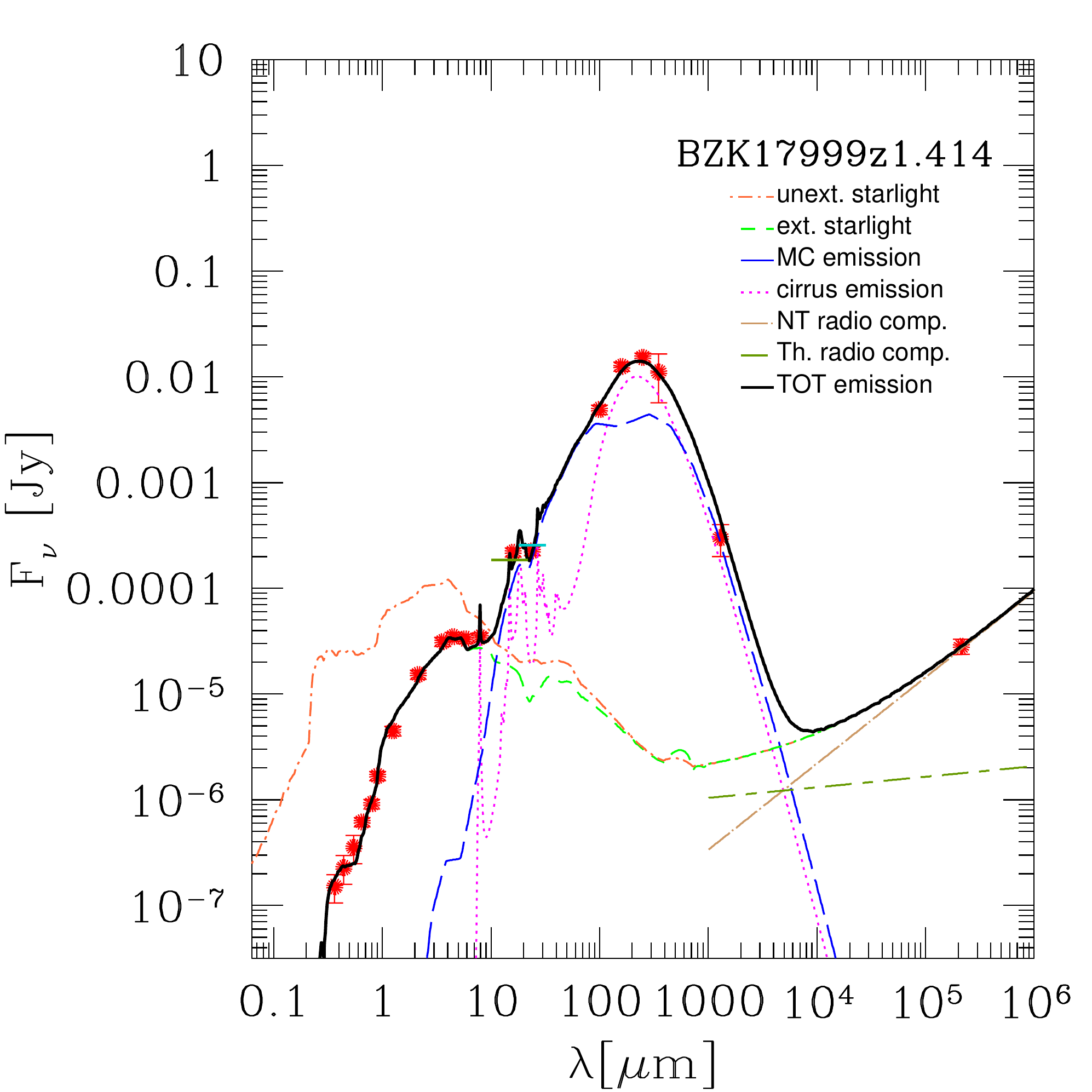}
\includegraphics[width=6.cm]{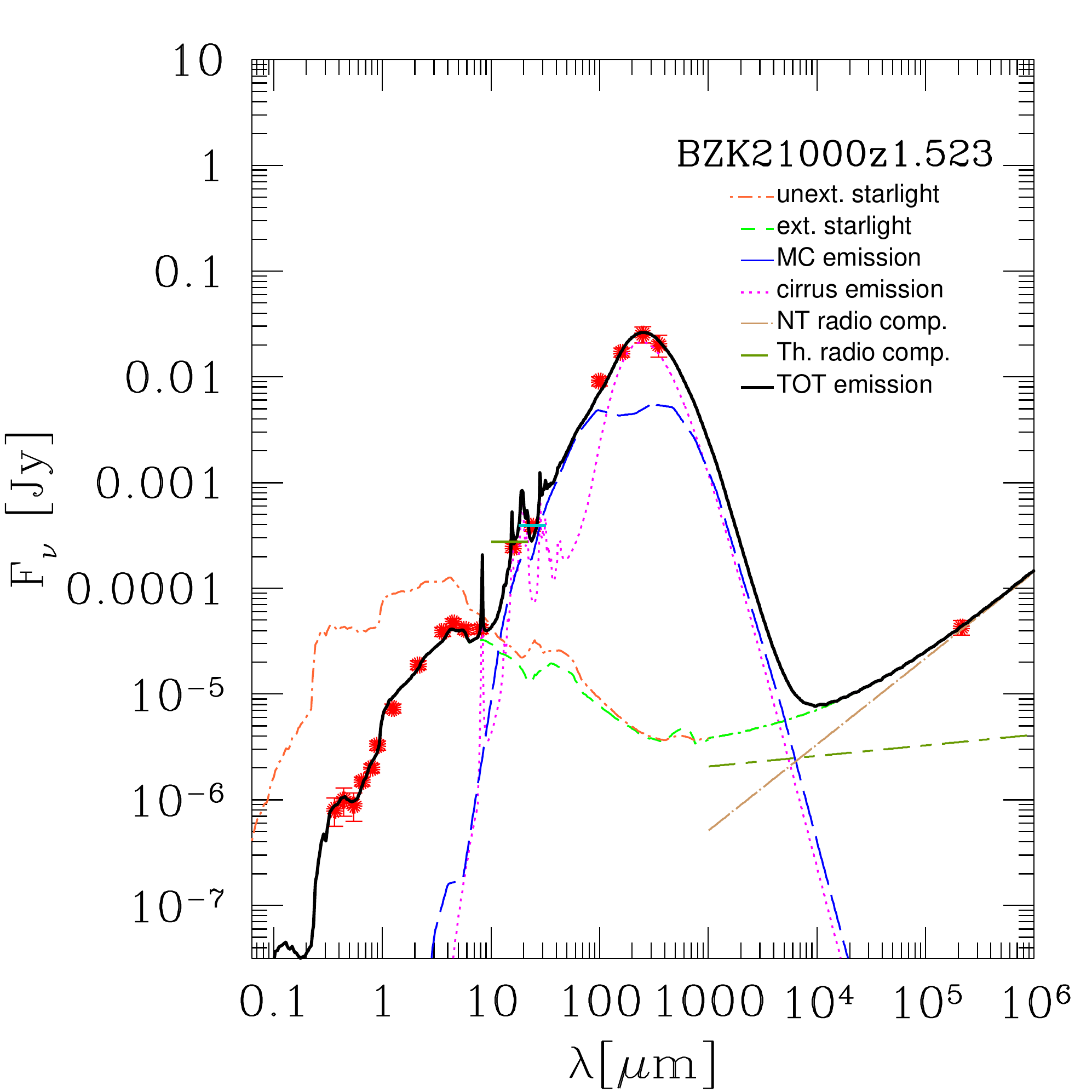}
\includegraphics[width=6.cm]{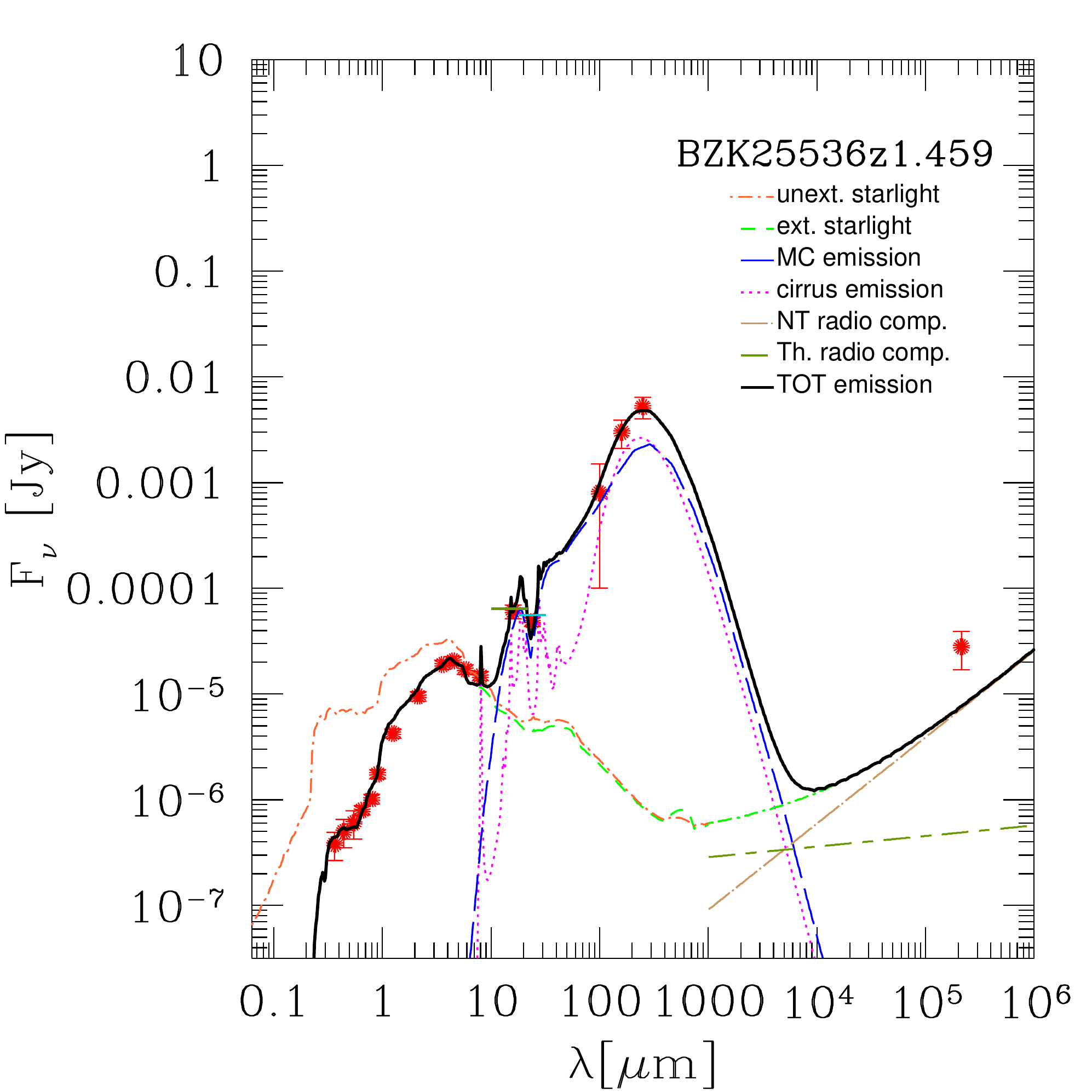}}
\caption[Best-fits to the observed SEDs of $z\,\sim$\,1.5 BzK star forming galaxies]{Best-fits to the observed SEDs (red filled circles) of $z\,\sim$\,1.5 BzK star forming galaxies. The short green and cyan segments at 16 and 24 $\mu m$, respectively, represents the modelled fluxes integrated over the filter bands. These are required by the presence of several spectral features ascribed to PAHs falling in these wavelength range.}
\label{bestfitbzk1}
\end{figure*}

From the detailed shape of the broad-band best-fit SED of these
6 BzK galaxies we have derived the SFHs shown in
Figure~\ref{sfhBzK}. 3/6 objects
(top-left and top- and bottom-centre panels) show relatively
long infall timescales ($\tau_{inf}\,\sim\,4$\,Gyr) typical of
gradually evolving star forming disks at high redshift. Two of
them also have high SF efficiencies ($\nu_{Sch}\,=1.4$\,
Gyr$^{-1}$) while one of them shows a lower SF efficiency of
0.6 Gyr$^{-1}$ typical of normal SF spiral galaxies. The
remaining three BzK show SFHs characterized by shorter infall
timescales ($\tau_{inf}\,\sim\,0.5$ Gyr) and moderate SF
efficiencies ($\nu_{Sch}=0.8$\,Gyr$^{-1}$), typical of objects
in earlier phases of SF. The latter also show smaller ages with
respect to the former, typically in the range between $1.78$ and
$2.66$ Gyr. All of them are observed within 2 Gyr from the peak
of the SF activity, showing a moderate ongoing activity of star
formation with SFRs\,$\la$\,107\,$M_{\odot}$/yr typical of MS
star forming galaxies \citep{Daddi2010}. Anyway the statistics
of the sample is still too low to draw a self-consistent
picture of their SFHs. 
%
\begin{table*}
\centering
\begin{tabular}{|c|c|c|c|c|c|c|c|c|c|c|c|c|c|c|}
\hline
  \multicolumn{1}{|c|}{ID} &
  \multicolumn{1}{c|}{z} &
  \multicolumn{1}{c|}{$\chi^{2}$} &
  \multicolumn{1}{c|}{L$_{IR}$} &
  \multicolumn{1}{c|}{SFR$_{10}$} &
  \multicolumn{1}{c|}{SFR$_{K}$} &
  \multicolumn{1}{c|}{SFR$_{D10}$} &
  \multicolumn{1}{c|}{SFR$_{M12}$} &
  \multicolumn{1}{c|}{LogM$^{GR}_{\star}$} &
  \multicolumn{1}{c|}{LogM$^{D10}_{\star}$} &
  \multicolumn{1}{c|}{A$_{V}$} &
  \multicolumn{1}{c|}{A$_{FUV}$} &
  \multicolumn{1}{c|}{q$^{GR}_{TIR}$} &
  \multicolumn{1}{c|}{q$^{obs}_{TIR}$}\\
           &       & & [L/L$_{\odot}$] & [M$_{\odot}$/yr] &  [M$_{\odot}$/yr] &  [M$_{\odot}$/yr] & [M$_{\odot}$/yr] & [M$_{\odot}$] & [M$_{\odot}$] & & & \\
\hline
  BzK16000 & 1.522 & 0.96 & 8.10E11 & 75 & 81 & 152 & 74 & 11.25 & 10.63 & 1.42 & 3.41 & 2.59 & 2.54\\
  BzK12591 & 1.600 & 4.85 & 2.95E12 & 205 & 295 & 400 & 275 & 11.65 & 11.04 & 1.78 & 4.42 & 2.65 & 2.06\\
  BzK21000 & 1.523 & 1.35 & 2.40E12 & 163 & 240 & 220 & 209 & 11.35 & 10.89 & 2.36 & 4.06 & 2.65 & 2.66\\
  BzK17999 & 1.414 & 2.22 & 1.27E12 & 96 & 127 & 148 & 115 & 11.43 & 10.59 & 2.60 & 5.19 & 2.64 & 2.64\\
  BzK25536 & 1.459 & 2.00 & 3.58E11 & 28 & 36 & 62 & 29 & 10.90 & 10.52 & 1.18 & 2.84 & 2.62 & 2.07\\
  BzK4171  & 1.465 & 2.66 & 1.22E12 & 71 & 122 & 103 & 95 & 11.28 & 10.60 & 2.57 & 5.10 & 2.75 & 2.51\\
\hline\hline
\end{tabular}
\caption[Best-fit physical parameters of BzK galaxies: I]{Best-fit physical parameters of BzK galaxies compared to estimated values from \cite{Daddi2010} (D10) and \cite{Magdis2012} (M12).}
\label{BzKpar1}
\end{table*}

Table~\ref{BzKpar1} summarizes the main results
of our analysis and compare them to the estimates
provided by D10 and \citeauthor{Magdis2012}~(2012, M12
hereafter). Differently from D10, whose SFR estimates rely only
on 24 $\mu\,m$, the SFRs from M12 have been computed using the
information coming from the full coverage from MIR to sub-mm
offered by Herschel. As already stated above, the $L_{\mathrm{IR}}$ from
24 $\mu\,m$ can be up to a factor $\sim$\,2 higher than the
`real' $L_{\mathrm{IR}}$ measured with \textit{Herschel} (see e.g.
\citealt{Oliver2012}, Canalog et al.~2013 in prep.). As shown
in Tab.~\ref{BzKpar1} our SFRs, averaged over the last 10
Myr, well agree with those derived by M12. For comparison we
also report the SFRs based on the Kennicutt calibration. These
differ with respect to our estimates by a factor lower than
$\sim$\,1.4, on average, well within the typical uncertainties
for this kind of measure.

The stellar masses derived from the best-fit SED (SFH) of these
objects are also listed in Tab.~\ref{BzKpar1} and compared to
the estimates provided by D10.


Stellar masses in D10 (column n. 9 of Tab.~\ref{BzKpar1}) are
derived by fitting the
\citeauthor{Maraston2005}~(2005, M05 hereafter) SSP models to
the UV-optical-NIR (up to 5.8\,$\mu m$) band of each galaxy.
M05 models include a particularly strong contribution of AGB emission,
much stronger than e.g. B02 and \citeauthor{BruzualCharlot2003}~(2003, BC03 hereafter),
both the latter are in fact based on the same Padova stellar isochrones.

Given that all the galaxies in the
sample are star forming, a constant SFR in combination with a
large range of ages, a \cite{Chabrier2003} IMF and different
metallicities from half solar to twice solar, is adopted by
D10. Dust effects are accounted for by assuming an homogeneous
foreground screen of dust and the Calzetti~(2000) reddening
law. The 1\,$\sigma$ error on D10 stellar masses is of the
order of 0.10-0.15 dex. As evident from Tab.~\ref{BzKpar1}, D10 $M_{\star}$ are much lower than our
estimates by a factor ranging between 2.4 up to $\sim$\,7 (for
\textit{BzK-17999}). As shown in M05 and \cite{Maraston2006}
the differences in the (1) Stellar Evolutionary models used to
construct the isochrones, (2) the treatment of the TP-AGB phase
and (3) the specific procedure used for computing the
integrated spectra can lead to large differences in terms of
stellar ages and masses when comparing the BC03 stellar models
to the M05 ones. In particular M05 SSPs are typically brighter
and redder than BC03 ones, this results in lower ages and lower
stellar masses with respect to BC03 by a factor $\sim$\,60\%.
We believe that, once corrected for the use of different
stellar population models, the combination of a different SFH
and dust attenuation treatment is again the major source of
discrepancy as demonstrated by BLF13.
%

\section{FIR-radio correlation: comparing models to observations}
\label{qtirsection}

Despite the FIR/radio correlation is now well established up to
high redshifts (e.g. \citealt{Ivison2010,Sargent2010,Mao2011,Pannella2014}), its
physical origin is still debated.

B02 found the tightness of the FIR/radio correlation to be
natural when the synchrotron mechanism dominates over the
inverse Compton, and the electron cooling time is shorter than
the fading time of the SN rate. Both these conditions are met
in star forming galaxies, from normal spirals to obscured
starbursts. However, since the radio NT emission is delayed,
deviations in the correlation are expected both in the early
phases of a starburst, when the radio thermal component
dominates, and in the post-starburst phase, when the bulk of
the NT component originates from less massive stars.

By taking advantage of the full FIR coverage provided by both
Herschel PACS and SPIRE instruments together with the radio
detection at 1.4 GHz, we have estimated for each galaxy in our
sample the ratio $q_{\mathrm{TIR}}$, of the rest-frame 8-1000~$\mu m$
luminosity to the rest-frame radio luminosity at 1.4 GHz
directly derived from our model. The logarithmic total IR
(TIR)/radio flux ratio $q_{\mathrm{TIR}}$ (\citealt{Helou1985}, in its original definition they were using instead $L_{\mathrm{FIR}}$ defined in the range from 42.5 to 122.5 microns) has been
computed according to the following relation:
\begin{equation}
q_{TIR}=log\left(\frac{L_{TIR}}{3.75\,\times\,10^{12} W}\right) - log\left(\frac{L_{1.4 GHz}}{W Hz^{-1}}\right)
\label{qtir_eq}
\end{equation}
We have then compared our predictions to the observational
estimates by \cite{Sargent2010} for galaxy samples selected at
far-infrared and radio wavelengths at the same redshift and in
the same luminosity range as our high-$z$ (U)LIRGs.
\begin{figure}
\centerline{
\includegraphics[width=9.cm]{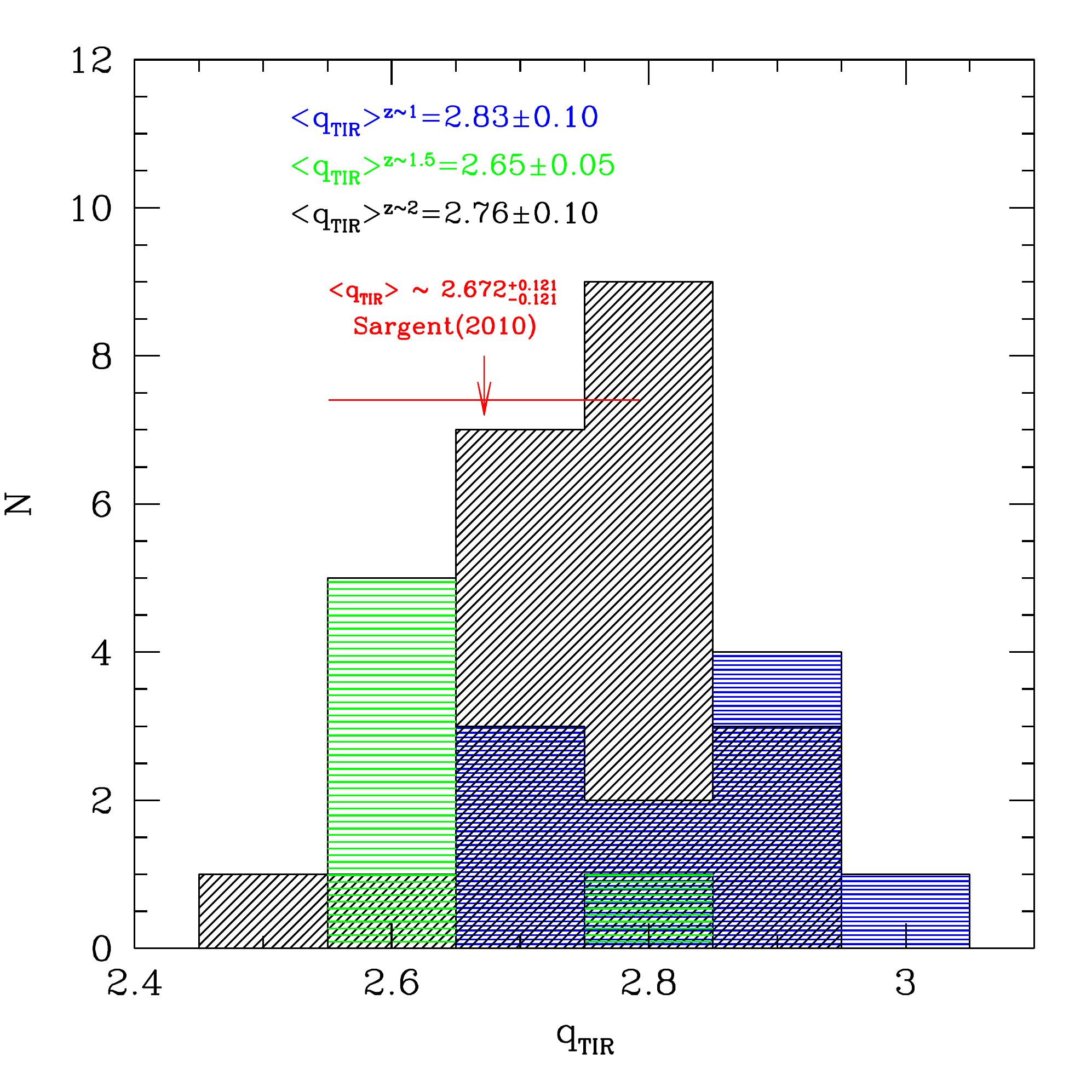}} 
\caption{Distributions of the predicted $q_{\mathrm{TIR}}$ values for the $z\sim\,1$ LIRGs (blue horizontal lines) $z\sim1.5$ BzK (green wide spaced horizontal lines) and $z\sim\,2$ (U)LIRGs (black 45$^{\circ}$ angled solid lines). Mean values and standard deviations are also highlighted in the figure. The red arrow and horizontal line indicate the typical $q_{\mathrm{TIR}}$ of $z\,\sim\,1-2$ (U)LIRGs, as computed by \citet{Sargent2010}, and its range of values.
}
\label{qtirdist}
\end{figure}

Figure~\ref{qtirdist} shows the distribution of the predicted $q_{\mathrm{TIR}}$ values for our $z\sim\,1$ LIRGs (blue horizontal lines) $z\sim1.5$ BzKs (green wide spaced horizontal lines)
and $z\sim\,2$ (U)LIRGs (black 45$^{\circ}$ angled solid lines). The two distributions of LIRGs and (U)LIRGs are quite similar with the lower redshift LIRGs reaching higher values
of $q_{\mathrm{TIR}}$. The $q_{\mathrm{TIR}}$ distribution of BzK is shifted instead towards lower values. Mean values and standard deviations are also highlighted in the figure and compared to the
observational estimates provided by \cite{Sargent2010} (in red).

\cite{Sargent2010} studied the evolution of the IR/radio relation out to $z\sim\,2$ for a statistically significant volume-limited sample of IR-luminous galaxies selected in the COSMOS field.
Their sample includes $\sim$\,1,692 star forming ULIRGs and $\sim$\,3000 SF `IR bright' ($L_{\mathrm{TIR}} \geq L^{\mathrm{knee}}_{\mathrm{TIR}}(z)$) COSMOS sources up to $z\sim\,2$.
They found no evolution of the median TIR/radio ratio among the ULIRG sample, with a median value at $z\,\sim\,2$ of $2.672^{+\infty}_{-0.121} (2.892)$.
The value within brackets represents the median in high-redshift bins before the correction ($\Delta\,q_{\mathrm{TIR}}\,\sim\,0.22$), needed in order to compensate for the relative offset between medians at high and low redshift that arises artificially due to the increased scatter ($\sigma_{\mathrm{TIR}}$) in the data at $z \ga 1.4$ \citep{Sargent2010}.


Our predicted $q_{\mathrm{TIR}}$ for $z\,\sim\,2$ (U)LIRGs (2.76\,$\pm$\,0.10), well agrees, within the errors, with the results from \cite{Sargent2010} discussed above and
also with the median $q_{\mathrm{TIR}}$ measured by the authors for local (U)LIRGs ($\langle\,q_{\mathrm{TIR}}\,\rangle\,2.703_{-0.050}^{+0.050}$).
Our results thus confirms, from a modellistic point of view and in line with theoretical and numerical simulation expectations,
that ULIRGs should follow the local IR-radio relation until at least $z\,\sim\,2$ \citep{LackiThompson2010, Murphy2009}.
This implies that that magnetic fields are sufficiently strong to ensure cosmic-rays electrons to predominantly lose energy through synchrotron radiation rather than inverse Compton scattering off the CMB.

Good agreement, within the errors, is found also between our predicted $q_{\mathrm{TIR}}$ for $z\,\sim\,1$ LIRGs and the median $q_{\mathrm{TIR}}$ derived by \cite{Sargent2010} for $z\,\sim\,1$ SF IR-bright sources ($2.672^{+0.069}_{-0.061}$).
These are actually very similar to the median $q_{\mathrm{TIR}}$ measured for $z\,\sim\,2$ SF ULIRGs. In Figure~\ref{qtirdist} this is emphasized by the two samples showing very similar $q_{\mathrm{TIR}}$ distributions.
An even better agreement with the observational estimates provided by Sargent is found for the 6 BzKs which show an average $q_{\mathrm{TIR}}$ of 2.65$\pm$0.05.
Of course given the low statistics of this sample we cannot draw strong conclusion from this comparison. It gives anyway important hints about our solutions.
No strong evolution of the FIR/radio correlation is thus observed also between our $z\sim$1 and $\sim$\,2 $L\,\ga\,10^{11}\,L_{\odot}$ objects.

Our physical model thus seems to be able to reproduce the radio properties of high-$z$ (U)LIRGs including the FIR-radio correlation up to $z\,\sim$\,1-2.
This provides a further important constraint to model the SFR and SFHs underlying the observed SED.

\section{Radio constraints on the current SFR of galaxies}
\label{radioSFR}

\begin{figure}
\centerline{
\includegraphics[width=9.cm]{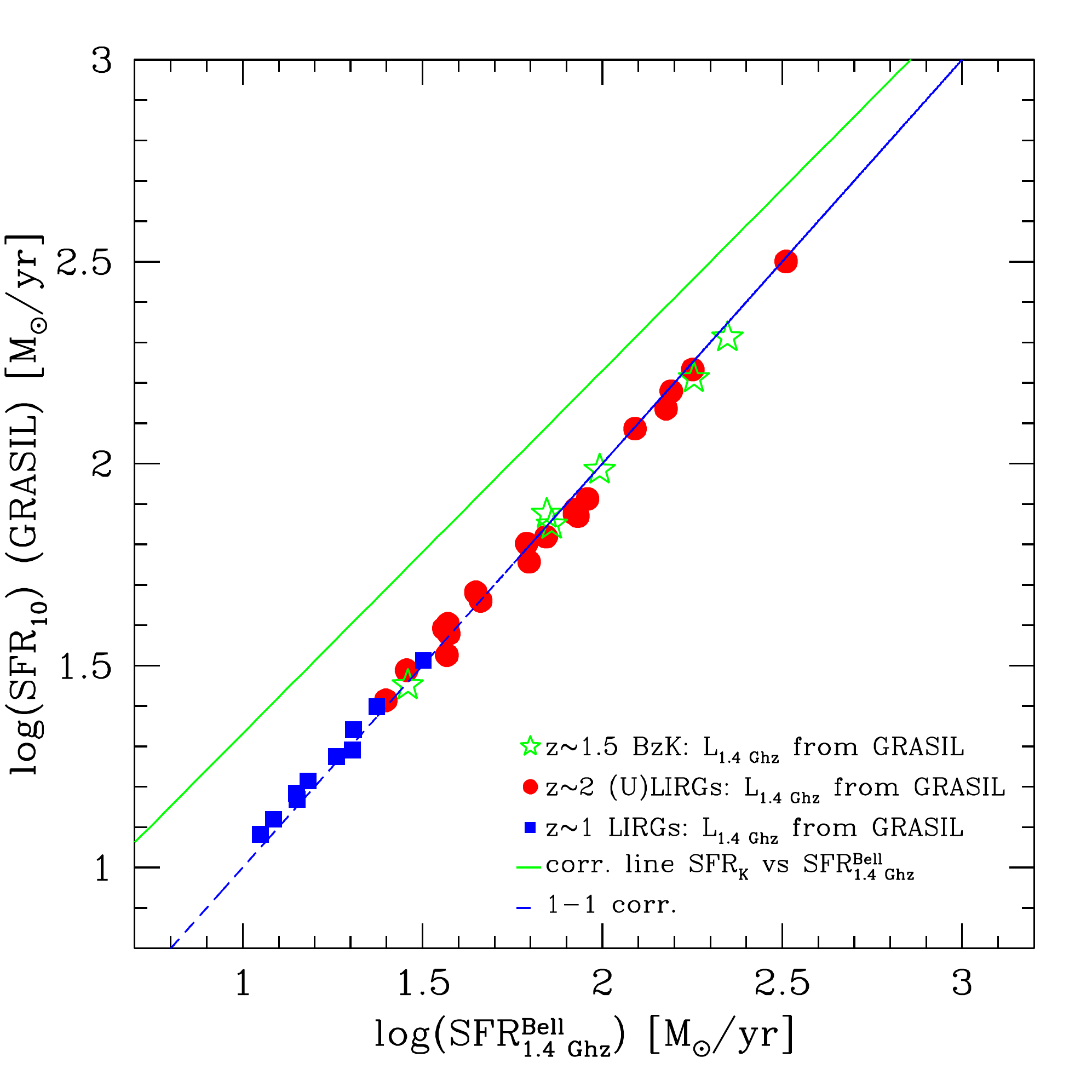}} 
\caption{Comparison of our GRASIL-estimated $SFR_{10}$, averaged over the last 10 Myr, with the SFR derived from the modelled (filled points) rest-frame $L_{1.4 GHz}$ luminosity using the Bell~(2003) calibration (eq. \ref{Bell_eq}). The $z\sim$1.5 BzK galaxies are shown as green open stars. The 1-1 correlation is shown as blue dashed line. The green solid line represents the SFR estimates provided by the Kennicutt~(1998) calibration plotted as a function of the radio SFRs derived from the modelled $L_{1.4 GHz}$. The slope of the green line is $\sim$\,0.90. The yellow shaded area highlights the region of the plot where the SFR estimates are within a factor of $\sim$\,2 from the 1-1 corr. line. While a very good agreement is clearly visible between our $SFR_{10}$ and the best-fit radio based $SFR^{Bell}_{1.4 GHz}$, the SFRs based on the Kennicutt relation appear to be systematically off-set with respect to those provided by radio luminosity. In particular we observe that the discrepancy between the FIR- and radio-based SFRs seems to be larger at lower redshifts (see text for details).}
\label{sfrcomp}
\end{figure}
Recent works by \cite{Daddi2007a, Daddi2010}, have found very good agreement, within a factor of $\sim$ 2, between different SFR indicators (UV dust-corrected, MIR and radio 1.4 GHz) for a
GOODS sample of BzK-selected galaxies including both individual sources and stacked sources. All these SFR estimates, however, rely on calibrations based on similar assumptions, namely \cite{Kennicutt1998}.
This calibration assumes that the $L_{\mathrm{bol}}$ of a constant SF lasting 100 Myr is totally emitted in the IR (K98; \citeauthor{Leitherer1995},1995 LH95 hereafter).
For a constant SF, the $L_{\mathrm{bol}}$ after the first 10 Myr evolves relatively slowly because the rate of birth and death of the most massive stars (with lifetimes $\la$ 10 Myr and dominating the $L_{\mathrm{bol}}$) reaches a steady state (see Fig.~2 and 8 of LH95). The K98 SFR/$L_{\mathrm{IR}}$ calibration adopts the mean bolometric luminosity for a 10-100 Myr continuous SF, solar abundance, Salpeter IMF of the starburst synthesis models of LH95, and assumes that $L_{\mathrm{IR}}$=$L_{\mathrm{bol}}$.

For the radio band the SFR is usually estimated from the 1.4 GHz flux using the calibration of \cite{Bell2003}.
This calibration is based on the IR-radio correlation. It assumes that non-thermal radio emission directly tracks the SFR, and it is chosen so that the radio SFR matches the
IR SFR for L $\geqslant$ $L^{\star}$ galaxies. The SFR calibration is given by the following relation:
\begin{equation}
SFR^{Bell}_{1.4 GHz} (M_{\odot}/yr) = 5.52 \times 10^{-22} L_{\nu,\,1.4 GHz}
\label{Bell_eq}
\end{equation}
where L$_{\nu,\,1.4 GHz}$ is in units of W $\times$ Hz$^{-1}$ and a Salpeter IMF is assumed.
With respect to the original calibration by \citet{Condon1992} this one is found to be lower by a factor of two \citep{Kurczynski2012}.
Indeed the calibration of \citet{Condon1992} explicitly models the thermal and non thermal emission mechanisms, whereas the calibration of \citet{Bell2003} relies upon the IR-Radio correlation.
Thus we expect agreement between SFR$^{\mathrm{Bell}}_{1.4 GHz}$ and IR-based SFR estimates, if the IR-radio correlation continues to hold at high redshift, as it has indeed been suggested in the
literature \citep{Sargent2010,Ivison2010}. Very similar to the Bell~(2003) calibration for the SFR based on radio luminosity is that one found in Yun, Reddy, \& Condon~(2001) based on a large sample of nearby disk galaxies and using IRAS and NVSS data (hence empirical and based on the FIR-radio correlation). 

In BLF13 we have shown that, due to the significant contribution of cirrus emission to the total $L_{\mathrm{IR}}$ whose heating source includes already evolved stellar populations
(ages older than the typical timescale for young stars to escape from the parent MCs, ($t_{\mathrm{esc}}$) typically ranging between $\sim$\,3 Myr and $\sim$\,90 Myr),
our inferred SFR$_{10}$ are systematically lower than those based on the K98 calibration, by a factor $\sim$\,2-2.5.

Averaging out our results for LIRGs and ULIRGs we found the following calibration between the total IR luminosity and SFR:
\begin{equation}
SFR [M_{\odot}/yr] \simeq (5.46 \pm 0.6)\times 10^{-11}\ L_{\mathrm{IR}}/L_{\odot}.
\end{equation}
The calibration already includes the factor 1.7 used to pass from Salpeter to Chabrier. The important point here, given the dominant role of cirrus emission to $L_{\mathrm{IR}}$, is to understand if the predictions for the FIR are consistent with those from the radio emission.


The inclusion of radio emission into our procedure is therefore extremely important to understand if this conspicuous cirrus component contributing to the FIR
is an `effect' of a poor parameter exploration or if it is real, and required in order to reproduce the NIR-to-FIR properties of our galaxies.
One possibility in fact is that, forcing the contribution by intermediate age ($\ga 50$ Myr) stellar populations to reproduce the NIR peaked fluxes of our galaxies,
we get the right FIR emission, as the integrated luminosity (mass) of these stars is large, but we lack in SNII production. This increases the $L_{\mathrm{FIR}}$/$L_{\mathrm{1.4 GHz}}$ ratio of the model but fitting the FIR we underpredict the radio. It is important, therefore, to check any possible systematics within the model.
\begin{figure*}
\centerline{
\includegraphics[width=8.2cm]{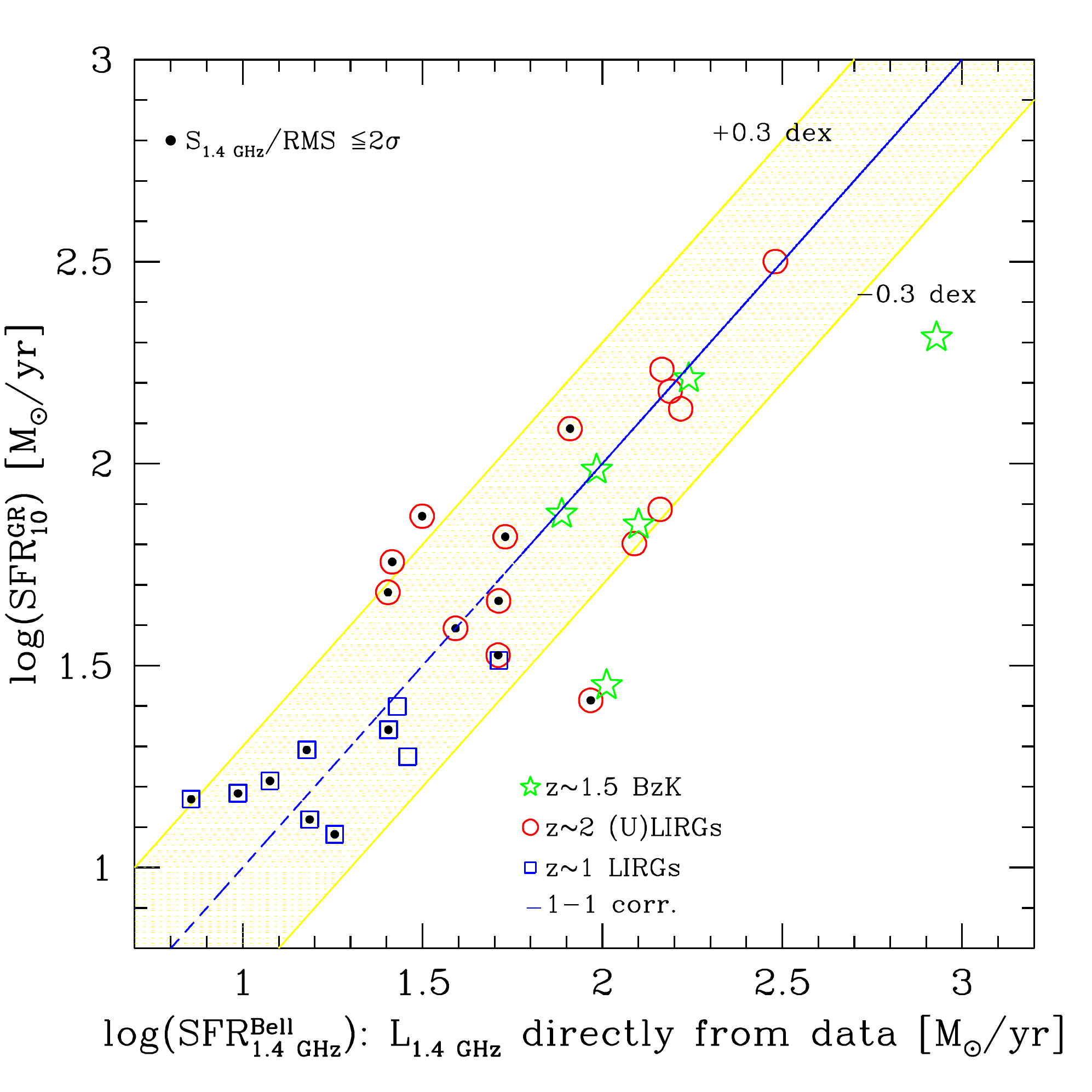} 
\includegraphics[width=8.2cm]{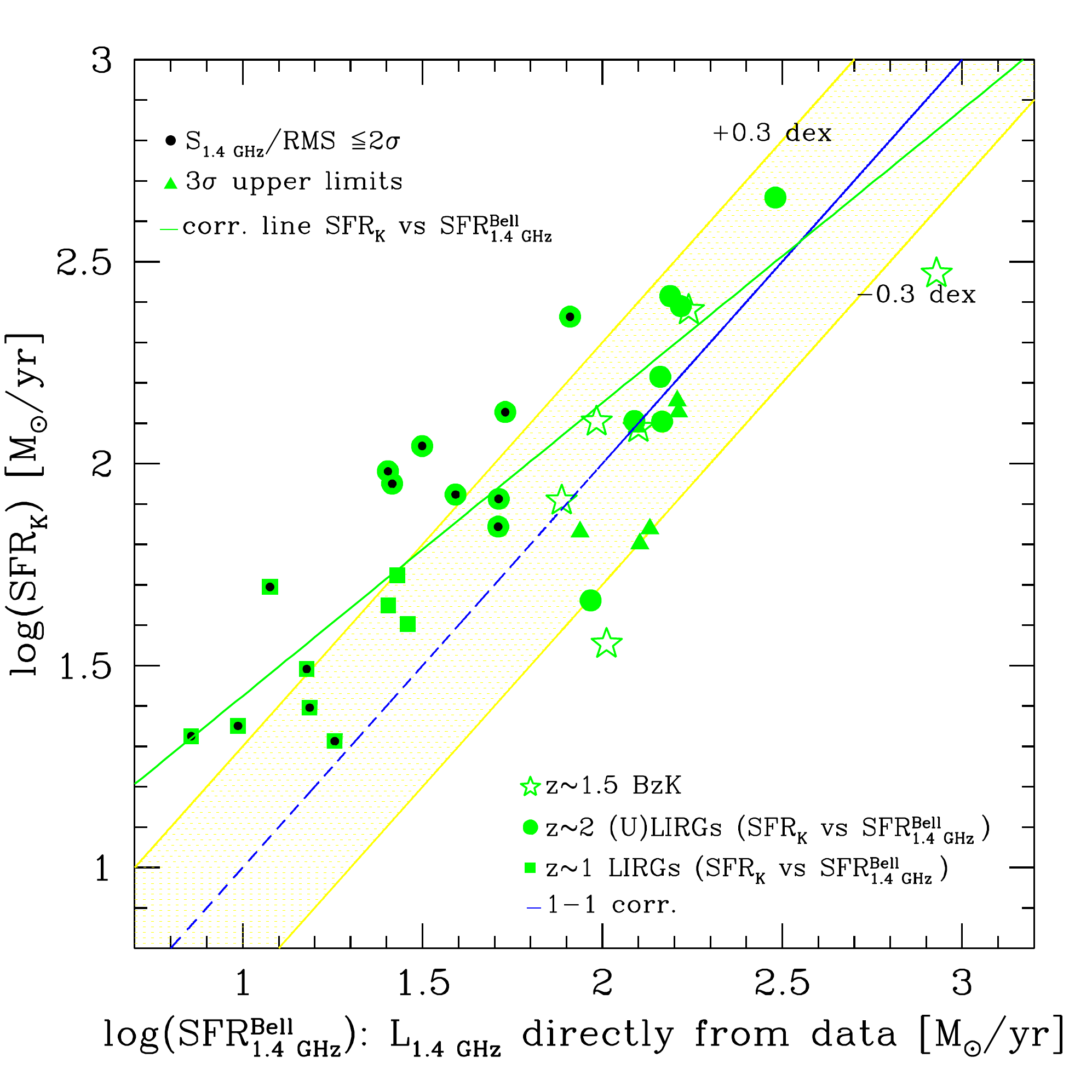}} 
\caption{Left: Comparison of our GRASIL-estimated $SFR_{10}$ with the SFR derived from the observed (open points) rest-frame radio luminosity, $L_{1.4 GHz}$, using the Bell~(2003) calibration (eq. \ref{Bell_eq}). The 1-1 correlation is shown as blue dashed line. In both panels the yellow shaded area identifies the region of the plot where the SFRs are within a factor of 2 from the 1-1 correlation line. Right: Comparison between the SFR estimates provided by the Kennicutt~(1998) calibration and the SFR$_{1.4 GHz}^{Bell}$ derived directly from the observed rest-frame radio luminosity. The green filled circles and squares represent, respectively, the $z\sim$\,2 and $z\sim$\,1 (U)LIRGs while the green open stars the $z\sim$1.5 BzKs of D10. The green solid line is the linear regression of the green points. Its slope is $\sim$\,0.73. A larger scatter around the 1-1 correlation line is present here. It reflects the scatter of our best-fit solutions already discussed in Fig.~\ref{deltaL} and pertains mostly to the faint detections. Anyway most of our predictions are well within a factor of $\sim$\,2 (yellow shaded area) from the 1-1 corr.\ line.}
\label{sfrcomp2}
\end{figure*}

Figure~\ref{sfrcomp} compares our model SFRs averaged over the
last 10 Myr, thus sampling the most recent SF activity, to the
SFR obtained from the rest-frame radio luminosities at 1.4 GHz,
as provided by the best-fit physical model, using the
Bell~(2003) calibration of Eq.~\ref{Bell_eq}. For the
comparison we have re-scaled this relation, originally computed
for a Salpeter IMF to a Chabrier one. Red filled circles and blue squares represent,
respectively, our $z\sim$\,2 and $\sim$\,1 (U)LIRGs while the
starred symbols mark the $z\sim$1.5 BzKs. The green solid line
represents the SFR estimates based on the Kennicutt~(1998)
calibration plotted as a function of the SFR$_{1.4 GHz}^{Bell}$
derived from the modelled rest-frame radio luminosity. Its
slope is $\sim$\,0.90. The dashed blue line is the 1-1
correlation line.

Our $SFR_{10}$ (from GRASIL) appear to be in perfect agreement with
the $SFR^{Bell}_{1.4 GHz}$ (from Bell~2003) derived from the best-fit $L^{GR}_{1.4 GHz}$.
For our model predictions we have derived the calibration factor between modelled SFR and radio
rest-frame luminosity, corresponding to the ratio $SFR_{10}^{Salp.}$/$L^{GR}_{1.4 GHz}$,
and compared it to the same factor as derived by \citet{Bell2003} (i.e. $5.52\,\times\,10^{-22}$).
For the filled points and starred symbols of Figure~\ref{sfrcomp} we have measured a factor of $(5.53\,\pm\,0.38) \times 10^{-22}$.

The estimates from the Kennicutt relation are, instead, systematically off-set with respect to those provided by radio luminosity,
by a factor ranging between 1.5 and $\ga$\,2. In particular we observe that the discrepancy between the FIR- and radio-based SFRs is $\ga$\,2 up to $z\sim$\,1.8 then larger
than 1.5 up to $z\sim$\,2.5 and lower than 1.5 at higher redshifts. In other words the discrepancy seems to be larger at lower redshifts.

In Figure~\ref{sfrcomp2} we compare our model SFR$_{10}$ (left)
and those based on the Kennicutt calibration, SFR$_{K}$
(right), to the radio estimates where the rest-frame radio
luminosity is computed directly from the observed flux density
using the relation specified in Eq.~\ref{eqL14} (Left: open
blue squares and red circles; Right: filled green circles and
squares). The blue dashed line represents, in both panels, the
1-1 corr. line, while the green solid line is the linear
regression for the filled green points.

As shown in Fig.~\ref{sfrcomp2} (Left) a larger scatter around
the 1-1 correlation line is evident when considering the
$SFR^{Bell}_{1.4 GHz}$ estimated from the observed rest-frame
radio luminosity. This scatter reflects the scatter of our
best-fit solutions already discussed in Fig.~\ref{deltaL} and
pertains mostly to the $z\sim$\,2 2\,$\sigma$ - $\la$
3\,$\sigma$ detections. It is, in fact, lower for the
$z\sim$\,1 LIRGs. As discussed above, only one ULIRG,
\textit{U5152}, and the two BzK dominated by an AGN show a
significant (larger than a factor of 2) discrepancy between our
model and the observed radio fluxes and most of our predictions are
well within a factor of $\sim$\,2 (yellow shaded area) from the
1-1 corr.\ line. Averaging out our results, we have
measured a calibration factor between our model SFRs and the
observed rest-frame radio luminosity of
$\sim\,(6.06\,\pm\,3.00) \times 10^{-22}$, with the 6/31 (19\%)
3\,$\sigma$ upper limits excluded from this computation. Also
in this case our solutions are in agreement, within the errors,
with the empirical calibration provided by \citet{Bell2003}.

The important point here is that there is no systematic effect
in our solutions which tend to scatter both above and below the
empirical calibration. On the contrary, compared to the SFRs
estimated from the FIR using the Kennicutt calibration, we were
systematically lower by a factor of $\sim$\,2. Therefore, this
seems to go in the direction of confirming our physical
predictions and it gives an indication that the radio emission more
than the FIR is able to accurately predict the current SFR mostly
contributed by young massive stars. It doesn't seem to suffer
from contamination by intermediate-age stellar populations, as
it can happen for the FIR. Of course we have to take into account the large
errors associated to the radio data, but even with this in mind
our results are in good agreement.

In the right panel of Fig.~\ref{sfrcomp2} we compare the
SFR$_{K}$ based on the Kennicutt calibration and FIR luminosity
(filled green circles and squares), to the radio estimates
derived from the observed rest-frame radio luminosity. Also
here there is a large scatter of the data points mostly above
the 1-1 correlation line. The slope of the green line is
$\sim$\,0.73. What's evident here is that the FIR-based SFRs
tend to be higher with respect to the radio ones by a factor
larger than 2 up to $z\sim$\,1.5. Moreover, up to z$\sim$\,1.8
most of the points tend to lie above the 1-1 corr.\ line. Beyond
redshift 1.8 the discrepancy between the FIR- and radio-based
SFRs decreases up to a factor below 1.5. So there seems to be, in
the trend shown in this Figure, a redshift dependence of the
SFR based on Kennicutt calibration and on the radio flux,
already highlighted in Fig.\ref{sfrcomp}.
\begin{figure*}
\centerline{
\includegraphics[width=8.cm]{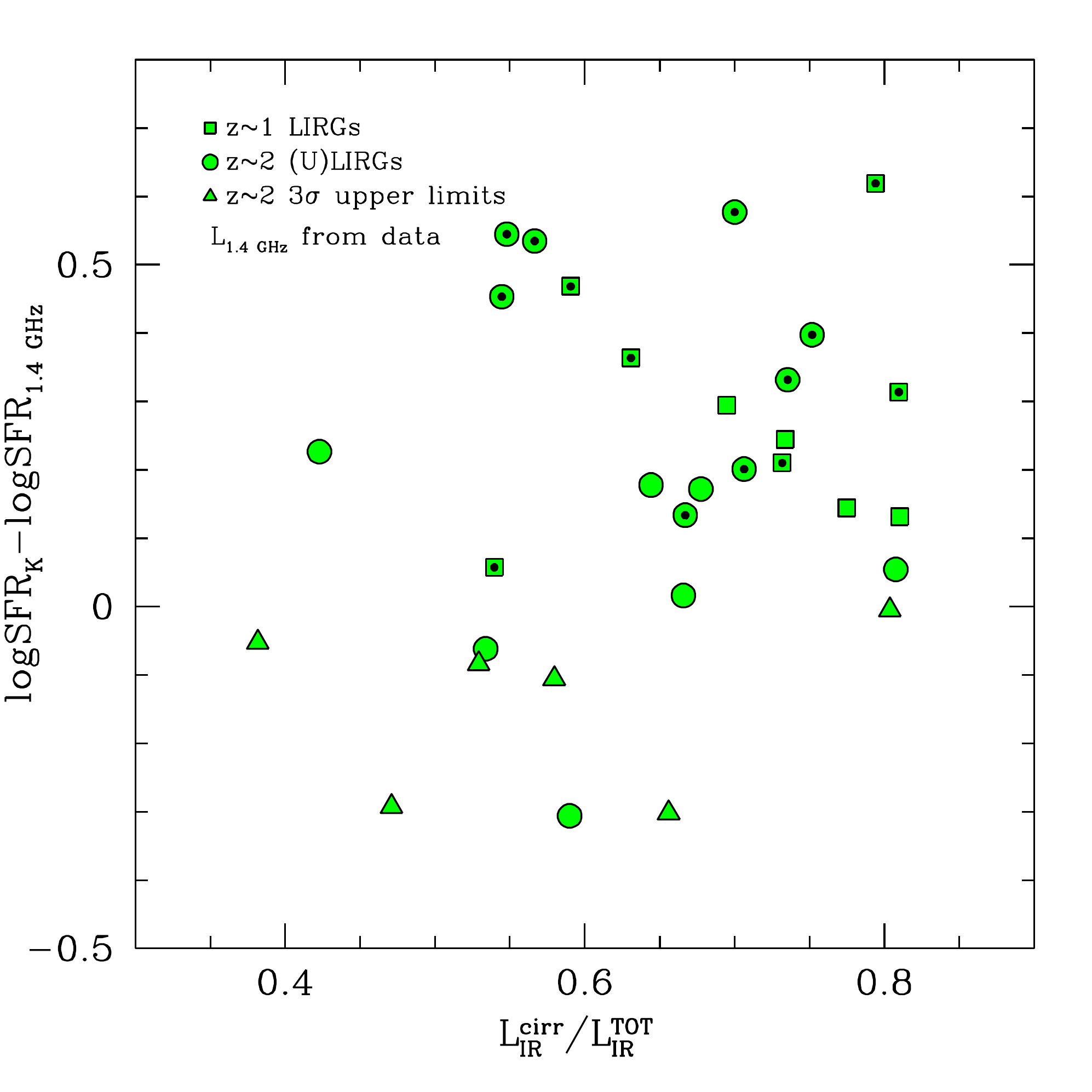}
\includegraphics[width=8.cm]{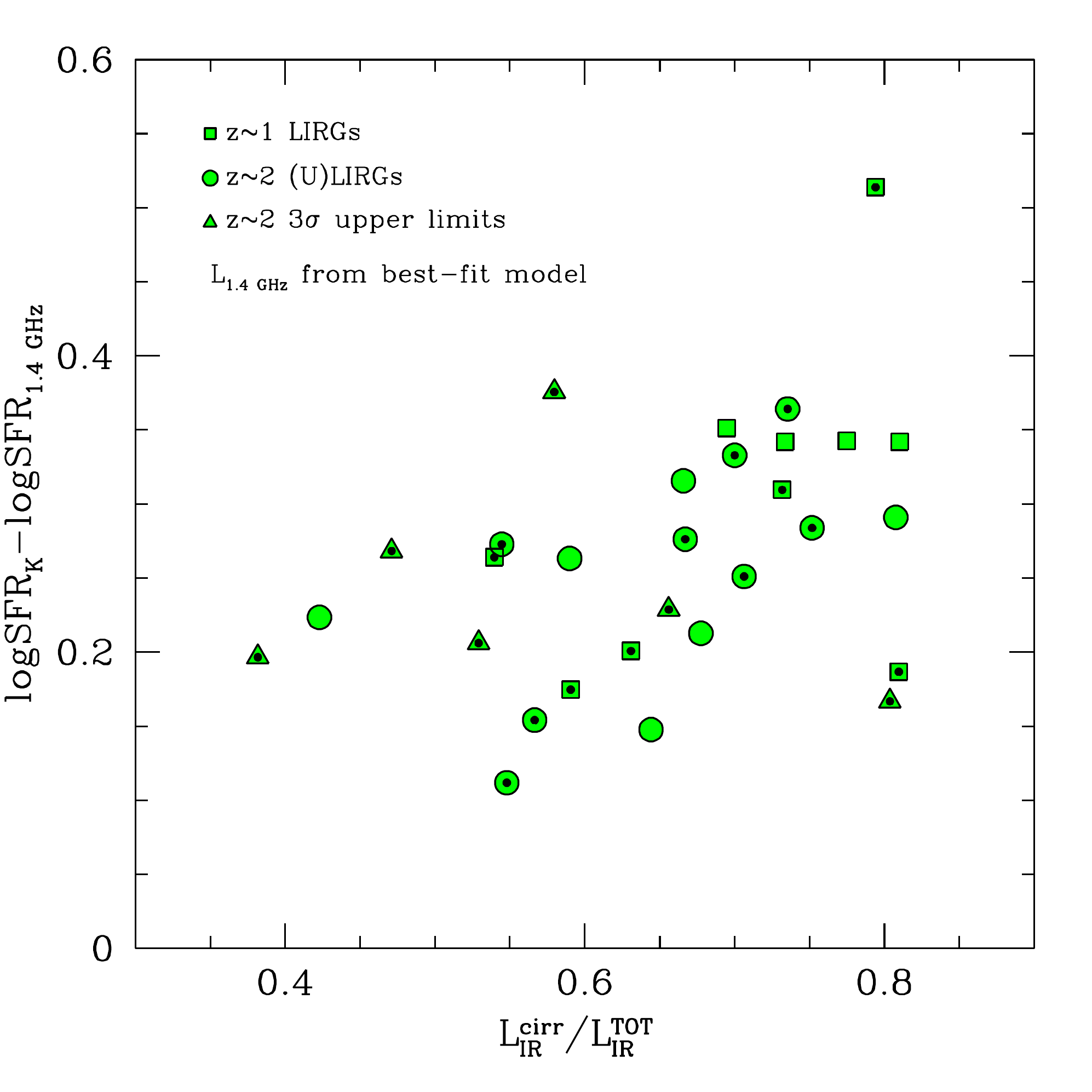}}
\caption{ Logarithmic difference of the SFR estimated from the rest-frame IR luminosity using the Kennicutt~(1998) calibration, SFR$_{K}$,
and that one derived from the rest-fame radio luminosity using the Bell~(2003) empirical relation, SFR$_{1.4 GHz}^{Bell}$, as a function of the
fractional contribution of cirrus emission to the total IR luminosity as provided by the best-fit model.
In the left panel $L_{1.4 GHz}$ is directly derived from the observed data using the relation~\ref{Bell_eq}, while on the right panel it comes from the best-fit model.
There is not a clear correlation in the left panel, due to the large scatter, rather a `trend' showing that the discrepancy between the two different
SFR estimates becomes larger, on average, for higher fractional contributions of cirrus emission to the IR luminosity. The correlation becomes, instead, clearer in the right panel where the scatter is strongly reduced as all the quantities are derived from the model.
There seems to be also a redshift dependence in the sense that being $z\sim$\,1 LIRGs (filled squares) characterized by higher cirrus fractions, on average, they also show the larger discrepancies. }
\label{sfrkrcirrus}
\end{figure*}
As we have discussed above, the FIR, differently from the radio
emission, depends on the dominant population of stars heating
the dust. We have shown
that when cirrus emission powered mostly by intermediate age stars
dominates at FIR wavelengths the
Kennicutt calibration can overestimate the current SFR of
galaxies. We have thus investigated the correlation between the
logarithmic difference of the FIR- and radio-based SFRs and the
fractional contribution of cirrus emission to the total IR
luminosity computed as $L_{\mathrm{IR}}^{\mathrm{cirr}}/L_{\mathrm{IR}}^{\mathrm{TOT}}$. It is
worth stressing again that in our model the sources responsible
for the cirrus heating are all the stars outside the MCs, i.e.
all the stars with ages older than the typical escaping time scale.
The results are shown in Figure~\ref{sfrkrcirrus} as filled
green circles ($z\sim$\,2) and squares ($z\sim$\,1). There is
not a clear correlation in the first (left) plot, due to the
large scatter, rather a `trend' showing that the discrepancy
between the two different SFR estimates becomes larger for
higher fractional contributions of cirrus emission to the IR
luminosity. A correlation appears to be more evident in the
second (right) plot where the scatter is strongly reduced as
all the quantities are derived from the model. There seems to
be also a redshift dependence in the sense that being
$z\sim$\,1 LIRGs (filled squares) characterized by higher
cirrus fractions, on average, they also show the larger
discrepancies.

The extension of our analysis also to low redshift main sequence galaxies would be necessary in order to
draw a more complete picture of the variation of the two different SFR estimators as a function also of redshift. This will be part of our future works.





%
%



\section{How does our physical analysis affect the SFR-M$_{\star}$ diagram?}
\label{sfr-m-section}

Many recent studies have found evidence that the SFR in
galaxies correlates with their stellar mass along a Main
Sequence (MS) relation which evolves with redshift and representing a
``steady'' mode of SF (e.g., \citealt{Guzman1997}; \citealt{BrinchmannEllis2000}; \citealt{Bauer2005}; \citealt{Bell2005}; \citealt{Papovich2006}; \citealt{Reddy2006}; \citealt{Noeske2007a}; \citealt{Elbaz2007}; \citealt{Daddi2007a}; \citealt{Pannella2009}; \citealt{Rodighiero2010a,Rodighiero2011}; \citealt{Karim2011}). 
Above the MS with higher sSFRs, (by a factor ranging between 4 and 10), 
there are the so-called outlier galaxies characterized by a ``starburst'' 
mode of SF generally interpreted as driven by mergers. These off-MS (or
outliers) galaxies have been found to contribute only 10\% of
the cosmic SFR density at $z \sim 2$ \citep{Rodighiero2011}.
This has been interpreted as a further indication that most of
the stellar mass forms in continuous mode of SF. Under this
picture high-$z$ LIRGs and (U)LIRGs seem to mostly reflect the
high SFR typical for massive galaxies at that epoch, so they
are not brief stochastic starbursts as their local counterpart.
They simply represent the early gas-rich phase of smoothly
declining SFH of $\geqslant L_{\star}$ galaxies, as
demonstrated by our analysis.
\begin{figure*}
\centerline{
\includegraphics[width=8.1cm]{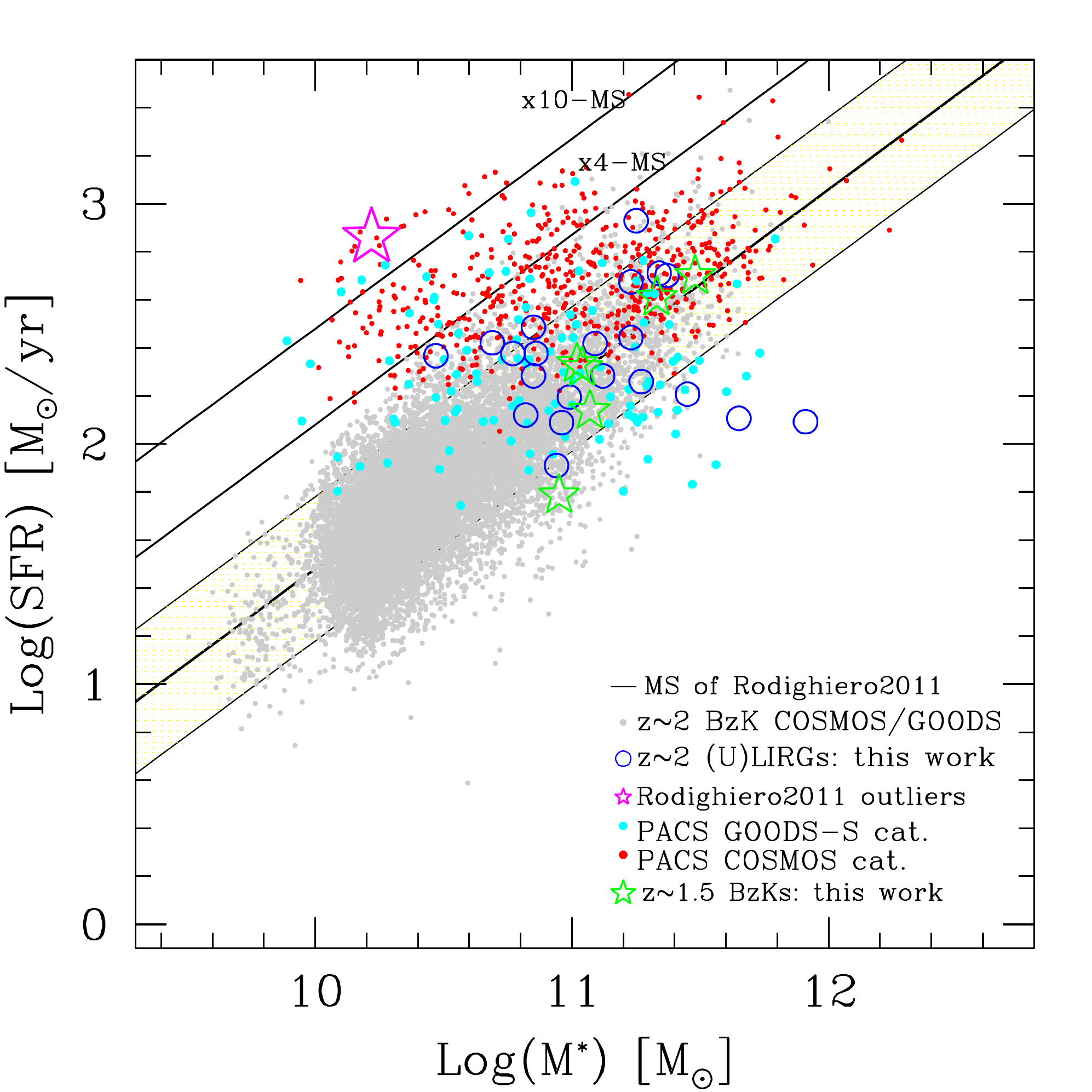}
\includegraphics[width=8.1cm]{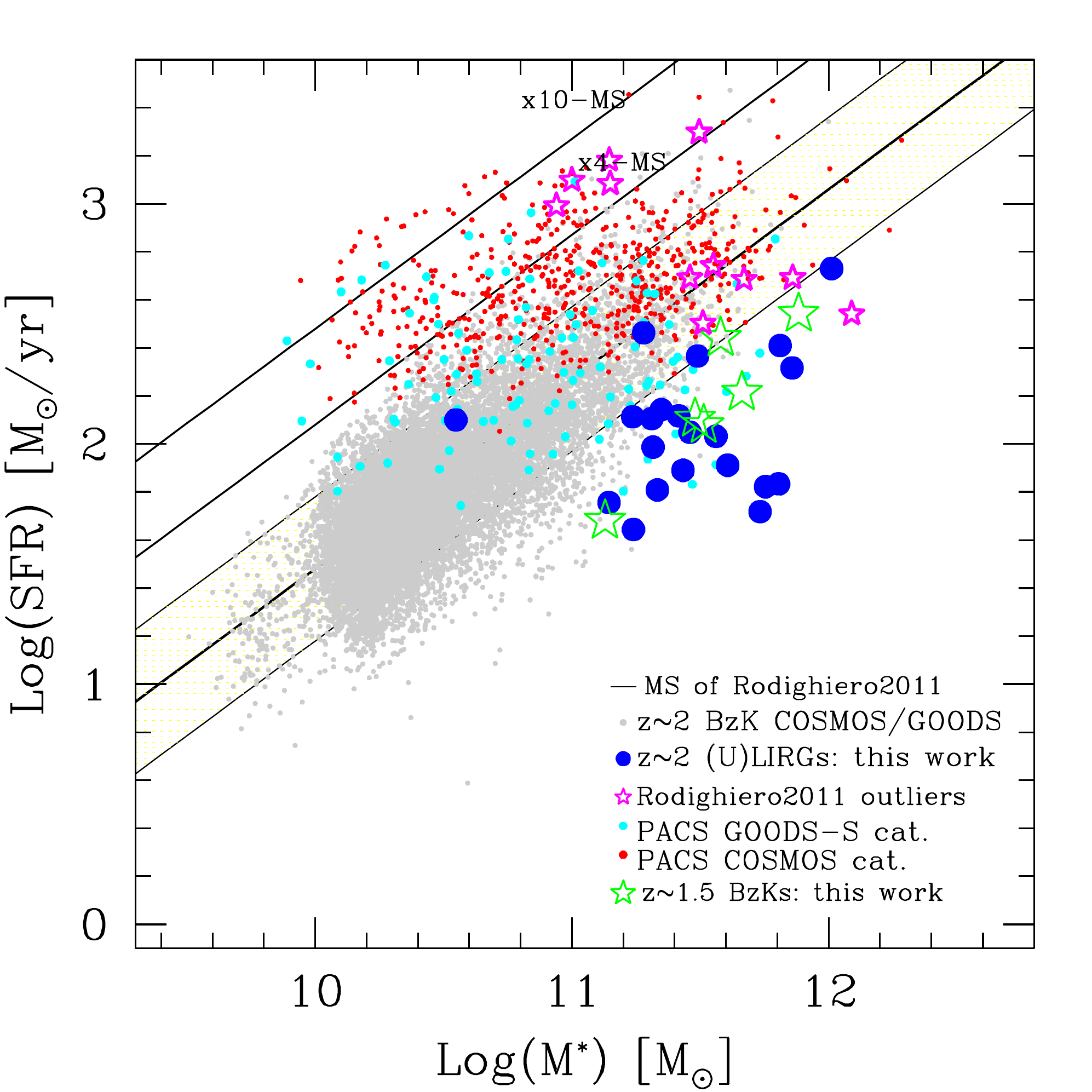}}
\caption{SFR-M$_{\star}$ diagram: GRASIL model predictions vs empirical estimates based on observations. In both panels small gray circles are $z\sim$1.5-2.0 BzK galaxies from GOODS and COSMOS survey while small red and cyan circles are, respectively, PACS-COSMOS and PACS-GOODS-S detected sources. The black solid line represents the MS relation computed by Rodighiero et al.~2011, while the yellow region defines the locus occupied by objects with sSFRs within a factor of two from the MS.
The other two black lines highlight the x4 and x10 off-MS regions. The big open blue circles and green stars shown in the left panel represent, respectively, our $z\sim$2 (U)LIRGs and $z\sim$1.5 BzKs whose stellar masses have been computed with the Hyperz code (thus based on optical-only SED-fitting procedure), and SFRs have been derived by using the empirical Kennicutt~(1998) calibration. The big blue filled circles and green thick stars shown in the right panel represent, instead, the $z\sim$2 (U)LIRGs and the six $z\sim$1.5 BzKs whose M$_{\star}$ and SFRs have been self-consistently derived here through our physical analysis. The magenta stars identify, a subsample of the x10-off-MS galaxies of Rodighiero et al.~2011. The big magenta star shown in the left-panel highlights the region where the outliers have been selected with their SFR and M$_{\star}$ derived by using classical approaches. For the small magenta stars shown in the right-panel the stellar masses and SFR come from our physical analysis.}
\label{SFR-Mstar}
\end{figure*}

The availability of radio data has allowed us to put stronger
constraints on our solutions, particularly in terms of the SFH
and current SFR of these objects. Moreover being the SFR and
M$_{\star}$ of a galaxy related one to each other and dependent
on the age distribution of stellar populations and giving the
radio constraints also on the fraction of very young stars
contributing to the NT radio luminosity through core-collapse
SNe explosions, the analysis performed in this work goes in the
direction of strengthening our predictions also in terms of the
stellar content of these galaxies. Of course the results
presented here rely on the specific choice of the IMF being a
Salpeter then converted to a Chabrier IMF by assuming the
conservative value specified in \S\,~\ref{sfhcheevo}.

We investigate here the effects of our physical analysis on the SFR-$M_{\star}$ relation described above. While our derived galaxy SFRs have been discussed thoroughly in this paper, we refer the reader to BLF13 for the details of stellar mass determination. However as the entire discussion about the placement of galaxies in the main sequence depends on how our SFR and M$_{\star}$ have been estimated, it is worth summarizing here the key results of our previous work on M$_{\star}$ measurements.

The stellar masses of all the galaxies in our sample have been self-consistently derived by BLF13 by coupling the chemical evolution models with the radiative transfer based spectral synthesis code GRASIL. The stellar mass is computed by accounting for the contribution of both still surviving stars and dead remnants.
The fit is performed taking into account the information coming from the entire SED from far-UV to radio and, as the dust extinction and reprocessing are computed by solving the differential equations for radiative transfer, the energy balance is always naturally preserved in the code. As already highlighted in Sec.~\ref{interpretation}, when comparing our stellar mass estimates to those provided by more classical SED-fitting procedures we found systematically larger values by a factor up to $\sim$6 for the most dust obscured (U)LIRGs at $z\sim2$. We also found for the same objects higher extinctions by $\Delta A_{V} \sim$ 0.81 (z$\sim$1) and 1.14 (z$\sim$2). 

We investigated in BLF13 the possible origin of the larger differences in our M$_{\star}$ estimates by considering different configurations for the factors which are supposed to be the main drivers of these discrepancies, that is to say, the different SFHs, (continuous vs starburst), evolutionary models, wavelength coverage and dust extinction and reprocessing parametrization. \cite{Michalowski2012} found in the adopted SFH, (double component vs $\tau$-model), and evolutionary models the major factor ($\sim$ 2.5 in M$_{\star}$) affecting the stellar mass estimates of high-$z$ dust obscured sub-mm galaxies when a $\tau$-model is considered. We tested all these hypothesis (see discussion in Sec.~5.1 of BLF13)
%
%
%
 and concluded that the different SFH and SSPs cannot account for the larger discrepancies we find among the most dust obscured ULIRGs at $z\sim2$; for these the dominant factor is the dust exinction. We found in fact, the stellar mass discrepancy to be larger for higher values of A$_{V}$ concluding that the stellar mass which is missed by fitting the optical data alone is hidden in dust. With our analysis we showed how extremely idealized approaches, not accounting for the information coming from the entire observed SED and thus based on optical-only SED-fitting procedures, may produce highly degenerate model solutions and galaxy SEDs unable to energetically balance the dust reprocessed IR emission from the galaxy (see e.g. Fig.7 in BLF13). For the most dust obscured objects this can result in severely underestimated stellar masses. Similar conclusions have been obtained recently also by \cite{Mitchell2013}. 

Our results, in terms of larger stellar masses and lower SFRs with respect to classical estimates, seem to have a strong impact on the SFR-$M_{\star}$ diagram.

Figure~\ref{SFR-Mstar} shows the Stellar mass-star formation
rate relation for different galaxies at $1.5 < z < 2.5$. Most
of the points are the same ones discussed in Fig.~1 of
\cite{Rodighiero2011} work (see left panel). The main thick
black solid line represents the MS relation from
\cite{Rodighiero2011}. The light yellow region (right-panel)
defines the locus occupied by objects having sSFR within a
factor of 2 from the MS while the two solid lines above the
yellow region mark, respectively, the loci 4 and 10 times above
the MS (along the SFR axis). The small gray circles are BzK
galaxies from both COSMOS and GOODS catalogues. The small red
and cyan circles represent, respectively, PACS-COSMOS and
PACS-GOODS-S detected sources. The open blue circles and green
stars shown in the left-panel represent our sample of
$z\sim$\,2 (U)LIRGs and $z\sim$\,1.5 BzKs. For all these
data-points the stellar masses have been measured by using BC03
templates in combination with the Hyperz code of
\cite{Bolzonella2000}. The SFRs are derived from the total IR
luminosity using the \citet{Kennicutt1998} calibration.

In BLF13 we have applied our physical analysis also to a sample
including some of the most extreme outliers of
\cite{Rodighiero2011} (those having sSFR above a
factor of 8-10 with respect to MS) in order to explore its effects also on
starburst classified galaxies. The big open magenta star shown
in the left-panel of Fig.~\ref{SFR-Mstar} highlights the region
of the diagram where the outliers have been selected with their
$M_{\star}$ and SFRs computed using classical approach.

The big blue filled circles and green thick stars shown in the
right-panel are, respectively, the $z\,\sim$\,2 (U)LIRGs and
the six BzKs analyzed here. Their stellar masses and SFRs have
been self-consistently computed using our physical model. The
same holds for the small magenta stars representing the
high-$z$ and highly SF outliers analyzed in BLF13.


Our physical analysis applied to high-$z$ dusty star forming (U)LIRGs and BzKs thus has the effect of
shifting downwards, below the MS, these objects with the
exception of the two (U)LIRGs showing lower $A_{V}$. The effect
strongly depends on the level of dust obscuration of the galaxy
and on the nature of SF (starburst vs quiescent SF). We have
seen in BLF13 that for the most dust obscured objects the
stellar mass can be significantly larger than that derived from
`optical-only' SED fitting approaches. As for the SFRs, we have
found that they can be systematically lower, by a factor
$\sim$\,2-2.5, with respect to those based on
\citet{Kennicutt1998} calibration due to significant cirrus
contribution to IR luminosity, powered by intermediate age
($t_{\star}\,\ga\,10-90\,Myr$) stellar populations. The
combination of these two effects on the SFR-$M_{\star}$
relation is clearly visible when we look at the so-called
outliers (magenta stars). The starting point is represented by
the big open magenta star lying above the x10-off MS black line
in the left-panel of Fig.~\ref{SFR-Mstar}. After applying our
analysis we get the magenta stars shown in the
right-panel. We can clearly distinguish in the figure two
different situations, corresponding to the two well-distinct
groups of magenta stars. For the most dust obscured objects
whose IR SEDs are dominated by an ongoing burst of star
formation in the MC (fig.~5 of BLF13), the SFRs derived from
the model are consistent with the \citet{Kennicutt1998}
calibration, within the errors, while the stellar masses are
still larger than classical predictions and for some objects
reach a factor of $\sim$\,7. The effect here is that of
shifting these most extreme cases to the region between the x4-
and x10-MS solid lines, still above the MS. For the dust
obscured objects whose IR SED is dominated by cirrus component,
thus characterized by gradual evolving SFHs, we measure SFRs
lower by a factor $\sim$\,2-2.5 than Kennicutt calibration, in
addition to large stellar masses. Here the effect of our
analysis is that of bringing back to the MS these peculiar
galaxies. Similar results, for a sample of high-$z$ sub-mm
galaxies, have been recently obtained also by
\cite{Michalowski2012}.

The question we would like to address now is:  ``Are these
``outliers'' a real population of off-MS galaxies or due to a
severe mass underestimate related to their high dust
obscuration, are they ``simply'' highly star forming MS
galaxies ?''. Moreover combining the two results discussed
above (relative to normal star forming (U)LIRGs and starburst
galaxies), it seems that the global effect of our analysis is
that of simply changing the normalization of the relation. Of
course to investigate more quantitatively this possibility we
need to apply the same physical analysis to a larger
statistical sample of both MS and starburst galaxies to see
whether there is a shift downwards (to larger stellar masses)
of the entire sequence or only a compression due to the fact
that only the most dust obscured objects have stellar masses
which have been underestimated. Moreover another important
aspect that should be taken into consideration is the `real'
effect of assuming a different IMF on the estimates of stellar
mass and SFR. We have adopted here a conservative value to scale
from one IMF to another but assuming a different stellar mass
distribution for the stars in the chemical evolution code could
affect in different way, according to their SFH, the
contribution by old and young stars to the UV-to-FIR SED. And
this could be different from the simple assumption of a fixed
scaling factor for all the stellar populations. All these
aspects will be dealt in forthcoming works.

\section{Summary \& Conclusions}
\label{conclusions}

In this paper we have extended the thorough SED analysis performed by BLF13, by including 	also the modelling of radio emission for a sample of high-$z$ (U)LIRGs and BzK-selected galaxies for which a full multiwavelength dataset from far-UV to sub-mm was already available.  

In BLF13 we have found that the values of M$_{\star}$ and SFR estimated through our approach can be, in some cases, significantly different with respect to previous works based on more classical methods (e.g. hyperz-mass). In particular we have found higher stellar masses for the most dust obscured galaxies and lower SFR by a factor $\sim$\,2 for the best fits requiring a significant contribution by intermediate-age stars (with ages older than 10-90 Myr on average) to the NIR and the FIR (through ‘cirrus’ emission).

We have investigated here the origin and relevance of the cirrus component contributing to the FIR emission of our $z\sim1-2$ (U)LIRGs by complementing their far-UV to sub-mm data with VLA radio observations at 1.4 GHz carried out in the GOODS-South field. The addition of radio luminosity to the spectral multi-band fitting is crucial to further constrain our inferred SF rate and history considering
that the radio emission is strictly connected to the most recent SF activity.

The main points of our work are summarized below:

\begin{itemize}
\item[-] We have collected for the sample of $z\sim1-2$ (U)LIRGs of BLF13 both published and unpublished 1.4 GHz VLA observations in the GOODS-S field. Given the low number of (U)LIRGs with high S/N radio data, and also to extend our analysis to a different population of galaxies, we have included in our study 6 well sampled near IR-selected BzK galaxies at $z\sim1.5$ by D10 in GOODS-N. 
\item[-] For each galaxy we have searched the best fit to their far-UV to radio SEDs, within a GRASIL-generated model library, with $\sim 10^{6}$ models, covering a large SF and dust-related parameter space.
The modelling of the radio emission includes both the thermal and non-thermal components, as described in B02.
\item[-] We are able to reproduce, well within a factor of two, the far-UV to radio emission for all the (U)LIRGs with RMS $\ga$ 3-4 $\sigma$, and 4/6 BzKs in our sample. The inclusion of the radio data in the SED fitting has required a different best fit for the sources \textit{U4812} and \textit{U5152} with respect to BLF13, with a different geometrical distribution for stars and dust for the former, and the addition of a late burst of SF for the latter, to enhance the radio emission. 

The extensive spectro-photometric coverage of our high-z galaxy sample up to radio allows us to set important constraints on the star formation history (SFH) of individual objects. For essentially all (U)LIRG and BzK objects we find evidence for a rather continuous SFR and a peak epoch of SF preceding that of the observation by a few Gyrs. This seems to correspond to a formation redshift of z$\sim$ 5-6.

We underestimate the radio data of the two BzKs, \textit{BzK-12591} and \textit{-25536}. The first one has confirmed AGN \citep{Cowie2004} and shows low value of $q_{TIR}=2.06$. The latter does not have confirmed AGN contributing to its SED but also shows the same $q$ value. The low $q$ observed for this galaxy may be indicative of a hidden AGN.
\item[-] We have compared our predicted SFR averaged over the last 10 Myr, $SFR_{10}$, to the values derived from radio luminosity using the Bell et al.~(2003) calibration, and from the IR luminosity using the
Kennicutt~(1998) calibration. We find very good agreement between our $SFR_{10}$ and the $SFR^{Bell}_{1.4 GHz}$ derived from the rest-frame radio luminosity. 
The estimated $SFR_{K}$ are instead systematically off-set by a factor 1.5-2, with respect to those provided by the radio luminosity. 
We find that the discrepancy between the FIR- and radio-based SFRs is larger at lower redshifts.
We also find that the agreement between the two different indicators seems to be, on average, better for the BzK-selected galaxies, as also found by \cite{Daddi2010}. 
We do not observe any systematic effect in our solutions (in terms of SFRs), which tend to scatter both above and below the empirical calibration. 
On the contrary, compared to the SFRs estimated from the IR calibration, we are systematically lower by a factor of 2.
\item[-] The discrepancy between the IR-and Radio-based SFRs can be explained by the fact that differently from the radio emission, the FIR depends on the dominant population of stars heating the dust.
We have shown that when an important contribution to the dust heating is provided by relatively evolved stars, usually via the cirrus emission, the Kennicutt calibration can overestimate the SFR. 
Since $z\sim1$ LIRGs are characterized by higher cirrus fractions, they also show the larger discrepancies.
The extension of our analysis also to low redshift main sequence galaxies is required in order to draw a more complete picture of the variation of the two different SFR estimators as a function of both redshift and cirrus content.

This comparison of the SFRs derived from different indicators (and methods) thus seem to go in the direction of confirming the BLF13 results, and provide an indication that radio emission more than FIR is able to accurately predict the SFR. It doesn’t seem to suffer from contamination by intermediate-age stellar populations, as it happens for the FIR where, in our solution, cirrus dominates. Of course we have to take into account the large errors associated to the radio data, but even with this in mind our results are in good agreement.
\item[-] We finally show how our results, in terms of larger stellar masses (BLF13) and lower SFRs with respect to previous estimates, can significantly affect the SFR-M$_{\star}$ diagram. 
The amount of the shift in mass or SFR strongly depends on the level of dust obscuration of the galaxy and on the nature of star formation (starburst vs gradually evolving SF). 

At high dust obscurations we estimate larger masses, but we find two different situations for the SFR: (a) for the most dust obscured objects whose IR SEDs are dominated by an ongoing burst of star formation in MCs, the SFRs derived from the model are consistent with the K98 calibration. These objects are shifted in the region between the x4- and x10-MS, still above the MS; (b) for the dust obscured objects whose SED fit does not require a starburst, our estimated SFRs are a factor $\sim$\,2 lower than the K98 calibration. We shift these objects to or below the MS. Similar results, for a sample of high-z sub-mm galaxies, have been recently obtained also by \cite{Michalowski2012}.

To quantitatively investigate if the net effect of our analysis is that of simply changing the normalization of the relation, or a compression due to the fact that only the most dust obscured
objects have stellar masses which may have been underestimated, we will need to analyse a larger statistical sample of both MS and starburst galaxies. Another important aspect now under investigation is the `real' effect of assuming a different IMF on the estimates of stellar mass and SFR. In fact assuming a different stellar mass distribution for the stars in the chemical evolution code could affect in different way, according to their SFH, the contribution by old and young stars to the UV-to-FIR SED. And this could be different from the simple assumption of a same scaling factor for all the stellar populations.
\end{itemize}
The results discussed in this paper mostly rely on VLA observations that were not completely matched in sensitivity to our Herschel and Spitzer data, and consequently only a few (U)LIRGs had
reliable radio detections. Our analysis points towards the radio flux as an essential information for interpreting star-forming galaxies at high redshifts and for recovering reliable SFH. 
The currently operational and much improved sensitivity of J-VLA promises to add
important data as soon as the relevant cosmological fields will be surveyed at faint enough limits. And, of course, forthcoming facilities like Meerkat, or future ones like SKA, will provide the full
extensive radio coverage needed for all the relevant use. 

\section*{Acknowledgments}

This paper uses data from Herschel`s photometers SPIRE
and PACS. PACS has been developed by a consortium of institutes
led by MPE (Germany) and including UVIE (Austria); KU
Leuven, CSL, IMEC (Belgium); CEA, LAM(France);MPIA (Germany);
INAF-IFSI/OAA/OAP/OAT, LENS, SISSA (Italy); IAC
(Spain). This development has been supported by the funding agencies
BMVIT (Austria), ESA-PRODEX (Belgium), CEA/CNES
(France), DLR (Germany), ASI/INAF (Italy), and CICYT/MCYT
(Spain). SPIRE has been developed by a consortium of institutes
led by Cardiff Univ. (UK) and including: Univ. Lethbridge
(Canada); NAOC (China); CEA, LAM (France); IFSI, Univ. Padua
(Italy); IAC (Spain); Stockholm Observatory (Sweden); Imperial
College London, RAL, UCL-MSSL, UKATC, Univ. Sussex (UK);
and Caltech, JPL, NHSC, Univ. Colorado (USA). This development
has been supported by national funding agencies: CSA
(Canada); NAOC (China); CEA, CNES, CNRS (France); ASI (Italy); MCINN (Spain); SNSB (Sweden); STFC,UKSA (UK); and NASA (USA). We acknowledge support from ASI (Herschel Science Contract
I/005/07/0). We thank Denis Burgarella, Veronique Buat, Alessandro Boselli, Giulia Rodighiero and Eduardo Ibar for the useful discussions which contributed to improve the paper and Maurilio Pannella to have provided us the optical-to-MIR photometry for the six BzK galaxies analysed here. We also thank the anonymous referee for his/her helpful comments which have contributed to increase the clarity of the paper. 




\bibliographystyle{mn2e}
\bibliography{Bibliography}

\bsp

\label{lastpage}
\end{document}
\bye